\documentclass[a4paper,11pt]{article}
\pdfoutput=1 
\usepackage{jheppub}
\usepackage[T1]{fontenc} 
\usepackage{braket}
\usepackage{verbatim}
\newcommand{\om} {\omega}
\newcommand{\ob} {\bar{\omega}}

\newcommand{\nr} {N}
\newcommand{\tm} {\text{TM}_\text{1}}

\newcommand{\fc} {\phi_{\scriptscriptstyle C}}
\newcommand{\fcx} {\phi_{\scriptscriptstyle C1}}
\newcommand{\fcy} {\phi_{\scriptscriptstyle C2}}
\newcommand{\fcz} {\phi_{\scriptscriptstyle C3}}
\newcommand{\fd} {\phi_{\scriptscriptstyle D}}
\newcommand{\fdx} {\phi_{\scriptscriptstyle D1}}
\newcommand{\fdy} {\phi_{\scriptscriptstyle D2}}
\newcommand{\fdz} {\phi_{\scriptscriptstyle D3}}
\newcommand{\fm} {\phi_{\scriptscriptstyle M}}
\newcommand{\fmx} {\phi_{\scriptscriptstyle M1}}
\newcommand{\fmy} {\phi_{\scriptscriptstyle M2}}
\newcommand{\fmz} {\phi_{\scriptscriptstyle M3}}
\newcommand{\nd} {\eta_{\scriptscriptstyle D}}
\newcommand{\nm} {\eta_{\scriptscriptstyle M}}

\newcommand{\yds} {y_{\scriptscriptstyle D1}}
\newcommand{\ydt} {y_{\scriptscriptstyle D3}}
\newcommand{\yms} {y_{\scriptscriptstyle M1}}
\newcommand{\ymt} {y_{\scriptscriptstyle M3}}
\newcommand{\vc} {v_{\scriptscriptstyle C}}
\newcommand{\vds} {v_{\scriptscriptstyle D1}}
\newcommand{\vdt} {v_{\scriptscriptstyle D3}}
\newcommand{\vms} {v_{\scriptscriptstyle M1}}
\newcommand{\vmt} {v_{\scriptscriptstyle M3}}
\newcommand{\kd} {k}
\newcommand{\km} {z}
\newcommand{\mc} {M_{\scriptscriptstyle C}}
\newcommand{\md} {M_{\scriptscriptstyle D}}
\newcommand{\mm} {M_{\scriptscriptstyle M}}
\newcommand{\ms} {M_{ss}}
\newcommand{\da} {d_1}
\newcommand{\db} {d_2}
\newcommand{\fa} {f_1}
\newcommand{\fb} {f_2}

\newcommand{\mw} {\mathcal{M}_w}
\newcommand{\mg} {\mathcal{M}_f}

\newcommand{\rs} {\boldsymbol{1}}

\newcommand{\rsp} {\boldsymbol{1'}}
\newcommand{\rd} {\boldsymbol{2}}
\newcommand{\rt} {\boldsymbol{3}}
\newcommand{\rtp} {\boldsymbol{3'}}

\title{ Predictive $S_4$ flavon model with $\text{TM}_1$ mixing and baryogenesis through leptogenesis }

\author[a]{Mainak Chakraborty,}
\author[b]{R. Krishnan}
\author[b]{and Ambar Ghosal}

\affiliation[a]{Department of Physics, University of Calcutta,\\92 Acharya Prafulla Chandra Road, Kolkata
700009, India }
\affiliation[b]{Saha Institute of Nuclear Physics, HBNI,\\1/AF Bidhannagar,
Kolkata 700064, India}

\emailAdd{mainak.chakraborty2@gmail.com}
\emailAdd{krishnan.rama@saha.ac.in}
\emailAdd{ambar.ghosal@saha.ac.in}

\abstract{
We use $S_4$ discrete group to construct a neutrino flavour model which leads to $TM_1$ mixing and is consistent
with the neutrino oscillation data. Using the model's constrained parameter space, we predict the values of Dirac
$CP$ phase and the light neutrino mass as $-1<\sin \delta <-0.9$ and $1.7<m_1 (\text{meV})<5.5$ respectively. 
We thoroughly examine the usefulness of this model in explaining the observed baryon asymmetry of the Universe. 
Near-maximal breaking of CP symmetry (arising due to the $\text{TM}_1$ constraint) helps us in generating adequate baryon 
asymmetry through leptogenesis. We study the evolution of the asymmetry (generated due to the decay of the heavy Majorana neutrinos) starting from the primordial Universe in two different ways (i)explicitly solving network of Boltzmann equations, 
(ii) using approximate analytic solution and we have shown the extent of their equivalence. Nearly accurate analytical
fits are used thereafter to evaluate baryon asymmetry for the whole parameter space allowed by $3\sigma$ global fit of
oscillation data and to impose a constraint on the yet unbounded mass scale parameter of Dirac neutrino mass matrix. 
Furthermore, significant contribution of $N_2$ decay in the context of flavoured leptogenesis is also estimated. 
}

\begin{document}
\maketitle
\flushbottom

\section{Introduction}
\label{sec:intro}

The study of the nature and the properties of neutrinos is key to extending our understanding of particle physics beyond the
Standard Model (SM). The phenomenon of neutrino oscillations\cite{Pontecorvo:1967fh,Gribov:1968kq} is one of the handles to learn about neutrinos. 
The story of neutrino oscillations started with the discovery of solar neutrino deficit by the Homestake experiment\cite{Davis:1968cp} 
in 1970s, which was also reported by a number of experiments in the following decades. A way to resolve the problem was 
through the neutrino oscillations under which electron neutrinos produced in the sun changed into other flavours causing the
deficit. In early 2000s, the SNO experiment\cite{Ahmad:2001an} gave the irrefutable confirmation of oscillations by showing that even
though the flux of electron neutrinos showed the deficit, it disappeared when the total flux of all neutrino flavours was taken 
into account. \\

The theoretical foundation for neutrino oscillations was laid in 1957 by Bruno Pontecorvo\cite{Pontecorvo:1967fh} 
who showed that if neutrinos had mass, they would oscillate from one flavour to another. As we know today, 
the SM fermions exist in three families or 
flavours. In the quark sector as well as in the neutrino sector, the flavour mixing arises as a consequence of the 
flavour eigenstates being a superposition of the mass eigenstates. In the neutrino sector, this superposition is described 
in terms of a unitary matrix called the PMNS matrix\cite{Maki:1962mu,Kobayashi:1973fv}. The PMNS matrix,
\begin{equation}\label{eq:pmns}
U_\text{PMNS}=\left(\begin{matrix}
c_{12} c_{13} &s_{12} c_{13} &s_{13}e^{-i\delta} \\
-s_{12} c_{23}-c_{12} s_{23} s_{13} e^{i\delta} &c_{12}c_{23} -s_{12}s_{23} s_{13} e^{i\delta} &s_{23} c_{13} \\
s_{12} s_{23}-c_{12} c_{23} s_{13}e^{i\delta} &-c_{12}s_{23}-s_{12}c_{23}s_{13}e^{i\delta} &c_{23}c_{13}
\end{matrix}\right) \left(\begin{matrix}
1 &0 &0 \\
0 &e^{i\frac{\alpha_{1}}{2}} &0 \\
0 &0 &e^{i\frac{\alpha_{2}}{2}}
\end{matrix}\right)
\end{equation}
is parameterized by three mixing angles ($\theta_{12}$, $\theta_{23}$ and $\theta_{13}$), a Dirac CP phase ($\delta$) and 
two Majorana phases\cite{Adhikary:2013bma,Samanta:2015hxa}. Neutrino oscillation probabilities depend on the three mixing angles and the Dirac phase along with
the neutrino mass-squared differences, $\Delta m^2_{21}$ and $\Delta m^2_{31}$. Starting with the SNO experiment, 
the last couple of decades saw tremendous progress in detecting various oscillation modes by a large number of solar, 
atmospheric and reactor neutrino experiments and measuring the three mixing angles and the two mass-squared differences. 
To have a clear idea (qualitatively as well as quantitatively) about the neutrino oscillation phenomenon, precision 
measurements of the above mentioned six parameters are necessary. It has to be remembered that the Majorana phases 
$(\alpha_1,\alpha_2)$ do not show up in the ordinary neutrino flavour oscillation probabilities. Their effect is reflected
only in neutrino-antineutrino oscillations\cite{deGouvea:2002gf,Xing:2013ty,Delepine:2009qg} which have not been observed yet.
While we have quite robust as well as precise 
results on mixing angles and mass squared differences, the ambiguity in Dirac CP phase $(\delta)$ and mass ordering 
(normal: $m_1<m_2<m_3$ or inverted: $m_3<m_1<m_2$) of the light neutrinos still persists. Recently, significant advancement 
has been made in measuring this $\delta$ by NO$\nu$A\cite{Adamson:2017gxd,NOvA:2018gge} and T2K\cite{Abe:2017vif,Abe:2018wpn} experiments which seems to shed light on the 
yet unsolved issue of CP violation in the Dirac neutrino mass matrix. Determination of the octant of $\theta_{23}$ is also
one of the major goals of the ongoing experiments. A global analysis\cite{Esteban:2018azc} including the latest data coming out from
the experiments MINOS\cite{Adamson:2013whj}, T2K\cite{Abe:2017vif,Abe:2018wpn}, 
NO$\nu$A\cite{Adamson:2017gxd,NOvA:2018gge}, Daya Bay\cite{Adey:2018zwh}, RENO\cite{Bak:2018ydk}, Double CHooz\cite{Abe:2014bwa}
shows that the Inverted mass ordering (IO) is disfavoured with a $\Delta \chi^2 =4.7$ ($9.3$ with SK atm) and that for $\theta_{23}$, the second octant is preferred (first octant disfavoured with $\Delta \chi^2=4.5$). The best fit value of $\delta$
has been found to be $215^\circ$ (normal ordering) whereas the CP conserving case $(\delta=180^\circ)$ is disfavoured 
with $\Delta \chi^2=1.5$. Although this global analysis\cite{Esteban:2018azc} is quite rigorous and points towards
normal ordering (NO), second octant for $\theta_{23}$ and non-zero Dirac type CP violation, more precise results are expected in the coming years. If the neutrinos are Majorana particles, it would result in double beta decays without the emission of neutrinos. Several experimental searches (EXO-200\cite{Anton:2019wmi}, GERDA\cite{Agostini:2018tnm}, KamLAND-Zen\cite{Gando:2018kyv}, CUORE\cite{Alduino:2017ehq}, CUPID\cite{Azzolini:2019tta}) seek to detect and observe such neutrinoless double beta decays ($0\nu\beta\beta$). These experiments have not yet detected $0\nu\beta\beta$, but they set upper bounds on the effective Majorana neutrino mass, $m_{\beta\beta}$. \\

Cosmological observations have already proven the overwhelming predominance of matter over antimatter in the Universe. To
the best of our knowledge any trace of appreciable amount of antimatter hasn't been found yet. The baryon asymmetry observed
in the present day Universe is assumed to be generated dynamically from a baryon symmetric state, rather than taking it as 
an initial condition (the asymmetry would have been highly fine tuned in that case). More importantly, any primordial asymmetry
would have been diluted very much during the inflationary epoch. Therefore, to generate the baryon asymmetry the other approach 
is followed, i.e the asymmetry is generated dynamically starting from a baryon symmetric era which is known as 
baryogenesis\cite{Riotto:1998bt,Cline:2006ts,Dine:2003ax}. Among the various alternatives of baryogenesis 
(Affleck Dyne baryogenesis\cite{Affleck:1984fy,Dine:1995kz}, GUT baryogenesis\cite{Ignatiev:1978uf,Ellis:1978xg})
the widely accepted and compatible to our present model is baryogenesis through 
leptogenesis\cite{Fukugita:1986hr,Riotto:1999yt,Buchmuller:2004nz,Davidson:2008bu,Bertuzzo:2009im,DiBari:2012fz,Adhikary:2014qba,Samanta:2016hcj,Samanta:2018hqm,Samanta:2018efa}. Here the asymmetry is 
created in the leptonic sector first, later which is converted into baryon asymmetry through sphaleron process\cite{Kuzmin:1985mm}.
There are three very important conditions known as Sakharov conditions\cite{Sakharov:1967dj} which are inevitable for successful 
leptogenesis to take place. They are (i) Baryon number violation, (ii) C and CP violation, (iii) departure from thermal 
equilibrium. Here, the baryon number is violated through lepton number violation which is again ensured by the presence of the Majorana
mass term. The Majorana mass matrix involves the couplings among the Right-Handed (RH) neutrinos which are theorised to 
be present in addition to the SM fermions. The RH neutrinos are SM gauge singlets and assumed to be much heavier than than the
SM fermions. The Dirac neutrino mass matrix (couplings between the RH neutrinos and the left-handed SM neutrinos) is complex
in general which acts as the source of CP violation. 
It is to be noted that non zero CP asymmetry in the leptonic sector may arise due to low energy CP phases (Dirac type denoted 
by $\delta$  or Majorana type denoted by $\alpha_1,\alpha_2$ ) or 
high energy CP phases or both. This phase dependence is best understood if the Dirac neutrino Yukawa couplings are expressed in terms of 
the Casas-Ibarra\cite{Casas:2001sr} parametrization. With the use of this parametrization we have shown in the Appendix \ref{CSI} that the most general flavoured CP
asymmetry parameter in our case is related with low energy as well as high energy CP phases and both contribute nontrivially to the 
asymmetry parameter.
The out of equilibrium condition is satisfied naturally at some epoch of
evolution of the Universe when the temperature of the thermal bath
falls below the mass of the decaying particle. So, the CP violating (and $L$ violating) out of 
equilibrium decays of heavy RH neutrinos to SM Higgs and leptons give rise to lepton asymmetry, the evolution of which
with temperature (from the scale of RH neutrino decay to the present epoch) is tackled by the set of Boltzmann equations. The 
sphaleron induced $(B+L)$ violating\cite{tHooft:1976rip,tHooft:1976snw} processes are in thermal equilibrium between the temperature range from $10^{2}$ GeV to
$10^{12}$ GeV. During this process the change in the baryon number and the lepton number is the same, i.e $\Delta B=\Delta L= N_f$ 
($N_f$ is the number of fermion generations). So the sphalerons violate $(B+L)$ keeping $(B-L)$\footnote{In case of flavoured
leptogenesis the conserved quantity is $B/3-L_\alpha$.}  conserved and as a result, a fraction of the $L$ (or $B-L$) asymmetry
is converted into baryon asymmetry. This asymmetry is expressed by $\eta_B$ (or $Y_B$) which is the measure of excess of
baryons over antibaryons scaled by photon density ($n_\gamma$) (or comoving entropy density $(s)$), i.e.
\begin{eqnarray}
&& (\eta_B )_0= \frac{n_B-n_{\bar{B}}}{n_\gamma} \Bigg |_0 = (6-6.6) \times10^{-10}, {\rm or ~equivalently}\nonumber\\
&& (Y_B )_0= \frac{n_B-n_{\bar{B}}}{s} \Bigg |_0 = (8.55-9.37) \times10^{-11}.
\end{eqnarray}
where $n_B$ ($n_{\bar{B}}$) are the number density of baryons (antibaryons) and the subscript zero denotes the value of 
the corresponding asymmetry parameter\cite{Aghanim:2016yuo} at the present epoch. 
\\


In the present work, we aim to build a model which simultaneously tackles the issues of the construction of the mass matrices
in the lepton sector as well as the generation of the baryon asymmetry through leptogenesis. The model should be consistent
with the observed neutrino mass-squared differences and the mixing. Unlike the case of the quarks, the neutrino mixing angles
are quite large. Early measurements of solar and atmospheric oscillation probabilities have pointed towards
$\sin^2 \theta_{12}\approx \frac{1}{3}$ and $\sin^2 \theta_{23}\approx \frac{1}{2}$. This led to the tri-bimaximal mixing 
(TBM)\cite{Harrison:2002er} which theorized
$\ket{\nu_2}=\frac{1}{\sqrt{3}}\ket{\nu_e}+\frac{1}{\sqrt{3}}\ket{\nu_\mu}+\frac{1}{\sqrt{3}}\ket{\nu_\tau}$ and 
$\ket{\nu_3}=\frac{1}{\sqrt{2}}\ket{\nu_\mu}-\frac{1}{\sqrt{2}}\ket{\nu_\tau}$. Even though the observation of non-zero 
reactor angle in 2012 ruled out TBM, often model builders use it as a starting point. A large number of ansatze which modify
TBM, by preserving some of its features or symmetries while breaking some others, have been proposed. One of the most 
promising among these is the $\tm$\cite{Albright:2008rp, Albright:2010ap, Xing:2006ms, Rodejohann:2012cf} mixing which preserves the first
column of TBM and leads to constraints among the mixing angles and the CP phase. A number of 
papers~\cite{Antusch:2011ic, Ge:2011qn, Varzielas:2012pa, Li:2013jya, Luhn:2013vna, King:2013xba, King:2013vna, Zhao:2015bza, King:2016yvg, King:2015dvf, Shimizu:2017fgu, Gautam:2018izb, Krishnan:2019xmk} had been published in which 
the $\tm$ mixing arises as a result of an underlying discrete symmetry. In our paper, we construct a flavon model based 
on the $S_4$ discrete symmetry. Flavons are scalar fields and they transform as multiplets under the discrete group. 
They acquire Vacuum Expectation Values (VEVs) through spontaneous symmetry breaking. These VEVs form the building blocks 
of the lepton mass matrices. The structure of the discrete group as well as the residual symmetries of the VEVs result
in specific textures of the mass matrices. The mass matrices constructed in our model lead to $\tm$ mixing and we use them 
in our study of the leptogenesis.

The plan of the paper is as follows. In Section~\ref{sec:model}, we first describe $\tm$ mixing and briefly outline the properties
of the flavour group considered, i.e. $S_4$. The SM fermions as well as the theorised flavon fields are assigned as 
multiplets under $S_4$. The Lagrangian is constructed and the flavons are given specific VEVs. Then we construct various
mass matrices and extract the lepton masses along with generating the $\tm$ mixing. In Section~\ref{baryo}, we start by expressing our mass 
matrices in the standard basis employed in the leptogenesis calculations. We then move on to describe the basic techniques 
of these calculations such as the CP asymmetry parameters and the Boltzmann equations. We discuss two versions of  
approximate analytic solutions of the Boltzmann equations. We also discuss the phenomenon of flavour decoherence and the 
significance of next to lightest RH neutrino $(N_2)$ in the context of our model. In Section~\ref{num}, the neutrino
oscillation data are fitted with the model parameters 
and their allowed ranges are found. The neutrino masses and the CP phase are predicted from the model constraints. 
The baryon asymmetry parameter is calculated in both the unflavoured and the $\tau$-flavoured regimes. This calculation 
is carried out by numerically solving the Boltzmann equations and also using the analytic approximations. This analysis is 
used to further constrain the parameter space. Finally, our results are summarised in Section~\ref{summ}.

\section{The Flavon Model}
\label{sec:model}

\subsection{$\tm$ mixing}

$\tm$ mixing can be expressed as a perturbation on the tri-bimaximal mixing, 
{\renewcommand*{\arraystretch}{1.4}
\begin{equation}\label{eq:tm1}
|U_{\tm}|^2 =  \left(\begin{matrix} \frac{2}{3} \,\,\,& \frac{1}{3}-\epsilon_1 & \epsilon_1\\
\frac{1}{6} \,\,\, & \frac{1}{3}+\frac{\epsilon_1}{2}+\frac{\epsilon_2}{2} \,\,\,& \frac{1}{2}-\frac{\epsilon_1}{2}-\frac{\epsilon_2}{2} \\
\frac{1}{6} \,\,\, & \frac{1}{3}+\frac{\epsilon_1}{2}-\frac{\epsilon_2}{2} \,\,\,& \frac{1}{2}-\frac{\epsilon_1}{2}+\frac{\epsilon_2}{2}
\end{matrix}\right),
\end{equation}}
\hspace{-1.5mm}using two parameters $\epsilon_1$ and $\epsilon_2$. By comparing the above matrix with the standard
PMNS mixing (PDG convention) matrix, Eq.~(\ref{eq:pmns}), we obtain the parameters $\epsilon_1$ and $\epsilon_2$ in terms of the 
mixing angles $\theta_{13}$ and $\theta_{23}$,
\begin{align}
\epsilon_1&=\sin^2 \theta_{13},\label{eq:epsilon1}\\
\epsilon_2&=\cos^2 \theta_{13} \cos 2\theta_{23}.\label{eq:epsilon2}
\end{align}
The reactor mixing angle is parameterized by $\epsilon_1$, since we have $\theta_{13}\rightarrow 0$ when $\epsilon_1\rightarrow 0$. Since the reactor mixing angle is observed to be non-zero, $\epsilon_1$ should not vanish. On the other hand, $\epsilon_2$ parameterizes the breaking of $\mu\text{-}\tau$-reflection
symmetry\cite{Harrison:2002kp,Harrison:2002et,Harrison:2004he,Babu:2002dz,Ma:2002ce,Grimus:2003yn, Ghosal:2004qb}, since  $\epsilon_2= 0$ leads to the $\mu$ and the $\tau$ rows of the mixing matrix, Eq.~(\ref{eq:tm1}), to become equal. Conservation of this symmetry implies $\theta_{23}=\frac{\pi}{4}$, as is evident by the vanishing of Eq.~(\ref{eq:epsilon2}). The latest experimental results point towards non-maximal atmospheric mixing and indicate the breaking of $\mu\text{-}\tau$-reflection symmetry and thus a non-vanishing $\epsilon_2$. In the $\text{TM}_1$ mixing, the solar angle can be obtained in terms of the reactor angle,
\begin{equation}\label{eq:solartm1}
\sin^2 \theta_{12}=\frac{1-3\sin^2 \theta_{13}}{3\cos^2 \theta_{13}}.
\end{equation}

The Jarlskog's $CP$ violation parameter\cite{Jarlskog:1985ht,Jarlskog:1985cw,Jarlskog:2004be} can be expressed in terms of the elements of $|U_{PMNS}|^2$ using
\begin{equation}\label{eq:jcpform}
J_{CP}^2 =\frac{1}{4}\left( \sum_i P_{\alpha i} P_{\beta i}\right)^2- \frac{1}{2} \sum_i P_{\alpha i}^2 P_{\beta i}^2\quad  \text{with} \quad \alpha\neq \beta,
\end{equation}
where $P_{\alpha i}=|U_{PMNS\,\alpha i}|^2$\cite{Harrison:2006bj}. Substituting (\ref{eq:tm1}) in (\ref{eq:jcpform}) we obtain 
\begin{equation}\label{eq:jcp2}
J_{CP}^2=\frac{1}{144}(8 \epsilon_1 -24 \epsilon_1^2-\epsilon_2^2).
\end{equation}
In terms of the mixing angles and the CP phase, we have 
\begin{equation}\label{eq:jcp}
J_{CP}=\sin \delta \sin\theta_{13}\sin\theta_{12}\sin\theta_{23}\cos^2\theta_{13}\cos \theta_{12}\cos \theta_{23}.
\end{equation}
Using Eqs.~(\ref{eq:epsilon1}, \ref{eq:epsilon2}, \ref{eq:solartm1}, \ref{eq:jcp2}, \ref{eq:jcp}), we obtain the expression for the CP phase in the $\text{TM}_1$ scenario,
\begin{equation}\label{eq:sindelta}
\sin^2 \delta = \frac{8 \sin^2 \theta_{13}(1-3\sin^2 \theta_{13})-\cos^4 \theta_{13}\cos^2 2\theta_{23}}{8 \sin^2 \theta_{13}\sin^2 2\theta_{23} (1-3\sin^2 \theta_{13})}~.
\end{equation}
From the above expression, we can see that when $\theta_{23}=\frac{\pi}{4}$ we obtain $\sin^2 \delta = 1$, i.e. $\delta = \pm \frac{\pi}{2}$. Therefore, for $\tm$ mixing $\mu\text{-}\tau$-reflection symmetry implies maximum CP violation.

The $\tm$ mixing obtained in this paper has non-vanishing values of both $\epsilon_1$ and $\epsilon_2$. Therefore, we have non-zero mixing angle, breaking of $\mu\text{-}\tau$-reflection symmetry (non-maximal atmospheric mixing) and non-maximal CP violation.

\subsection{$S_4$ group}

We construct the model in the framework of the discrete group $S_4$ which has been studied extensively in the literature \cite{Brown:1984dk, Lee:1994qx, Mohapatra:2003tw, Ma:2005pd, Hagedorn:2006ug, Zhang:2006fv, Caravaglios:2006aq, Koide:2007sr, Varzielas:2012pa, Krishnan:2012me, Krishnan:2012sb, Luhn:2013vna, Li:2013jya, Shimizu:2017fgu, King:2016yvg, King:2015dvf}.
Here we briefly mention the essential features of this group in the context of model building. $S_4$ is the rotational symmetry group of the cube. $S_4$ has 24 elements which 
fall under four conjugacy classes. Its conjugacy classes and irreducible representations are listed in Table~\ref{tab:char}. 
{\renewcommand{\arraystretch}{1.2}
\begin{table}[h]
\begin{center}
\begin{tabular}{|c|c c c c c|}
\hline
&$(1)$ & $(12)(34)$	& $(12)$ & $(1234)$ & $(123)$\\
\hline
$\rs$ & $1$ & $1$ & $1$ & $1$ & $1$ \\
$\rsp$ & $1$ & $1$ & $-1$ & $-1$ & $1$  \\
$\rd$ & $2$ & $2$ & $0$ & $0$ & $-1$ \\
$\rt$ & $3$ & $-1$ & $1$ & $-1$& $0$ \\
$\rtp$ & $3$ & $-1$ & $-1$ & $1$& $0$ \\
\hline
\end{tabular}
\end{center}
\caption{The character table of $S_4$ group.}
\label{tab:char}
\end{table}}

$S_4$ group can be generated using 
\begin{equation}\label{eq:gens}
S =  \left(\begin{matrix} 1 & 0 & 0\\
0 & -1 & 0 \\
0 & 0 & -1
\end{matrix}\right), \quad T =  \left(\begin{matrix} 0 & 1 & 0\\
0 & 0 & 1 \\
1 & 0 & 0
\end{matrix}\right), \quad U =  \left(\begin{matrix} -1 & 0 & 0\\
0 & 0 & -1 \\
0 & -1 & 0
\end{matrix}\right).
\end{equation}
The above matrices represent a convenient basis for the triplet ($\rt$) of $S_4$. The tensor product of two triplets $(x_1, x_2, x_3)$ and $(y_1, y_2, y_3)$ leads to 
\begin{equation}
\rt \times \rt = \rs+\rd+\rtp+\rt
\end{equation}
with
\begin{align}
\rs& \equiv x_1 y_1 + x_2 y_2 + x_3 y_3\,, \label{eq:tp1}\\
\rd& \equiv \left(2 x_1 y_1 - x_2 y_2 - x_3 y_3, \sqrt{3}(x_2 y_2 -x_3 y_3)\right), \label{eq:tp2}\\
\rtp& \equiv (x_2 y_3+x_3 y_2, x_3 y_1+x_1 y_3, x_1 y_2+x_2 y_1)\,, \label{eq:tp3p}\\
\rt& \equiv (x_2 y_3-x_3 y_2, x_3 y_1-x_1 y_3, x_1 y_2-x_2 y_1)\,. \label{eq:tp3}
\end{align}
For the triplet ($\rtp$), the generators are given by $S'=S$, $T'=T$ and $U'=-U$.

\subsection{The Model Lagrangian}
\label{subsec:lagrangian}

Table~\ref{tab:flavourcontent} provides the field content of our model. The three families of the left-handed-weak-isospin lepton doublets, $L$, and the three right-handed heavy neutrinos, $\nr$, form triplets ($\rt$) under $S_4$. The flavons $\fc$, $\nd$, $\fd$, $\nm$ and $\fm$, are scalar fields and are gauge invariants. They transform as singlets ($\nd, \nm$) and triplets, $\rt$ ($\fc$) and $\rtp$ ($\fd, \fm$), under $S_4$. 
The $C_3$ and the $C_6$ groups are introduced so that the various flavons couple only in the intended mass terms. The flavon $\fc$ 
transforms as $\om$ under $C_3$ and it couples in the charged-lepton mass terms. $C_3$ also helps in defining the vacuum alignment 
of $\fc$. The flavons $\nd$, $\fd$ and $\nm$, $\fm$ transform as $-1$ and $\om$ respectively under $C_6$ and they couple in the
Dirac and the Majorana sectors respectively of the neutrino mass terms. 
{\renewcommand{\arraystretch}{1.4}
\begin{table}[h]
\begin{center}
\begin{tabular}{|c|c c c c c c c c c c c|}
\hline
&$L$ & $e_R$	&$\mu_R$&$\tau_R$	&$\nr$&$\fc$&$\nd$&$\fd$&$\nm$&$\fm$&$H$\\
\hline
$S_4$ &$\rt$ & $1$&$1$&$1$&$\rt$&$\rt$&$\rs$&$\rtp$&$\rs$&$\rtp$&$\rs$\\
$ C_3$ &$ 1$ & $1$ &$\om$&$\ob$&$1$&$\om$&$1$&$1$&$1$&$1$&$1$\\
$ C_6$ &$1$ & $1$ &$1$&$1$&$-\om$&$1$&$-1$&$-1$&$\om$&$\om$&$1$\\
\hline
\end{tabular}
\end{center}
\caption{The flavour structure of the model. The complex cube roots of unity, $e^{i\frac{2\pi}{3}}$ and $e^{-i\frac{2\pi}{3}}$, are represented by $\om$ and $\ob$ respectively.}
\label{tab:flavourcontent}
\end{table}}

Given the field assignments, Table~\ref{tab:flavourcontent}, we write the Lagrangian,
\begin{align}\label{eq:lagr}
\begin{split}
{\mathcal{L}}=\,\,\,&y_\tau \bar{L} \frac{\fc}{\Lambda} \tau_R H+y_\mu \bar{L} \frac{\fc^*}{\Lambda} \mu_R H+y_e \bar{L} \frac{(\fc^*\fc)_{\rt}}{\Lambda^2} e_R H+\yds \bar{L} \nr \frac{\nd}{\Lambda} \widetilde{H}+ +\ydt \left( \bar{L} \nr \right)_{\rtp} \frac{\fd}{\Lambda} \widetilde{H}\\
&+\yms  \bar{\nr^c} \nr \,\nm + \ymt \left( \bar{\nr^c} \nr \right)_{\rtp} \fm,
\end{split}
\end{align}
where $y_i$ with $i=\tau, \mu, e, D_1, D_3, M_1, M_3$ are the coupling constants. $\left(\right)_{\rtp}$ and $\left(\right)_{\rt}$ denote the symmetric and the antisymmetric tensor products which transform as $\rtp$ and $\rt$, Eqs.~(\ref{eq:tp3p}, \ref{eq:tp3}). Note that $\fc$ couples in the charged-lepton mass term, $\nd$, $\fd$ couple in the neutrino Dirac mass term and $\nm$, $\fm$ couple in the neutrino Majorana mass term. We assume that the coupling constants in the model, $y_i$, are real numbers, i.e. we do not introduce CP violation explicitly. Rather, CP is broken spontaneously by the flavon vacuum alignments. Because of the complex $C_6$ assignments, Table~\ref{tab:flavourcontent}, the flavons $\fc$, $\nm$, $\fm$ have complex degrees of freedom. In contrast, $\nd$ and $\fd$ are real fields.

The scalar fields acquire VEVs through Spontaneous Symmetry Breaking (SSB). The Higgs VEV is the familiar $(0,v)$ where $v\approx170$~GeV. For the flavon fields, we assign the following VEVs,
\begin{align}
\langle\fc\rangle&= \vc (1,\om,\ob)\,, \label{eq:vc}\\
\langle\nd\rangle&= \vds \,, \label{eq:vd1}\\
\langle\fd\rangle&= \vdt (-1,-1,1)\,, \label{eq:vd3}\\
\langle\nm\rangle&= \vms  e^{i\xi_1}\,, \label{eq:vm1}\\
\langle\fm\rangle&= \vmt  e^{i\xi_3}(0,1,0)\,. \label{eq:vm3}
\end{align}
The alignments of $\langle\fc\rangle$, $\langle\fd\rangle$ and $\langle\fm\rangle$ in their respective triplet flavour spaces are fully determined by their residual symmetries which we will discuss in the next section. We also obtain the charged-lepton and the neutrino mass matrices in terms of the coupling constants and the VEVs of the scalar fields. The residual symmetries of the flavon VEVs manifest as the symmetries of the mass matrices.
In Appendix~\ref{pots}, we construct the flavon potentials whose minimisation leads to these VEVs.
\subsection{The mass matrices}\label{sec:massmatrices}

\subsubsection{Charged-lepton mass matrix}

In the charged-lepton sector, $\bar{L}$ which transforms as $\rt$ couples with the flavon triplet $\fc$ which also transforms as $\rt$. Since $\fc$, $\fc^*$ and $(\fc^*\fc)_{\rt}$ transform as $\om$, $\ob$, and $1$ respectively under $C_3$, they couple with $\tau_R$, $\mu_R$ and $e_R$ respectively. Note that the quadratic term, $(\fc^*\fc)_{\rt}$, needs to be included because neither $\fc$ nor $\fc^*$ can couple in the electron sector. The vacuum alignment of this term is calculated by taking the antisymmetric product of $\langle \fc^* \rangle = \vc (1,\ob,\om)$ and $\langle \fc\rangle = \vc (1,\om,\ob)$ using Eq.~(\ref{eq:tp3}),
\begin{equation}
\langle (\fc^*\fc)_{\rt} \rangle = i \sqrt{3} v_C^2 \,(1, 1, 1).
\end{equation}
Substituting the Higgs VEV, $\langle H\rangle=(0,v)$, and the flavon VEVs, $\langle\fc^*\rangle$, $\langle\fc\rangle$, $\langle (\fc^*\fc)_{\rt} \rangle$ in
\begin{equation}
y_\tau \bar{L} \frac{\fc}{\Lambda} \tau_R H+y_\mu \bar{L} \frac{\fc^*}{\Lambda} \mu_R H+y_e \bar{L} \frac{(\fc^*\fc)_{\rt}}{\Lambda^2} e_R H,
\end{equation}
we obtain the charged-lepton mass term, 
\begin{equation}\label{eq:leptcontrib}
\bar{l_L}\mc l_R
\end{equation}
where 
\begin{equation}\label{eq:leftright}
l_L = (e_L, \mu_L, \tau_L)^T, \quad l_R = (e_R, \mu_R, \tau_R)^T,
\end{equation}
\begin{equation}\label{eq:clmassmatrix}
\mc =i\sqrt{3} v\frac{v_C^2}{\Lambda^2}\left(\begin{matrix} y_e & 0 & 0\\
y_e & 0 & 0\\
y_e & 0 & 0
\end{matrix}\right)+ v\frac{v_C}{\Lambda}\left(\begin{matrix} 0 & y_\mu & y_\tau \\
0 & \ob y_\mu & \om y_\tau \\
0 & \om y_\mu & \ob y_\tau 
\end{matrix}\right).
\end{equation}
A charged-lepton mass matrix of the same form as Eq.~(\ref{eq:clmassmatrix}), was recently obtained in Refs.~\cite{Krishnan:2019ftw, Krishnan:2019xmk}. 

$\mc$, has the cyclic symmetry generated by the group element $T$, Eq.~(\ref{eq:gens}),
\begin{equation}\label{eq:mcsym}
\begin{split}
&T \mc \text{Diag}(1, \om, \ob) = \mc,\\
&T \mc \mc^\dagger T^\dagger = \mc \mc^\dagger.
\end{split}
\end{equation}
$T$ has eigenvectors $(1,1,1)$, $(1, \om, \ob)$, $(1, \ob, \om)$ corresponding to the eigenvalues $1$, $\om$, $\ob$. Therefore, $\ob T$ generates the residual symmetry of $\langle \fc \rangle$,
\begin{equation}
\ob T \langle \fc \rangle = \langle \fc \rangle.
\end{equation}
The $C_3$ group generated by $\ob T$, i.e.~$\{\ob T, \om T^2, I \}$ is a subgroup of our flavour group, $S_4\times C_3 \times C_6$. $\langle\fc\rangle$ can be uniquely defined (upto a scale factor) as the alignment that uniquely breaks the flavour group into the above mentioned $C_3$ subgroup. The cyclic symmetry of $\mc$, Eq.~(\ref{eq:mcsym}), is in fact the consequence of the $C_3$ residual symmetry of $\langle\fc\rangle$.

Using the unitary matrix,
\begin{equation}\label{eq:utbm}
V =  \frac{1}{\sqrt{3}}\left(\begin{matrix} 1 & 1 & 1\\
1 & \om & \ob \\
1 & \ob & \om
\end{matrix}\right),
\end{equation} 
we diagonalize the charged-lepton mass matrix,
\begin{equation}\label{eq:mcdiag}
V \mc \,\text{diag}(-i, 1, 1) = \text{Diag(}m_e, m_\mu, m_\tau\text{)}
\end{equation}
where
\begin{equation}
m_e  = 3 y_e v \frac{v_C^2}{\Lambda^2}, \quad m_\mu =\sqrt{3}y_\mu v \frac{v_C}{\Lambda}, \quad m_\tau=\sqrt{3}y_\tau v \frac{v_C}{\Lambda},
\end{equation}
are the charged-lepton masses. Compared to the muon and the tau masses, the electron mass is suppressed by an additional factor of $\frac{v_C}{\Lambda}$. This is similar to the Froggatt-Nielsen Mechanism of obtaining the mass hierarchy. The matrix $V$, Eq.~(\ref{eq:utbm}), is often referred to as the $3\times3$ trimaximal matrix or the magic matrix. The absolute value of every element of this matrix is equal to $\frac{1}{\sqrt{3}}$. The middle column of the magic matrix coincides with the middle column of the TBM mixing matrix and often the magic matrix plays an important role in neutrino model building.

\subsubsection{Neutrino Dirac mass matrix}

Substituting the Higgs VEV, $\langle H\rangle=(0,v)$, and the flavon VEVs, $\langle\nd\rangle= \vds$, $\langle\fd\rangle= \vdt (-1,-1,1)$, in
\begin{equation}
\yds \left( \bar{L} \nr \right)_{\rs} \frac{\nd}{\Lambda} \widetilde{H}+\ydt \left( \bar{L} \nr \right)_{\rtp} \frac{\fd}{\Lambda} \widetilde{H},
\end{equation}
we obtain the neutrino Dirac mass term, 
\begin{equation}\label{eq:dircontrib}
\bar{\nu}_L\md \nr
\end{equation}
where 
\begin{equation}\label{eq:leftrightneutrino}
\nu_L = (\nu_e, \nu_\mu, \nu_\tau)^T, \quad 
\nr = (N_3, N_2, N_1)^T,
\end{equation}
\begin{equation}\label{eq:md}
\md =v\frac{1}{\Lambda} \left(\begin{matrix} \yds \vds &  \ydt \vdt & -\ydt \vdt \\
\ydt \vdt  &  \yds \vds & -\ydt \vdt \\
-\ydt \vdt  & -\ydt \vdt  &  \yds \vds
\end{matrix}\right).
\end{equation}
The three right-handed neutrino components in Eq.~(\ref{eq:leftrightneutrino}) are named in the reverse order in anticipation of the leptogenesis calculations where the lightest component is named $N_1$ and so on. $\md$, Eq.~(\ref{eq:md}), is symmetric even though, in general, the neutrino Dirac mass matrix need not be so. The antisymmetric term $\left( \bar{L} \nr \right)_{\rt}$ does not appear in our Lagrangian because a flavon transforming as $\rt$ which can couple in the neutrino sector does not exist in our model. As a result, $\md$ turns out to be symmetric. We rewrite $\md$ as 
\begin{equation}\label{eq:md2}
\md =\mw \left(\begin{matrix} 1 &  \kd & -\kd \\
\kd  &  1 & -\kd \\
-\kd  & -\kd &  1
\end{matrix}\right), 
\end{equation}
where
\begin{equation}\label{eq:kdkm}
\mw= \yds v \frac{\vds}{\Lambda}, \quad  \kd=\frac{\ydt \vdt}{\yds \vds}.
\end{equation}
$\mw$ has the dimension of mass and is at the scale of the SM fermion masses while $\kd$ is dimensionless and is of the order of one. Both $\mw$ and $\kd$ are real numbers because the flavons, $\nd$, $\fd$, are real fields and the coupling constants are assumed to be real.

Consider the group elements,
\begin{equation}\label{eq:d6gen}
T'U'T'^2 =\left(\begin{matrix} 0 &  1 & 0 \\
1  &  0 & 0 \\
0  & 0 &  1
\end{matrix}\right), \quad S'T' =\left(\begin{matrix} 0 &  1 & 0 \\
0  &  0 & -1 \\
-1  & 0 &  0
\end{matrix}\right).
\end{equation}
They generate the $C_2$ and the $C_3$ groups, $\{T'U'T'^2, I \}$ and $\{S'T', (S'T')^2, I \}$, respectively. They correspond to the following residual symmetries of $\langle\fd\rangle$: 
\begin{equation}\label{eq:d6resvev}
T'U'T'^2\langle\fd\rangle= \langle\fd\rangle, \quad S'T' \langle\fd\rangle= \langle\fd\rangle.
\end{equation}
These residual symmetries uniquely define the alignment of $\langle\fd\rangle$. It can be shown that the group elements given in Eq.~(\ref{eq:d6gen}), when taken together, generate the dihedral group $D_6$, 
\begin{equation}\label{eq:d6full}
\{T'U'T'^2, S'T', (S'T')^2,  (T'U'T'^2)(S'T'), (S'T')(T'U'T'^2), I\},
\end{equation}
which forms a subgroup of $S_4$. In other words, $\langle\fd\rangle$ breaks $S_4$ into one of its $D_6$ subgroups which in turn uniquely defines $\langle\fd\rangle$. Note that the residual symmetries of the VEV, Eqs.~(\ref{eq:d6resvev}), manifests as the symmetries of the mass matrix, Eq.~(\ref{eq:md2}), as well,
\begin{equation}\label{eq:mdsym}
(TUT^2) \md (TUT^2)^T = \md, \quad (ST) \md (ST)^T = \md.
\end{equation}

\subsubsection{Neutrino Majorana mass matrix}

Substituting the flavon VEVs $\langle\nm\rangle= \vms e^{i\xi_1}$ and $\langle\fm\rangle= \vmt e^{i\xi_3}(0,1,0)$ in 
\begin{equation}
\yms \left( \bar{\nr^c} \nr \right)_{\rs} \nm + \ymt \left( \bar{\nr^c} \nr \right)_{\rtp} \fm,
\end{equation}
we obtain the neutrino Majorana mass term,
\begin{equation}\label{eq:majcontrib}
\bar{\nr^c}\mm \nr,
\end{equation}
where 
\begin{equation}\label{eq:mm}
\mm =\left(\begin{matrix} \yms \vms  e^{i\xi_1} &  0 & \ymt \vmt  e^{i\xi_3}\\
0  &  \yms \vms e^{i\xi_1}& 0 \\
\ymt \vmt  e^{i\xi_3} & 0  &  \yms \vms  e^{i\xi_1}
\end{matrix}\right)
\end{equation}
is the Majorana mass matrix. We rewrite $\mm$ as 
\begin{equation}\label{eq:mm2}
\mm =\mg  \left(\begin{matrix} 1 &  0 & \km \\
0  &  1 & 0 \\
\km & 0 &  1
\end{matrix}\right), 
\end{equation}
where
\begin{equation}\label{eq:kdkm}
\mg= \yms \vms  e^{i\xi_1}, \quad \km= \frac{\ymt \vmt}{\yms \vms} e^{i(\xi_3-\xi_1)}.
\end{equation}
$\mg$ has the dimension of mass and is at the scale of flavon VEV. We assume that this scale is quite high $ \approx 10^{12}$~GeV. This leads to the suppression of the light neutrino masses through the 
Type-1\cite{Minkowski:1977sc,GellMann:1980vs,Yanagida:1980xy,Mohapatra:1979ia}
seesaw mechanism. The parameter $\km$ is dimensionless and is of the order of one. 

Consider the group elements,
\begin{equation}\label{eq:c22gen}
T'^2U'T' =\left(\begin{matrix} 0 &  0 & 1 \\
0  &  1 & 0 \\
1 & 0 &  0
\end{matrix}\right), \quad T'^2S'T' =\left(\begin{matrix} -1 & 0 & 0 \\
0  &  1 & 0 \\
0  & 0 &  -1
\end{matrix}\right).
\end{equation}
They generate the $C_2$ groups, $\{T'^2U'T', I \}$ and $\{T'^2S'T', I \}$ respectively. When taken together, they generate the $C_2\times C_2$ group,
\begin{equation}\label{eq:c22full}
\{T'^2U'T', T'^2S'T', (T'^2U'T')(T'^2S'T'), I \}.
\end{equation}
This group represents the residual symmetries of $\langle\fm\rangle$,
\begin{equation}
T'^2U'T' \langle\fm\rangle= \langle\fm\rangle, \quad T'^2S'T' \langle\fm\rangle= \langle\fm\rangle,
\end{equation}
which uniquely defines $\langle\fm\rangle$. Correspondingly, we have the following symmetries for the Majorana mass matrix, Eq.~(\ref{eq:mm2}):
\begin{equation}\label{eq:mmsym}
(T^2UT) \mm (T^2UT)^T = \mm, \quad (T^2ST) \mm (T^2ST)^T = \mm.
\end{equation}
The second $C_2$ symmetry in Eqs.~(\ref{eq:mmsym}) is responsible for the off-diagonal zeros in $\mm$.

\subsubsection{Effective seesaw mass matrix}

The effective seesaw mass matrix is given by
\begin{equation}\label{eq:ms}
\ms= -\md \mm^{-1} \md^T.
\end{equation}
Substituting Eqs.~(\ref{eq:md2}, \ref{eq:mm2}) in Eq.~(\ref{eq:ms}), we obtain
\begin{equation}\label{eq:msfinal}
\ms =- \frac{\mw^2}{\mg} \left(\begin{matrix} \da &  -\fa & \fb \\
-\fa  &  \db & \fa \\
\fb & \fa &  \da
\end{matrix}\right), 
\end{equation}
where
\begin{align}
\da&= \frac{1 + 2 \kd \km + \kd^2 (2 - \km^2)}{1 - \km^2} , &\db&= \frac{1-\km+2 \kd^2}{1-\km},  \label{eq:dab}\\
\fa&= \frac{-\kd (2-\km)-\kd^2}{1-\km}, &\fb&=  \frac{-\km-2\kd-\kd^2(1+\km-\km^2)}{1 - \km^2}.  \label{eq:fab}
\end{align}
In the previous discussion, we have shown that the Dirac mass matrix and the Majorana mass matrix break the original $S_4$ group into the $D_6$ group, Eq.~(\ref{eq:d6full}), and the $C_2\times C_2$ group, Eq.~(\ref{eq:c22full}), respectively. Besides the identity, the only common element in Eq.~(\ref{eq:d6full}) and  Eq.~(\ref{eq:c22full}) is $(T'U'T'^2)(S'T') = (T'^2U'T')(T'^2S'T')$ which is nothing but
\begin{equation}
T'^2U'S'T' = \left(\begin{matrix} 0 &  0 & -1 \\
0  &  1 & 0 \\
-1 & 0 &  0
\end{matrix}\right).
\end{equation}
This element generates the $C_2$ group, $\{T'^2U'S'T', I\}$. Hence, the seesaw mass matrix, Eq.~(\ref{eq:msfinal}), constructed from both the Dirac and the Majorana mass matrices possesses only the $C_2$ residual symmetry, 
\begin{equation}
(T'^2U'S'T') \ms (T'^2U'S'T')^T = \ms.
\end{equation}

Using the $(1,3)$-bimaximal unitary matrix,
\begin{equation}\label{eq:ubm}
U_{\text{BM}}=\left(\begin{matrix} \frac{1}{\sqrt{2}} &  0 & \frac{-1}{\sqrt{2}} \\
0  &  1 & 0 \\
\frac{1}{\sqrt{2}} & 0 &  \frac{1}{\sqrt{2}}
\end{matrix}\right),
\end{equation}
we block diagonalize the seesaw mass matrix,
\begin{equation}
U_{\text{BM}}^\dagger \ms U_{\text{BM}}^\ast = -\frac{\mw^2}{\mg} \left(\begin{matrix} \da+\fb & 0 & 0 \\
0  &  \db & \sqrt{2}\fa \\
0 & \sqrt{2}\fa &  \da-\fb
\end{matrix}\right).
\end{equation}
This matrix is diagonalized using a unitary matrix with vanishing entries in $(12)$, $(13)$, $(21)$,$(31)$ positions. We call this unitary matrix $U_{23}$. Therefore, the complete diagonalization of the seesaw mass matrix is given by
\begin{equation}\label{eq:msdiag}
U_{23}^\dagger U_{\text{BM}}^\dagger \ms U_{\text{BM}}^\ast U_{23}^\ast = \text{Diag(}m_1, m_2, m_3\text{)},
\end{equation}
where $m_1$, $m_2$, $m_3$ are the light neutrino masses.

The neutrino mixing matrix, $U_\text{PMNS}$, is obtained as the product of the unitary matrices that diagonalize the charged-lepton mass matrix, $V$, Eq.~(\ref{eq:mcdiag}), and the effective seesaw mass matrix, $U_{\text{BM}}U_{23}$, Eq.~(\ref{eq:msdiag}),
\begin{equation}
U_{\text{PMNS}}=VU_{\text{BM}}U_{23}.
\end{equation}
$VU_\text{BM}$ is nothing but the tri-bimaximal mixing matrix, $U_\text{TBM}$. The multiplication of $U_\text{TBM}$ with  $U_{23}$ mixes the $2^\text{nd}$ and the $3^\text{rd}$ columns of $U_{\text{TBM}}$ giving rise to $U_\text{TM1}$, Eq.~(\ref{eq:tm1}).\\

Having discussed the $S_4$-symmetry-motivated neutrino mass model followed by the diagonalization of the light neutrino
mass matrix, now it is evident that this model is capable of generating large mixing. Whether the numerical values of these observables (as predicted by our model) are consistent with the latest global fit data of neutrino oscillations will be explored in Section~\ref{num}. The existence of the heavy RH neutrino and the presence
of its Majorana type mass term ensure lepton number violation. The complex mass matrix in the neutrino Dirac sector \footnote{Even though the mass matrix given in Eq.~(\ref{eq:md}) is real, it becomes complex in the standard basis where the charged-lepton and the Majorana neutrino mass matrices are diagonal. The corresponding basis transformation is explained in Section 3.} 
acts as the source of CP violation which is a necessary ingredient of the asymmetry generation. The detailed theoretical
framework to account for this asymmetry and its evolution to the present day since the primordial era of Universe is presented
in the following section. The problem of finding the final asymmetry has been dealt with rigorous kinetic equations 
(known as the Boltzmann equations) as well as with some useful analytical approximations. The usefulness of our model in the simultaneous explanation
of the oscillation data and the observed baryon asymmetry will be examined numerically in Section~\ref{num}.

\section{Baryogenesis through Leptogenesis}\label{baryo}
As discussed earlier,
we know that one of the convenient and effective ways to generate light neutrino mass is through Type-I seesaw mechanism
where three right handed heavy neutrinos ($N_i$) are added to the SM. These physical right handed neutrinos with definite mass can decay 
both to a charged lepton with a charged scalar and a light neutrino with a neutral scalar. Due to the Majorana character
of $N_i$, conjugate process is also possible. If out-of-equilibrium decay of $N_i$ in conjugate process occur 
at different rate from the actual process, net lepton asymmetry will be generated. 
As a first step to study the leptogenesis we have to compute the CP asymmetry\cite{Covi:1996wh}
parameter which depends upon the structure of the Dirac neutrino mass matrix and hence, depends upon the specific model
under consideration. The leptogenesis phenomenon may be flavour dependent or independent according to the temperature regime.
Accordingly, the CP asymmetries too may be flavour dependent. The CP asymmetries are then plugged into the Boltzmann
equations to get the final value of the lepton asymmetry (flavour independent/dependent) which is further converted
into baryon asymmetry by sphaleron process. Both flavour dependent and independent phenomena are explored in detail in 
the following subsections.

\subsection{Standard basis for leptogenesis calculation}
In the standard basis utilized for leptogenesis calculations, the charged-lepton mass matrix, Eq.~(\ref{eq:clmassmatrix}), as well as the Majorana mass matrix for the neutrinos, Eq.~(\ref{eq:mm2})  should be diagonal. In our model, both these matrices are non-diagonal. To diagonalize the charged-lepton mass matrix, we use the transformation, 
\begin{equation}\label{eq:transforml}
L\rightarrow V L,
\end{equation}
where $V$ is the $3\times3$ trimaximal matrix, Eq.~(\ref{eq:utbm}). The Majorana mass matrix for the heavy neutrinos, 
Eq.~(\ref{eq:mm2}), is diagonalized using the bimaximal matrix,
$U_{\text{BM}}$ Eq.~(\ref{eq:ubm}),
\begin{equation}\label{eq:mmasses}
U_p^T U_{\text{BM}}^T \mm U_{\text{BM}}U_p = |\mg| \text{Diag(} |1+\km|, 1 , |1-\km| \text{)} = \text{Diag(} M_3, M_2 , M_1 \text{)},
\end{equation}
where $M_3$, $M_2$, $M_1$ are the heavy neutrino masses and $U_p$ is the unitary diagonal matrix used to remove the phases from the masses,
\begin{equation}
U_p= e^{-\frac{i}{2}  \xi_1 }\text{Diag}\left( e^{-\frac{i\theta_1}{2}}, 1 , e^{-\frac{i\theta_2}{2}}\right) \quad \text{with} \quad \theta_1=\text{Arg}(1+\km), \,\, \theta_2=\text{Arg}(1-\km).
\end{equation}
The diagonalization, Eq.~(\ref{eq:mmasses}), corresponds to the following transformation
of the RH neutrino fields:
\begin{equation}\label{eq:transformnu}
\nr \rightarrow U_p^\dagger U_{\text{BM}}^\dagger \nr.
\end{equation}
In the basis in which both the charged-lepton mass matrix and the neutrino Majorana mass matrix are diagonal (hereafter referred to as the ``standard basis''), the neutrino Dirac mass matrix, Eq.~(\ref{eq:md}), becomes
\begin{equation}
\md \rightarrow \md^\prime = V \md U_{\text{BM}} U_p. \label{stndb}
\end{equation}

\subsection{CP asymmetry parameter}
In the present work, the smallness of the mass of active neutrinos is achieved through the well known Type-I seesaw mechanism where the
fermion sector of the SM is extended in a minimal fashion through the addition of three heavy right-chiral neutrinos ($N_i$) which 
are singlets under the SM gauge group. CP violating out-of-equilibrium decays of these neutrinos act as the source of lepton asymmetry. From the neutrino Dirac mass term, $y_{\alpha i}\bar{L}_\alpha N_i \tilde{H}$, it is clear that the decay product of the 
right handed neutrino is either the left-handed neutrino and the neutral scalar or charged lepton and the charged scalar. The basic quantity we are interested in is the 
flavour dependent CP asymmetry parameter\cite{Pilaftsis:2003gt,Adhikary:2014qba} which is expressed as 
\begin{eqnarray}
 \varepsilon^\alpha_i &=&\frac{\Gamma({N}_i\rightarrow
    l_\alpha^-H^+,\nu_\alpha H^0)-\Gamma({N}_i\rightarrow
    l_\alpha^+ H^-,\nu_\alpha^c H^{0*})}{\sum\limits_\alpha \left[ \Gamma({N}_i\rightarrow
    l_\alpha^- H^+,\nu_\alpha H^0)+\Gamma({N}_i\rightarrow
    l_\alpha^+ H^-,\nu_\alpha^c H^{0*}) \right] }, 
\end{eqnarray}
where $\Gamma$ denotes the decay width. Basically, CP asymmetry is a measure of the difference in decay widths of $N_i$ in a process and its conjugate process. At the tree level, these two are the same
\footnote{In the tree level, $\Gamma_{N_i \rightarrow ~l \phi}=\Gamma_{N_i \rightarrow ~l^c \phi^\dagger}$}
giving rise to vanishing CP asymmetry. Therefore, we have to investigate with higher order terms to obtain non-zero CP asymmetry.
Taking into account the one loop vertex and self energy diagrams, it is found that non-zero CP asymmetry arises due to the interference between the tree level and the one loop diagrams.
In the standard basis, the most general expression (keeping up to the fourth order of the Yukawa couplings) of the flavour dependent CP asymmetry parameter comes out to be
\begin{eqnarray}
\varepsilon^\alpha_i
&=&\frac{1}{8\pi v^2 H^\prime_{ii}}\sum_{j\ne i} Im\{H^\prime_{ij}({\md^\prime}^\dagger)_{i\alpha} (\md^\prime)_{\alpha j}\}
\left[f(x_{ij})+\frac{\sqrt{x_{ij}}(1-x_{ij})}
{(1-x_{ij})^2+\frac{{H^\prime}_{jj}^2}{64 \pi^2 v^4}}\right]\nonumber\\
&+&\frac{1}{8\pi v^2 {H^\prime}_{ii}}\sum_{j\ne i}\frac{(1-x_{ij})Im\{{H^\prime}_{ji}({\md^\prime}^\dagger)_{i\alpha} (\md^\prime)_{\alpha j}\}}
{(1-x_{ij})^2+\frac{{H^\prime}_{jj}^2}{64 \pi^2 v^4}},
\label{epsi_intro}
\end{eqnarray}
where $H^\prime={\md^\prime}^\dagger \md^\prime$, $x_{ij}=\frac{M_j^2}{M_i^2}$ and $f(x_{ij})$ is the loop function given by
\begin{equation}
f(x_{ij})=\sqrt{x_{ij}} \left\{1-(1+x_{ij})\ln \left(\frac{1+x_{ij}}{x_{ij}}\right)\right\}.
\end{equation}
It is worthwhile to mention that depending upon the temperature regime at which leptogenesis takes place,
the lepton flavours may be fully distinguishable, partly distinguishable or indistinguishable.
The flavours can not be treated separately when the leptogenesis process occurs above a temperature $T> 10^{12}~{\rm GeV}$.
The CP asymmetry parameter will also be flavour independent accordingly.
If leptogenesis occurs at a lower temperature $T \sim M_1$ ($M_1$ mass of the lightest right-handed neutrino)
then there are two possibilities: for $T< 10^9$ GeV all three ($e,\mu,\tau$) flavours are separately 
active\footnote{In this regime we need three CP asymmetry parameters $\varepsilon^e_i,\varepsilon^\mu_i,\varepsilon^\tau_i$
for each generation of RH neutrino.},
for $10^9<T ~(\rm GeV)< 10^{12}$ only $\tau$ flavour can be identified separately while $e$ and $\mu$ act indistinguishably
\footnote{Here we need two CP asymmetry parameters $\varepsilon^2_i=\varepsilon^e_i+\varepsilon^\mu_i$ and
$\varepsilon^\tau_i$ for each generation of RH neutrino.}. 
For $T>10^{12}$ GeV the flavour summed CP asymmetry parameter is given by
\begin{eqnarray}
\varepsilon_i &=& \sum\limits_{\alpha}\varepsilon^\alpha_i \nonumber\\
              &=& 
\frac{1}{8\pi v^2 H^\prime_{ii}}\sum_{j\ne i} Im\{{H^\prime_{ij}}^2\}
\left[f(x_{ij})+\frac{\sqrt{x_{ij}}(1-x_{ij})}
{(1-x_{ij})^2+\frac{{H^\prime}_{jj}^2}{64 \pi^2 v^4}}\right]
\label{sum_epsi_intro}.
\end{eqnarray}
For the model under consideration, it is straight forward to find the expression of the unflavoured CP asymmetry
parameters. The main ingredient $H^\prime$ in Eq.~(\ref{sum_epsi_intro}) can be represented in terms of the parameters 
of the mass matrix and the mixing matrix as
\begin{eqnarray}\label{eq:mddmd}
H^\prime & = & {\md^\prime}^\dagger \md^\prime \nonumber\\
         & = & U_p^\dagger U_{BM}^\dagger \md^\dagger V^\dagger V \md U_{BM} U_p \nonumber\\
         & = & \mw^2 \left(\begin{matrix} (\kd-1)^2 &  0 & 0 \\
0  &  2 \kd^2+1 & -\sqrt{2}(\kd^2+2\kd) e^{-\frac{i\theta_2}{2}} \\
0  & -\sqrt{2}(\kd^2+2\kd) e^{\frac{i\theta_2}{2}} &  3 \kd^2 + 2 \kd +1
\end{matrix}\right),
\end{eqnarray}
from which the unflavoured CP asymmetry parameters\footnote{It is to be noted that although 
$\varepsilon_1=0$ and $\varepsilon_2=-\varepsilon_3$, the total baryon asymmetry parameter ($Y_B/\eta_B$) turns out to be non zero since the 
efficiency factors corresponding to $N_1,N_2,N_3$ are different from each other. The explicit mathematical formula connecting baryon asymmetry
with CP asymmetries and efficiency factors is presented in Sec.\ref{boltz_approx}.} are computed to be
\begin{eqnarray}
&& \varepsilon_1=0~~~~~({\rm since}~H^\prime_{12}=0=H^\prime_{13}), \nonumber\\
&& \varepsilon_2=-\frac{\mw^2 (\kd^2+2\kd)^2 \sin \theta_2}{4 \pi v^2 (2\kd^2+1) }= -\varepsilon_3. \label{uncp}
\end{eqnarray}
Now, to calculate the flavoured CP asymmetry parameters we need individual elements of the Dirac neutrino mass matrix in the standard basis. Unlike the unflavoured case, here both the diagonalization matrices (of the charged-lepton mass matrix and the neutrino Majorana mass matrix) contribute to the CP asymmetry
parameters. The exact analytic expressions of the CP asymmetry parameters are 
too cumbersome to write here. We calculate them numerically and use in the appropriate formulas to calculate the final baryon asymmetry parameter. However,
different elements of $\md^\prime$ are spelt out in Appendix \ref{md}.

It is clear from Eq.~(\ref{uncp}) that the unflavoured CP asymmetry parameter depends only on one phase parameter $(\theta_2)$ apart from the 
modulus parameters and that phase comes from the RH Majorana neutrino mass matrix. Low energy phase parameters are absent here. On the other
hand the flavoured CP asymmetry parameters depends on the high energy as well as low energy CP phases in general, however certain symmetries
in the neutrino mass model may enable one to bring out the significance of high energy or low energy CP phases in generating the lepton 
asymmetry. An alternative parametrization  (known as the Casas-Ibarra parametrization\cite{Casas:2001sr}) of the Dirac neutrino mass matrix 
in terms of the  experimentally measurable low energy parameters (light neutrino mass eigenvalues, mixing angles, phases),
RH neutrino mass eigenvalues and an orthogonal matrix (containing three complex angles) may be helpful in understanding the dependence
of nonzero flavour asymmetries on the high energy and low energy CP phases. In Appendix \ref{CSI} we rewrite the CP asymmetry parameters
using Casas-Ibarra parametrization and examine whether we can conclusively say something regarding the dependence on low energy and high energy 
CP phases.

\subsection{Boltzmann equations for leptogenesis}
A particle species being coupled or decoupled with the thermal bath, depends roughly
on the rate of interaction of the particle\footnote{Precisely $\tilde{\Gamma}$ is the interaction rate per particle for the
reaction responsible for keeping the species under consideration in thermal equilibrium.} $(\tilde{\Gamma})$
and the Hubble parameter $(H)$, which can be expressed explicitly as\cite{Kolb:1990vq}
$\tilde{\Gamma} \gtrsim H ~~({\rm coupled}),
\tilde{\Gamma}\lesssim H ~~(\rm decoupled)$ .
It is interesting to find out the phase space distribution of the particle species near the epoch of decoupling.
The microscopic evolution of particle's phase space distribution ($f(p^\mu, x^\mu)$) is governed by the Boltzmann equation (BE) 
which is written as\cite{Luty:1992un}
\begin{equation}
\hat{{\bf L}} [f]=-\frac{1}{2}{\bf C}[f], 
\label{boltz_eq1}
\end{equation}
where $\hat{{\bf L}}$\footnote{The covariant form of the 
Liouville operator is 
$\hat{{\bf L}} = p^\alpha \frac{\partial}{\partial x^\alpha} - \Gamma^\alpha_{\beta \gamma} p^\beta p^\gamma \frac{\partial}{\partial p^\alpha}$
where $\Gamma^\alpha_{\beta \gamma}$ is the Christoffel symbol.}
is the Liouville operator and ${\bf C}$ is the collision operator which takes into account all such interactions that change
the number density of the particle species under consideration in the thermal bath. 
In simple words, through this equation the evolution of the number density of a particle species can be tracked from very
high temperature (early epoch) down to present temperature.
Our present model has ample scope for the generation of lepton asymmetry through CP violating decay of heavy Majorana neutrinos in the 
early Universe. Since we need to know the lepton asymmetry (to calculate the baryon asymmetry and compare it with the observed value) in the present 
epoch, we have to solve the Boltzmann equation for lepton number density which
in turn depends on the instantaneous value of the RH neutrino density. 
Thus the required set of classical kinetic equations\cite{Pilaftsis:2003gt} (which can be derived from Eq.~(\ref{boltz_eq1})) for the RH neutrino density and the lepton number density are given
by
\begin{eqnarray}
\label{BEN_i} 
\frac{d \eta_{N_i}}{dz} &=& \frac{z}{H(z=1)}\ \bigg[\,\bigg( 1
\: -\: \frac{\eta_{N_i}}{\eta^{\rm eq}_{N_i}}\,\bigg)\, \bigg(\,
\Gamma^{D\; (i)} \: +\: \Gamma^{SY\; (i)} +\:
\Gamma^{SG\; (i)}\, \bigg) \bigg]\nonumber\\
                        &=& -\bigg \{ D_{i}(z) + D^{SY}_i(z) +D^{SG}_i(z) \bigg \} \bigg (\eta_{N_i}(z)-\eta^{\rm eq}_{N_i}(z)  \bigg), \\
  \label{BEL_i}
\frac{d \eta_L}{dz} &=& -\, \frac{z}{H(z=1)}\, \bigg[
\sum\limits_{i=1}^3\, \varepsilon_i\ 
\bigg( 1 \: -\: \frac{\eta_{N_i}}{\eta^{\rm eq}_{N_i}}\,\bigg)\, \bigg(\,
\Gamma^{D\; (i)} \: +\: \Gamma^{SY\; (i)} +\:
\Gamma^{SG\; (i)}\, \bigg) \nonumber\\ 
&&+\, \frac{1}{2}\, \eta_L\, \bigg\{ \sum\limits_{i=1}^3\, 
\bigg(\, \Gamma^{D\; (i)} \: +\: 
\Gamma^{WY\;(i)}\: 
+\: \Gamma^{WG\; (i)}\,\bigg)\:
\bigg\}\bigg]\, \nonumber \\
&=&-\sum\limits_{i=1}^3\, \varepsilon_i\ \bigg \{ D_{i}(z) + D^{SY}_i(z) +D^{SG}_i(z) \bigg \} \bigg (\eta_{N_i}(z)-\eta^{\rm eq}_{N_i}(z)  \bigg)\nonumber\\ && -\frac{1}{2}\eta_L \sum\limits_{i=1}^3
\bigg\{ \frac{1}{2} D_i(z) z^2 \mathcal{K}_2(z) +D^{WY}_{i}(z) + D^{WG}_{i}(z) \bigg\},
\end{eqnarray}
respectively, where 
\begin{eqnarray}
&& z=\frac{{\rm Mass~of~lightest~RH~neutrino}}{{\rm temperature}}=\frac{M_1}{T}, \nonumber\\ 
&& \eta_a(z)=\frac{{\rm number~density~of~paticle~ species~}a}{{\rm photon~density}}=\frac{n_a(z)}{n_\gamma (z)} ,\nonumber\\
&& \eta^{\rm eq}_a(z)=\frac{n^{\rm eq}_a(z)}{n_\gamma(z)}~{\rm with}~n_\gamma(z)=\frac{2 M_1^3}{\pi^2 z^3}~~.
\end{eqnarray}
$H$ is the Hubble parameter, $\Gamma$s (detailed expression can be found in Ref.\cite{Pilaftsis:2003gt,Adhikary:2014qba})
are the different decay and scattering cross sections scaled by photon density.
Following Maxwell Boltzmann distribution, the number density of a particle species $a$ of mass $m_a$ with $g_a$ internal
degrees of freedom is given by
\begin{equation}
\label{na}
n_a (T)=  \frac{g_a\, m^2_a\,T\ e^{\mu_a (T)/T}}{2\pi^2}\
\mathcal{K}_2\bigg(\frac{m_a}{T}\bigg)\;, 
\end{equation}
from which the equilibrium density is obtained by setting the chemical potential to be zero as
\begin{equation}
n^{\rm eq}_a (T)=  \frac{g_a\, m^2_a\,T\ }{2\pi^2}\
\mathcal{K}_2\bigg(\frac{m_a}{T}\bigg)\;,
\end{equation}
where $\mathcal{K}_2$ is the modified Bessel function of 2nd kind with order 2.
It is to be noted that the first superscript $D(S)$ on $\Gamma$ denotes decay (scattering) and the other superscript designates whether the interaction is
Yukawa mediated $(Y)$ or gauge boson $(G)$ mediated.
Different $D_i$ parameters are defined as
\begin{eqnarray}
 && D_i(z) =\frac{z}{H(z=1)}\frac{\Gamma^{D\;(i)}}{\eta^{\rm eq}_{N_i}(z)} ~,\label{Di}\\
&&D^{SY/SG}_i(z) =\frac{z}{H(z=1)}\frac{\Gamma^{SY/SG\;(i)}}{\eta^{\rm eq}_{N_i}(z)} ~,\label{Ds}\\
&& D^{WY/WG}_i =\frac{z}{H(z=1)} \Gamma^{WY/WG \; (i)} ~.\label{Dw}
\end{eqnarray}
The above set of Boltzmann equations, Eqs.~(\ref{BEN_i}, \ref{BEL_i}), consider asymmetry generated by all the three generations of RH neutrinos, but they are valid in a temperature
regime ($T > 10^{12}$ GeV) where the lepton flavours are indistinguishable. If the RH neutrinos are strongly hierarchical $(M_1 \ll M_2,M_3)$, the asymmetry generated by heavier RH neutrinos is completely washed out due to $N_1$ interactions and the summation in the Boltzmann equation is required no more. We need quantities involving first generation only. 
In comparatively lower energies lepton flavours are partially 
distinguishable ($10^9<T(~{\rm GeV})<10^{12}$) or fully distinguishable ($T<10^9$ GeV). Accordingly the flavour effects 
have to be introduced in the Boltzmann equations (\ref{BEN_i}, \ref{BEL_i}). After incorporating the lepton flavour index in the suitable places, the modified Boltzmann equation for lepton number density is presented as
\begin{eqnarray}
\frac{d \eta^\alpha_L}{d z} =& - &\sum\limits_{i=1}^3\, \varepsilon^\alpha_i\ \bigg \{ D_{i}(z) + D^{SY}_i(z) +D^{SG}_i(z) \bigg \} \bigg (\eta_{N_i}(z)-\eta^{\rm eq}_{N_i}(z)  \bigg)\nonumber\\  
& - &\frac{1}{2}\eta^\alpha_L \sum\limits_{i=1}^3
\bigg\{ \frac{1}{2} D^\alpha_i(z) z^2\mathcal{K}_2(z) +D^{\alpha\;WY}_{i}(z) + D^{\alpha\;WG}_{i}(z) \bigg\} \label{BEL_f},
\end{eqnarray}
whereas that of the RH neutrino number density remains unaltered since it does not involve lepton flavour index. $D^\alpha_i(z),D^{\alpha\;WY}_{i}(z),D^{\alpha\;WG}_{i}(z)$ can be estimated using Eq.~(\ref{Di}), Eq.(\ref{Dw}) by introducing the flavour index $\alpha$ on different $\Gamma$s. It is to be noted that instead of $\eta$ parameter we can use an equivalent parameter $Y$ (particle number density/entropy density) to
express the abundance of a particle species at any instant of evolution.
In this type of representation, the quantity associated with the lepton
asymmetry is denoted as $Y_\alpha$ which is related to $\eta^\alpha_L$
as $Y_\alpha= (n_\gamma s^{-1}) \eta^\alpha_L$. It is well known that
$n_\gamma s^{-1}=1/\{1.8 g^\ast_s(T)\}$, where $g^\ast_s(T)$ is the effective number of massless degrees of freedom\cite{Kolb:1990vq} at temperature $T$. Although 
$g^\ast_s$ is a function of temperature, for sufficiently high temperature $(T>10^{12})$ GeV it becomes practically constants and its
value is 112 if we include the contribution of the three RH neutrinos. The lepton asymmetry created by the decay of the heavy Majorana neutrinos (before sphaleron processes set in) gets converted into baryon
asymmetry by the action by sphalerons. One interesting aspect of this sphaleron interactions is that both baryon number $(B)$ and lepton number $(L)$ are violated in this process keeping the difference $(B-L)$
(for flavoured regime this is $(B/3-L_\alpha)=\Delta_\alpha$) conserved. We now define a new asymmetry parameter $Y_{\Delta_\alpha}$
which is related to the $Y_\alpha$ through an asymmetry coupling matrix $A$ as $Y_\alpha=\sum\limits_{\beta} A_{\alpha\beta}Y_{\Delta_\beta}$. The Boltzmann equations describing the evolution of flavour asymmetry parameters can be represented in terms of $Y_{\Delta_\alpha}$ as
\begin{eqnarray}
\frac{d Y_{\Delta_\alpha}}{dz} = & - & \sum\limits_{i=1}^3 \,  \bigg[\varepsilon^\alpha_i \ \bigg\{D_i(z)+D^{\rm SY}_i(z)+D^{\rm SG}_i(z)\bigg\}\bigg(Y_{N_i}(z)-Y^{\rm eq}_{N_i}(z)\bigg)\bigg]  \nonumber\\
& + &\frac{1}{2}\sum\limits_{\beta}A_{\alpha\beta}Y_{\Delta_\beta} 
\sum\limits_{i=1}^3\ 
\bigg\{ \frac{1}{2}D^\alpha_i(z)z^2\mathcal{K}_2(z)+D^{\alpha ~\rm WY}_i(z)+ D^{\alpha ~\rm WG}_i(z) \bigg\}  .\label{BEL_Y1}
\end{eqnarray}
We are now in a position to estimate the final baryon asymmetry for which we have to solve the set of Boltzmann equations
for flavour asymmetry (Eq.~(\ref{BEL_Y1})) and RH neutrino density (Eq.~(\ref{BEN_i}))\footnote{Since $\eta_{N_i}$ and $Y_{N_i}$ 
are connected by a factor which is effectively constant in our working regime, the structure of Eq.~(\ref{BEN_i}) will remain 
identical in terms of the variable $Y_{N_i}$, we just have to replace $\eta \rightarrow Y$.} simultaneously upto
a large enough value of $z$ for which the asymmetry freezes to certain constant value. 

\subsubsection{Baryon asymmetry in different regimes}
The asymmetry creation takes place mostly around a temperature scale of the order  of lightest RH neutrino mass $(M_1)$. Accordingly, there are three different regimes of leptogenesis\cite{Abada:2006ea,Antusch:2006cw}.
\paragraph{$M_1 < 10^9~{\rm GeV}$:} In this regime all the three lepton flavours $(e,\mu,\tau)$ are distinguishable. So, the $3 \times 3$ asymmetry coupling matrix is given by
\begin{equation}
A=\left(\begin{array}{ccc}
-151/179 & 20/179 & 20/179\\
25/358 & -344/537 & 14/537\\
25/358 &  14/537 &  -344/537
\end{array}\right) . \label{a3f}
\end{equation}
The final value of baryon asymmetry ($Y_B=n_B/s$) is calculated by summing over the flavour asymmetry parameters (obtained by solving the Boltzmann equations) followed by multiplication with the sphaleron conversion factor, i.e.,
\begin{equation}
Y_B=\frac{28}{79}\sum\limits_{\alpha} Y_{\Delta_\alpha} .
\end{equation}
The equivalent $\eta_B$ parameter is connected to $Y_B$ as $\eta_B=\left.\frac{s}{n_\gamma}\right|_0Y_B=7.0394Y_B$ where the zero subscript denote its value at the present epoch.
\paragraph{{\bf $10^9<M_1({\rm GeV})<10^{12}$:}} In this regime only
$\tau$ flavour has separate identity, whereas $e$ and $\mu$ flavours
act indistinguishably as a single entity, i.e. effectively we have two flavours $\tau$ and $a (=e+\mu)$. Therefore, the $A$ matrix coupling 
$Y_\alpha$ and $Y_\Delta$ asymmetry is $2\times 2$ which is given by
\begin{equation}
A=\left(\begin{array}{cc}
  -417/589  & 120/589\\
   30/589 & -390/589
\end{array}\right). \label{a2f}
\end{equation}
The final baryon asymmetry parameter is given by
\begin{equation}
Y_B=\frac{28}{79}(Y_{\Delta_a}+Y_{\Delta_\tau}).
\end{equation}

\paragraph{$M_1>10^{12}~{\rm GeV}$:} In this regime all the three lepton flavours act indistinguishably and thus neither CP asymmetry nor the Boltzmann equations involve the flavour index $\alpha$. In the right hand side of Boltzmann equation (Eq.~(\ref{BEL_Y1})), the flavour dependent
terms have to the replaced by a sum over $\alpha$. The matrix $A$ is simply the negative identity. It is obvious that in this case we
have to solve for a single $Y_\Delta$ which is connected to the asymmetry parameter through the multiplicative sphaleron factor as
\begin{equation}
Y_B= \frac{28}{79} Y_\Delta. 
\end{equation}

\subsection{An approach towards analytic approximation of Boltzmann equations}\label{boltz_approx}
Lets start with the set of Boltzmann equations in the unflavoured regime, i.e.,
\begin{eqnarray}
&& \frac{d \eta_{N_i}}{dz}=-\bigg \{ D_{i}(z) + D^{SY}_i(z) +D^{SG}_i(z) \bigg \} \bigg (\eta_{N_i}(z)-\eta^{\rm eq}_{N_i}(z)  \bigg)~,\\
&&\frac{d \eta_{B-L}}{dz}= -\sum\limits_{i=1}^3\, \varepsilon_i\ \bigg \{ D_{i}(z) + D^{SY}_i(z) +D^{SG}_i(z) \bigg \} \bigg (\eta_{N_i}(z)-\eta^{\rm eq}_{N_i}(z)  \bigg) \nonumber\\ &&~~~~~~~~~~~~~ -\frac{1}{2}\eta_{B-L} \sum\limits_{i=1}^3
\bigg\{ \frac{1}{2} D_i(z) z^2\mathcal{K}_2(z) +D^{WY}_{i}(z) + D^{WG}_{i}(z) \bigg\} ~.
\end{eqnarray}
It has already been shown in the existing literature that for a strong washout scenario, if the active neutrinos follow hierarchical pattern, the $\Delta L=2$ terms and the scattering terms can be safely neglected. Thus we consider a simplified picture where only decays and inverse decays are taken into account, i.e. 
\begin{eqnarray}
&& \frac{d \eta_{N_i}}{dz}=- D_{i}(z)  \bigg (\eta_{N_i}(z)-\eta^{\rm eq}_{N_i}(z)  \bigg) ~, \label{bn}\\
&&\frac{d \eta_{B-L}}{dz}= -\sum\limits_{i=1}^3\, \varepsilon_i D_{i}(z) \bigg (\eta_{N_i}(z)-\eta^{\rm eq}_{N_i}(z)  \bigg) -\eta_{B-L} \sum\limits_{i=1}^3
\bigg\{ \frac{1}{4} D_i(z) z^2\mathcal{K}_2(z)  \bigg\} \nonumber\\
&&~~~~~~~~~=-\sum\limits_{i=1}^3\, \varepsilon_i D_{i}(z) \bigg (\eta_{N_i}(z)-\eta^{\rm eq}_{N_i}(z)  \bigg) - \sum\limits_{i=1}^3 W^i_{\rm ID}(z) \eta_{B-L} ~.\label{bl}
\end{eqnarray}
Now using Eq.~(\ref{Di}), $D_i(z)$ can also be expressed in the form
\begin{eqnarray}
D_i(z) &=& z \frac{\Gamma_{N_i}(T=0)}{H(z=1)} \frac{\mathcal{K}_1(z\frac{M_i}{M_1})}{\mathcal{K}_2(z\frac{M_i}{M_1})} \nonumber\\
       &=& z r_i \frac{\Gamma_{N_i}(T=0)}{H(z_i=1)} 
           \frac{\mathcal{K}_1(\sqrt{r_i}z)}{\mathcal{K}_2(\sqrt{r_i}z)}\nonumber\\
       &=& z r_i K_i
           \frac{\mathcal{K}_1(\sqrt{r_i}z)}{\mathcal{K}_2(\sqrt{r_i}z)},    
\end{eqnarray}
where $r_i=M_i^2/M_1^2$, $\Gamma_{N_i}$ is the decay width and $K_i$ is the decay parameter corresponding to the $i$th right 
handed neutrino. In Eq.~(\ref{bl}) the washout term gets maximum contribution from the inverse decay which is given by
\begin{eqnarray}
W^i_{\rm ID}(z)= \frac{1}{4} z_i^2 \mathcal{K}_2(r_i z) D_i(z) ~.
\end{eqnarray}
The set of Boltzmann equations, Eqs.~(\ref{bn},\ref{bl}), are solved up to a large enough value of $z$ to get the final value of the
$B-L$ asymmetry\cite{Pilaftsis:2003gt,Adhikary:2014qba,Buchmuller:2004nz} as 
\begin{equation}
\eta^f_{B-L}= \eta^{\rm in}_{B-L} ~e^{-\sum\limits_ i \int \limits^{z\rightarrow \infty}_{z_{in}}
W^i_{\rm ID}(z^\prime) dz^\prime} - \sum \limits_i \varepsilon_i \kappa^f_i, 
\end{equation}
where $\eta^{\rm in}_{B-L}$ is the pre-existing asymmetry. If it is assumed that there is no such kind of pre-existing asymmetry,
then the final value of $(B-L)$ asymmetry is simply obtained as
\begin{equation}
 \eta^f_{B-L}=-\sum \limits_i \varepsilon_i \kappa^f_i,
\end{equation}
where the final efficiency factor $\kappa^f_i$\cite{Blanchet:2006dq,Samanta:2019yeg} is given by
\begin{equation}
\kappa^f_i= -\int \limits^{z \rightarrow \infty}_{z_{in}} d z^{\prime} \frac{d N_i}{d z^{\prime}} ~
e^{-\sum \limits_i \int \limits^{z }_{z^\prime} W^i_{\rm ID}(z^{\prime\prime}) dz^{\prime\prime}} ~. \label{kf}
\end{equation}
It is well known that in the strong washout regime for hierarchical RH neutrino mass spectrum $(M_1 \ll M_2 \ll M_3)$
only the asymmetry produced by the lightest RH neutrino survives. Using the approximation 
$\frac{d \eta_{N_1}}{dz} \simeq \frac{d \eta^{\rm eq}_{N_1}}{dz}$, the analytic expression of final efficiency factor $(z\rightarrow\infty)$ 
corresponding to the lightest RH neutrino is obtained as
\begin{equation}
\kappa^f_1(K_1) \simeq \kappa (K_1)=\frac{2}{K_1 z_B(K_1)} \bigg ( 1- e^{-\frac{K_1 z_B(K_1)}{2}} \bigg ), \label{kp1}
\end{equation}
where $z_B$ is such a value of the independent variable $(z^\prime)$ around which the integrand of Eq.~(\ref{kf})
receives maximum contribution and for strong washout its analytical expression is given by\cite{Blanchet:2006dq,Samanta:2019yeg}
\begin{equation}
z_B(K_1) \simeq 2+ 4 K_1^{0.13} ~ e^{-\frac{2.5}{K_1}} ~. 
\end{equation}
Now, if the masses of the RH neutrinos are close to each other\footnote{In the present problem our neutrino mass model predicts
closeness of mass two RH neutrinos where as third one is order of magnitude higher than these two. So in this case it is 
sufficient to deal with two generations of RH neutrinos}, we can not neglect the contribution of the next to lightest
RH neutrino towards final asymmetry. Under the assumption that asymmetry production or washout by $N_2$ is not affected
by that of $N_1$, the efficiency factor corresponding to $N_2$ is given by a simple analytic expression as
\begin{equation}
\kappa^f_2  = \kappa(K_2) ~ e^{- \int \limits^\infty_0 W^1_{\rm ID}(z) dz} 
            = \kappa(K_2) ~e^{-\frac{3 \pi K_1}{8}} ~.\label{kp2}
\end{equation}
The final value of the baryon asymmetry parameter $\eta_B$ is obtained from $\eta^f_{B-L}$ after multiplication by the sphaleron conversion factor
$(a_{\rm sph})$\footnote{$a_{\rm sph}$ stands for the fraction of $(B-L)$ asymmetry being converted into baryon asymmetry through sphaleron process }  
and division by a dilution factor $f$\cite{Buchmuller:2002rq} (which is estimated assuming standard thermal history of the Universe as $f=N^{\rm rec}_{\gamma}/N^\ast_{\gamma}=2387/86$),
i.e.
\begin{equation}
\eta_B= \frac{a_{\rm sph}}{f} \eta^f_{B-L} =-0.96 \times 10^{-2} \sum \limits_i
\varepsilon_i \kappa^f_i~. \label{etab}
\end{equation}
For a definite value of the decay parameter $(K_i)$, the value of the efficiency parameter $\kappa^f_i$ can be found either by direct numerical integration (of  Eq.~(\ref{kf}) ) or by 
using approximate analytical formulas ( Eq.~(\ref{kp1}), Eq.~(\ref{kp2}) ). Goodness of this analytical approximation depends on the value of $K_1$ and 
$\delta_{12}$\footnote{A detailed analysis dealing with this issue has been carried out in Ref\cite{Samanta:2019yeg}. } 
where $\delta_{12}=(M_2 - M_1)/M_1$. Broadly, we can say that these two results match exactly with each other in the strong washout regime for hierarchical RH neutrinos. In the strong washout regime, an excellent fit to the efficiency factor for any value of $\delta_{12}$ has been found\cite{Samanta:2019yeg} to be
\begin{equation}
\kappa^{\rm fit}_1 =\frac{2 K_1}{z_B \bigg (K_1 + K_2^{(1-\delta_{12})^3}\bigg) \bigg ( K_1 +K_2^{(1-\delta_{12})}\bigg )}~. 
\end{equation}
In the present work, we will first show that the lagrangian parameters constrained by the $3\sigma$ oscillation data mostly favour strong washout regime.
Then for a bench mark point (we choose it to be the best fit point, i.e. the set of points having least $\chi^2$), we find $\eta_B$ using the analytical formulas (both $\kappa^\infty_i$ and $\kappa_{\rm fit}$) and also using the direct numerical solution of Boltzmann equations (Eq.~(\ref{bn}), Eq.~(\ref{bl}) ) to show the accuracy of the analytical fit. Upon showing satisfactory accuracy of the analytical formulas, we then proceed to find the baryon asymmetry for each and every point of the $3\sigma$ parameter space using the analytical formulas.

\subsection{Lepton flavour decoherence and flavoured leptogenesis}
It has been pointed out earlier that depending upon the mass of the lightest RH neutrino we can get unflavoured,
$\tau$-flavoured or fully flavoured leptogenesis. The distinguishability of lepton flavours in different temperature
regimes can be explained by flavour decoherence\cite{Blanchet:2006dq,Blanchet:2006be,Dev:2017trv,Samanta:2019yeg} phenomenon\footnote{An exhaustive discussion on the flavour decoherence 
issue is presented in Ref\cite{Blanchet:2006dq,Samanta:2019yeg}.}.
\paragraph{}
When $M_1>10^{12}$ GeV, the RH neutrino decays producing lepton doublets 
$( |l_i\rangle)$ which are nothing but a coherent superposition of corresponding flavour states, i.e
\begin{eqnarray}
&& |l_i\rangle = \sum \limits_\alpha  C_{i\alpha}| l_\alpha \rangle, \\
&& | \bar{l_i}\rangle = \sum \limits_\alpha \bar{C}_{i\alpha}| \bar{l}_\alpha \rangle,
\end{eqnarray}
where $(i=1,2,3)$ stands for generation index of RH neutrinos and $(\alpha=e,\mu,\tau)$ are lepton flavour indices. The coefficients are given by
\begin{equation}
C_{i\alpha} = \frac{(M_D^\dagger)_{i\alpha}}{\sqrt{(M_D^\dagger M_D)_{ii}}} ~.
\end{equation}
Since at this temperature regime there doesn't exist any fast interaction that can break the coherence between the flavour states before it inverse decays into $N_1$, the net 
asymmetry is produced along $| l_i \rangle$ in which we can not differentiate between flavours.
\paragraph{}
To carry out the analysis in the flavoured regime we define
branching ratio to the individual lepton flavours as
\begin{equation}
P_{i\alpha} =| C_{i\alpha} |^2 ~~{\rm and}~~ \bar{P}_{i\alpha} =| \bar{C}_{i\alpha}|^2~.
\end{equation}
The flavoured decay parameter is given by 
\begin{eqnarray}
K^\alpha_i= \frac{\Gamma^\alpha_i +\bar{\Gamma}^\alpha_i}{H(T=M_i)}=
\frac{|(M_D)_{\alpha i}|^2}{M_i m^\ast}, 
\end{eqnarray}
where $m^\ast \simeq 10^{-3}$ eV is the equilibrium neutrino mass. The flavoured efficiency factor is given by
\begin{equation}
\kappa^f_{i \alpha} = -\int \limits^{z \rightarrow \infty}_{z_{in}} d z^{\prime} \frac{d N_i}{d z^{\prime}} ~
e^{-\sum \limits_j \int \limits^{z }_{z^\prime}P_{j\alpha} W^j_{\rm ID}(z^{\prime\prime}) dz^{\prime\prime}} ~.\label{kff}
\end{equation}
The analytical form of the flavoured efficiency factor\cite{Blanchet:2006dq,Blanchet:2006be,Samanta:2019yeg} is obtained (using the same arguments as done in the unflavoured case) as
\begin{equation}
\kappa^f_{i\alpha} =\frac{2}{K^\alpha_i |A_{\alpha\alpha}| z_B(K^\alpha_i|A_{\alpha\alpha}|)} \bigg ( 1- e^{-\frac{K^\alpha_i |A_{\alpha\alpha}| z_B(K^\alpha_i |A_{\alpha\alpha}|)}{2}} \bigg ), \label{klf} 
\end{equation}
where $A_{\alpha\alpha}$ is the asymmetry coupling matrix (Eq.~(\ref{a3f}) or Eq.~(\ref{a2f}) ).
\paragraph{}
For the intermediate mass regime $(10^9 < M_1({\rm GeV})< 10^{12})$, the $\tau$
lepton Yukawa interactions reach equilibrium, i.e they become faster than the inverse decay process. Thus coherence between the lepton flavours (generated due to decay of $N_1$) is
broken projecting a portion of the lepton flavours along $\tau$, whereas the other portion being projected in a plane 
perpendicular to $\tau$\footnote{A diagrammatic representation of these flavour projections is given in \cite{Samanta:2019yeg}.} (which is actually a coherent 
superposition of $e$ and $\mu$). When $M_i<10^9$ GeV, $\mu$ lepton interaction rate becomes faster than the inverse decay process thereby breaking the coherence 
of $e$ and $\mu$ in $\tau_\perp$ direction. Therefore in this situation, flavour decoherence is completely achieved allowing us to track asymmetry 
along $e$, $\mu$ and $\tau$ directions separately (which is termed as the fully flavoured leptogenesis). We now discuss about the analytical formula to find the baryon asymmetry 
in the $\tau$-flavoured (or two flavoured) regime.
It is to be mentioned that in the present case it is sufficient to consider
asymmetry generated by the lightest and the next to lightest RH neutrinos (since the third one is orders of magnitude higher than these two). Now, in the $\tau$-flavoured regime when the partial flavour decoherence achieved, lepton doublet states $| l_1 \rangle$ and $| l_2 \rangle$ projected in $\tau_\perp$ plane are expressed by as linear superposition of $| e \rangle$ and $| \mu \rangle$
states as
\begin{eqnarray}
&& | l^{\tau_\perp}_1 \rangle = \frac{1}{\sqrt{|C_{1e}|^2 +|C_{1\mu}|^2}} \bigg [ C_{1e} |l_e\rangle + C_{1\mu} | l_\mu \rangle \bigg ], \\
&& | l^{\tau_\perp}_2 \rangle = \frac{1}{\sqrt{|C_{2e}|^2 +|C_{2\mu}|^2}} \bigg [ C_{2e} |l_e\rangle + C_{2\mu} | l_\mu \rangle \bigg ]~.
\end{eqnarray}
The asymmetry produced by $N_2$ along $\tau$ direction is washed out directly by $N_1$ interactions along $\tau$ through the exponential suppression factor
$\exp(-3 \pi K_{1\tau}/8)$. So, the total asymmetry generated by $N_1$ and $N_2$ along $\tau$ direction is given by
\begin{equation}
N_{\Delta_\tau} = -\varepsilon_{1\tau} \kappa^{f}_{1\tau} - \varepsilon_{2\tau} \kappa^{f}_{2\tau}~e^{-\frac{3\pi K_{1\tau}}{8}}. \label{kp1t}
\end{equation}
But this direct suppression of asymmetry doesn't hold for the asymmetry produced by $N_2$ along $\tau_\perp$ direction. To account for the washout of $N_2$ generated asymmetry 
along $\tau_\perp$ direction due to $N_1$ inverse decay, we have to calculate the probability of finding $| l^{\tau_\perp}_2 \rangle $ state along $| l^{\tau_\perp}_1 \rangle $, 
which is given by
\begin{eqnarray}
p_{12} &=& |\langle l^{\tau_\perp}_1 | l^{\tau_\perp}_2 \rangle |^2 \nonumber \\
       &=& \frac{K_1 K_2}{K_{{1\tau}^\perp}K_{{2\tau}^\perp}}
           \frac{|(M_D)_{e1}(M_D^\ast)_{e2} +(M_D)_{\mu 1}(M_D^\ast)_{\mu 2} |^2}{h_{11}h_{22}} ~. 
\end{eqnarray}
Therefore, the total asymmetry produced along $\tau_\perp$ direction due to the decay of $N_1$ and $N_2$ (after proper inclusion of the washout effects) is given by
\begin{equation}
N_{\Delta_{\tau^\perp_1}}=-\varepsilon_{1\tau^\perp} \kappa^f_{1\tau^\perp} -p_{12} \varepsilon_{2\tau^\perp} \kappa^f_{2\tau^\perp} ~e^{-\frac{3 \pi K_{1\tau^\perp}}{8}} ~. \label{kp1tp}
\end{equation}
The asymmetry produced by $N_2$\cite{Antusch:2010ms,DiBari:2015oca} along a direction which is again perpendicular to $\tau_1^\perp$ (being completely orthogonal to $| l^{\tau^\perp}_1 \rangle$) totally escapes the washout due to $N_1$ inverse decay. Thus, this asymmetry
along $| l^{\tau^\perp}_{1\perp} \rangle$ survives as a pure $N_2$ contribution which is mathematically expressed as
\begin{equation}
N_{\Delta_{\tau^\perp_{1\perp}}} = -( 1-p_{12} ) \varepsilon^\perp_{2\tau} \kappa^f_{2\tau^\perp} ~.
\end{equation}
Therefore in the flavour space, the asymmetry is assumed to be distributed along three direction, $\tau$, $\tau^\perp_1$, $\tau^\perp_{1\tau^\perp}$ from which the total final asymmetry is computed as
\begin{equation}
N^f_{B-L} = N_{\Delta_\tau} + N_{\Delta_{\tau^\perp_1}} + N_{\Delta_{\tau^\perp_{1\perp}}} ~. \label{ybf}
\end{equation}
The pure $N_2$ contribution $(N_{\Delta_{\tau^\perp_{1\perp}}})$ constitutes a significant part of the total $(B-L)$ asymmetry since it is not suppressed by any washout factor (which will be demonstrated clearly in the numerical analysis). This contribution is generally overlooked in the straightforward
solution of flavoured Boltzmann equations considering three generations of RH neutrinos. To account for this asymmetry, we have to solve the Boltzmann equations for $N_2$ separately in which the source term gets contribution from $\varepsilon_{2\tau^\perp}(~{\rm or}~ \varepsilon^{e+\mu}_2)$ only, whereas the washout term contains $N_2$ interactions only (washout due to $N_1$ inverse decay has to be neglected completely).
\section{Analysis of numerical results}\label{num}
The main objective of this numerical analysis is to find the allowed parameter space constrained by 
$3\sigma$ range\cite{Esteban:2018azc} of neutrino oscillation observables (mixing angles: $\theta_{12},\theta_{23},\theta_{13}$, mass 
squared differences: $\Delta m^2_{21},\Delta m^2_{31/32}$ ) as well as the bound\cite{Aghanim:2016yuo} on baryon asymmetry. 
Therefore, the whole analysis can be regarded as a two step process where in the first step we constrain the parameters 
with the $3\sigma$ range of oscillation data and thereafter the admissible parameter space gets a second round of restriction 
from the range of the observed baryon asymmetry.

\subsection{Fitting with neutrino oscillation data}

Along with the overall scale $(\mw^2/\mg)$, the seesaw mass matrix (\ref{eq:msfinal}) is parameterized by $\kd$ and $\km$ ($\km$ is a complex number), i.e.~we have a total of four real degrees of freedom. We fit these parameters with the experimental data on the neutrino mixing angles and the mass-squared differences. In this fit, the number of experimental degrees of freedom is six, i.e.~the three mixing angles ($\theta_{12}, \theta_{23}, \theta_{13}$), the Dirac CP phase ($\delta$) and the mass-squared differences $\Delta m^2_{21}$, $\Delta m^2_{31}$. As we have stated in Section~2, the $\tm$ mixing has only two degrees of freedom, $\epsilon_1$ and $\epsilon_2$. Since the oscillation mixing matrix is parametrised by four observables ($\theta_{12}, \theta_{23}, \theta_{13}, \delta$), the $\tm$ scenario leads two constraints among the mixing angles and the $CP$ phase, Eqs.~(\ref{eq:solartm1}, \ref{eq:sindelta}). It is straightforward to see that the solar mixing angle ($\theta_{12}$) given in Eq.~(\ref{eq:solartm1}) is smaller that the TBM value. We note that the global fit favours such a smaller $\theta_{12}$. The second constraint, Eq.~(\ref{eq:sindelta}), is also consistent with the global fit data. In $\tm$, a deviation from maximal atmospheric mixing ($\theta_{23}=\frac{\pi}{4}$) leads to a deviation from maximal CP violation ($\delta=\pm\frac{\pi}{2}$). The data shows a slight preference for non-maximal $\theta_{23}$ as well as an indication towards large negative CP phase ($\delta \approx -\frac{\pi}{2}$) which are consistent with $\tm$. The fact that the data naturally satisfies the two $\tm$ constraints helps us to fit the six observables with the four model parameters.

We scan the parameter space (having four real degrees of freedom - $\mw^2/\mg$, $k$, $\text{Re}(z)$, $\text{Im}(z)$ and generate the neutrino masses and the mixing observables. The parameter values which correspond to the observables lying outside their experimental $3\sigma$ ranges are omitted. For the allowed parameter values, we perform a chi-squared goodness of fit analysis. We evaluate 
\begin{equation}
\chi^2 = \sum_i \left(\frac{(\text{Obs}_i)_\text{model}-(\text{Obs}_i)_\text{expt}}{\sigma_i}\right)^2,
\end{equation}
where $(\text{Obs}_i)_\text{model}$ are the observables generated using the model parameters, $(\text{Obs}_i)_\text{expt}$ are the experimental best fit values taken from \cite{Esteban:2018azc} and $\sigma_i$ are the corresponding $1\sigma$ errors. The summation is made over the six observables: $\sin^2 \theta_{12}$, $\sin^2 \theta_{13}$, $\sin^2 \theta_{23}$, $\sin^2 \delta$, $\Delta m^2_{21}$ and $\Delta m^2_{31}$. By minimising $\chi^2$, we obtain the best fit values of the model parameters and also the corresponding values of the observables.

\begin{figure}[h]
\centering
\includegraphics[scale=0.5,trim={0 3.6cm 0 3.6cm},clip]{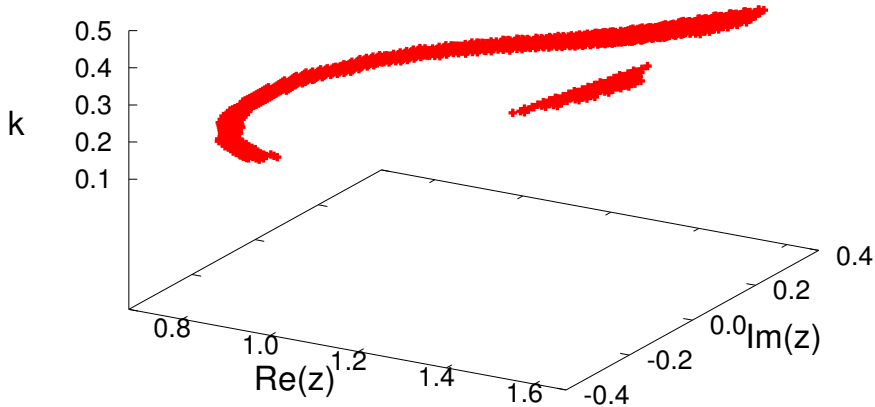}
\caption{The vertical axis represents the parameter $\kd$ ~and the horizontal plane represents the parameter $\km$. Im($\km$) indicates the breaking 
of $\mu\text{-}\tau$-reflection symmetry. The plot is not
symmetric about the Re($z$) axis because the deviation of the experimental data from $\mu\text{-}\tau$-reflection
symmetry is skewed towards a higher value for $\sin^2 \theta_{23}$ away from $0.5$, i.e.~$\sin^2 \theta_{23} =0.428 \leftrightarrow 0.624$.}
\label{fig:parameters}
\end{figure}

\begin{figure}[h]
\centering
\includegraphics[width=9.5cm,height=8.5cm,angle=0]{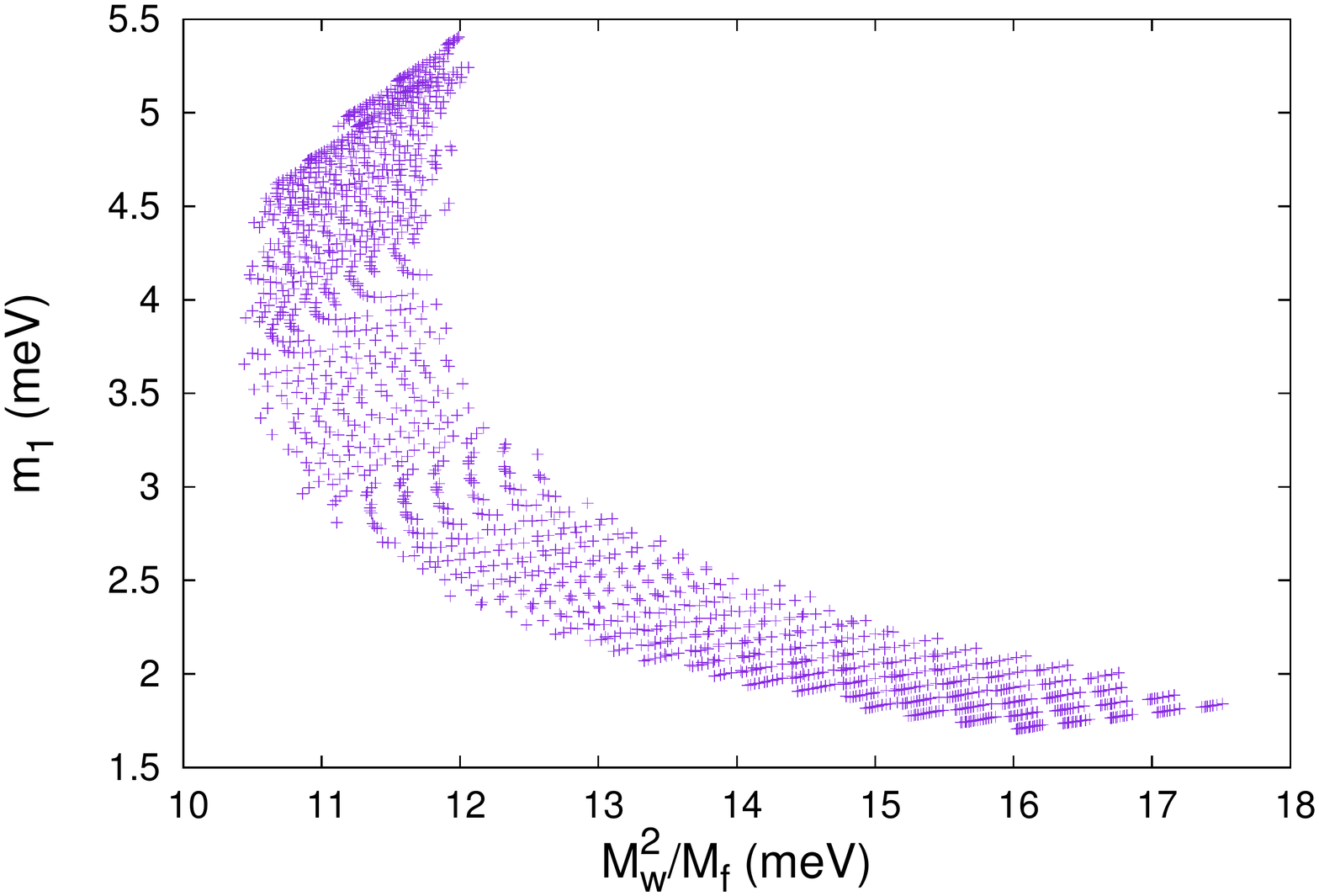}
\caption{The predicted ranges of $m_1$ and $\frac{\mw^2}{\mg}$.}
\label{fig:massscale}
\end{figure} 

\begin{figure}[h]
\centering
\includegraphics[width=9.5cm,height=8.5cm,angle=0]{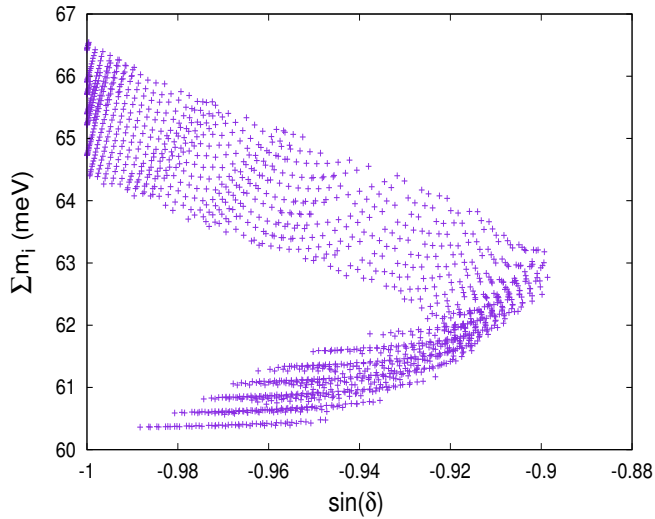}
\caption{The predicted ranges of $\Sigma m_i$ and $\sin\delta$.}
\label{fig:mdelta}
\end{figure}

The allowed parameter space for the dimensionless parameters $\kd$ and $z$ is given in Figure~\ref{fig:parameters}. The parameter $\kd$ generates non-zero reactor angle. When $\kd = 0$, we obtain TBM mixing which is experimentally ruled out. Non-zero reactor angle consistent with the experimental range is obtained with $0.1<\kd<0.5$. Im$(\km) = 0$ leads to $\mu\text{-}\tau$-reflection symmetry ($\theta_{23}=\frac{\pi}{4}$) which lies within the $3\sigma$ range of the oscillation data. The best fit values  for the parameters obtained by the $\chi^2$ analysis are $\kd=0.186$, $\km=0.994+i~0.332$ and $\mw^2/\mg = 11.36~\text{meV}$. They correspond to $\sin^2\theta_{12}=0.318$, $\sin^2\theta_{23}=0.583$, $\sin^2\theta_{13}=0.0224$, $\sin\delta=-0.93$ and $(m_1, m_2, m_3)=(3.67,9.34,50.38)$~meV. Because of the second $\tm$ constraint, Eq.~(\ref{eq:sindelta}), our analysis results in the prediction of the CP phase, $-1<\sin\delta<0.90$. 

Apart from the fact the model leads to $\tm$ mixing, it also helps us to calculate the overall mass scale of the effective seesaw matrix, i.e.~$\frac{\mw^2}{\mg}$ and thus predict the individual neutrino masses. Figure~\ref{fig:massscale} shows the prediction of the light neutrino mass $m_1$ as a function of the mass scale $\frac{\mw^2}{\mg}$. Cosmological observations provide the upper bound on the sum of the three light neutrino masses, $\sum\limits_i m_i = m_1+m_2+m_3$~\cite{Aghanim:2018eyx, Loureiro:2018pdz, Choudhury:2018byy, Vagnozzi:2017ovm}. The model predicts $60.3~\text{meV}<\sum m_i< 66.5~\text{meV}$ which is consistent with the current cosmological bounds. In Figure~\ref{fig:mdelta}, we show the predicted range of $\sum_i m_i $ against that of $\sin\delta$. We note that the model parameter space contains points corresponding to the inverted mass ordering of the light neutrinos also. However, we have omitted these points because the global oscillation analysis disfavours inverted ordering. Also, the inverted ordering in our model leads to $\sum_i m_i>150~\text{meV}$ which is disfavoured by the cosmological bound obtained under certain assumptions~\cite{Aghanim:2018eyx, Loureiro:2018pdz, Choudhury:2018byy, Vagnozzi:2017ovm}. The effective Majorana mass applicable in neutrinoless double-beta decays is given by $m_{\beta\beta}= \left|\sum_i U_{ei}^2 m_i\right|$ where $U_{ei}$ are the elements of the first row of the $U_\text{PMNS}$ matrix. From the allowed parameter space of the model, we calculate the effective mass and obtain $2.89~\text{meV}<m_{\beta\beta}<8.02~\text{meV}$. This range is well below the bounds set by the current $0\nu\beta\beta$ experiments~\cite{Anton:2019wmi, Agostini:2018tnm, Gando:2018kyv, Alduino:2017ehq, Azzolini:2019tta}.

\subsection{Constraining through baryon asymmetry bound} 
It is clear from the discussion of the previous section that, upon imposing the $3\sigma$ experimental bound of neutrino oscillation data, the scale factor sitting outside the effective light neutrino mass matrix (Eq.~(\ref{eq:msfinal})) is constrained to vary within a certain range. This factor contains $\mw^2$ in the numerator and $\mg$ in the denominator. 
The overall multiplicative factor to the Dirac neutrino mass matrix, i.e.
$\mw$, which consists of a Yukawa type coupling $(y_{D_1})$ with the Higgs VEV and the flavon VEV reduced by the scale $\Lambda$, is not an experimentally known quantity. However, we expect that $\mw$ is around the typical mass scale of a fermion i.e. it can be assigned values ranging from the top quark mass ($\sim 170$ GeV) down to the electron mass ($0.5$ MeV) (which are of course the two extremes). Therefore for a fixed value of the ratio $(\mw^2/\mg)$, we can get different values of $\mg$ (which is the scale of the RH neutrino mass) by varying $\mw$\footnote{or in other words, it can be said that for a fixed value of the ratio $(\mw^2/\mg)$ (taken from Figure~\ref{fig:massscale}), $\mg$ is proportional to the square of $\mw$}. This allows us to vary the RH neutrino masses over a wide range which thereby opens up the possibility
to study Leptogenesis in different regimes. Let us first proceed to analyze
the numerical results of unflavoured Leptogenesis and then examine whether we can put any constrain on $\mw$ (which was unbounded by oscillation data) by the baryon asymmetry bound.

\subsubsection{Unflavoured regime}
\begin{figure}[!h]
\centering
\includegraphics[scale=0.5,trim={0 3.6cm 0 3.6cm},clip]{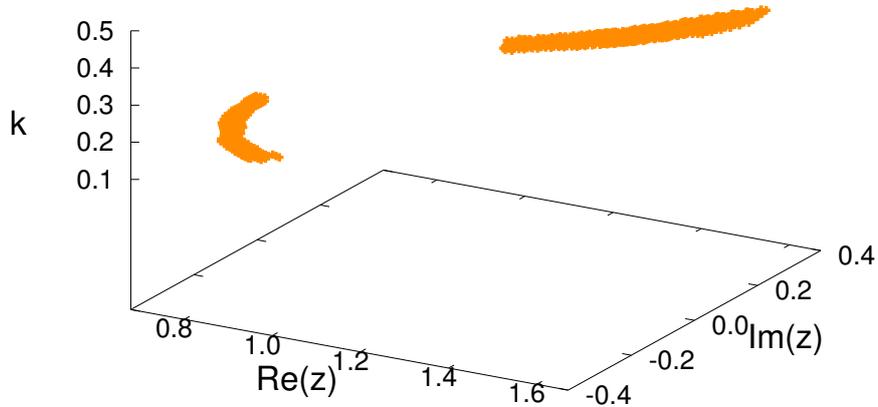}
\caption{3D parameter space after getting second round of restriction by the  requirement of positive baryon asymmetry.}
\label{unflav_para}
\end{figure}

\begin{figure}[!h]
\centering
\includegraphics[width=9.5cm,height=8.5cm,angle=0]{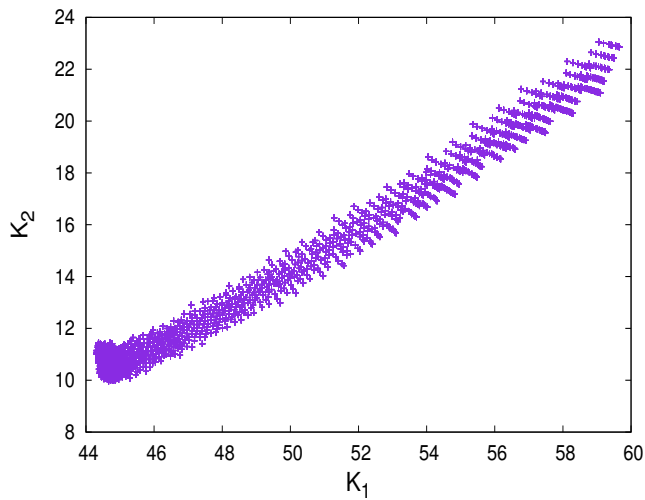}
\caption{Decay parameters corresponding to lightest $(N_1)$ and next-to-lightest
$(N_2)$ RH neutrinos for the parameter space allowed by $3\sigma$ global fit of oscillation data.}
\label{k2k3}
\end{figure}

It can be understood from the analytical formula of baryon asymmetry (Eq.~(\ref{etab})) that the sign of the final baryon asymmetry will depend on the sign of the CP asymmetry parameter (since the efficiency factor $\kappa$ is always positive). The expression of unflavoured CP asymmetry parameter (Eq.~(\ref{uncp})) shows that its sign is completely determined by the phase of $z$. Therefore the multiplicative factor $\mw^2/\mg$ of the $M_{ss}$ matrix
has no role in determining the sign of $\varepsilon_i$ (or equivalently $Y_B$). Therefore using the constraint $Y_B > 0$ we can further constrain the $3D$ parameter space which is already constrained by the $3\sigma$ global fit of oscillation data. The new parameter space after constraining by the condition of positive baryon asymmetry is shown in Figure~\ref{unflav_para}.
It is clear from Figure~\ref{unflav_para} that the parameter space has been reduced by a considerable amount with respect to that of Figure~\ref{fig:parameters}. The sign of $Y_B$ is determined by the set $\{k, \text{Re}(z), \text{Im}(z),\mw^2/\mg\}$, whereas the magnitude is controlled by the scale factor $\mw$. All the allowed points shown in Figure~\ref{unflav_para} has the potential to generate $Y_B$ within the experimentally allowed range.
For a definite set of values of $\{k, \text{Re}(z), \text{Im}(z), \mw^2/\mg\}$, $\mw$ can be constrained by the bound on $Y_B$ ($8.522<Y_B\times 10^{11}<9.375$\cite{Aghanim:2016yuo}). 
This exercise can be repeated for all the allowed points shown in Figure~\ref{unflav_para} and for each such set a bound on $\mw$ 
can be obtained with the imposition of baryon asymmetry bound. Now, $Y_B$ can be calculated either by direct numerical solution
of the Boltzmann equations or by using the analytical formulas of efficiency factors $(\kappa^f_i,\kappa_{\rm fit})$. In Figure~\ref{k2k3}, we show the $3\sigma$
allowed range of the decay parameters corresponding to the lightest $(N_1)$ and
next-to-lightest $(N_2)$ RH neutrinos. It shows their preference towards strong
washout regime. 

\begin{figure}[!h]
\centering
\includegraphics[width=9.5cm,height=8.5cm,angle=0]{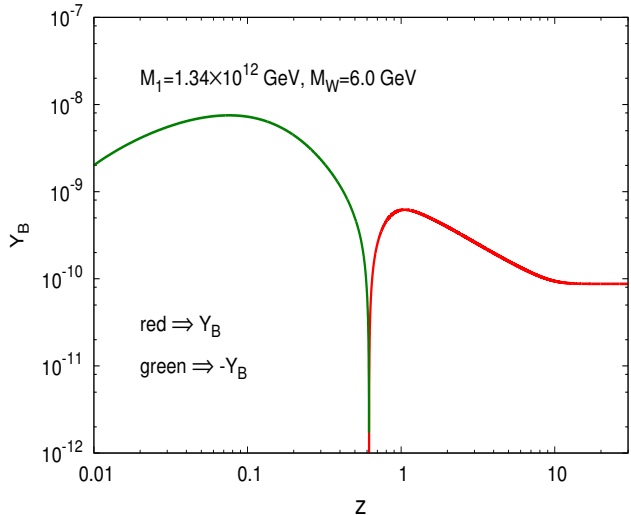}
\caption{variation of $Y_B$ with $z$ for best fit values of lagrangian parameters while the value of $\mw$ is chosen
such that final $Y_B$ freezes to a value in the experimental range.}
\label{YBunflav}
\end{figure}

\begin{figure}[!h]
\centering
\includegraphics[width=7cm,height=7cm,angle=0]{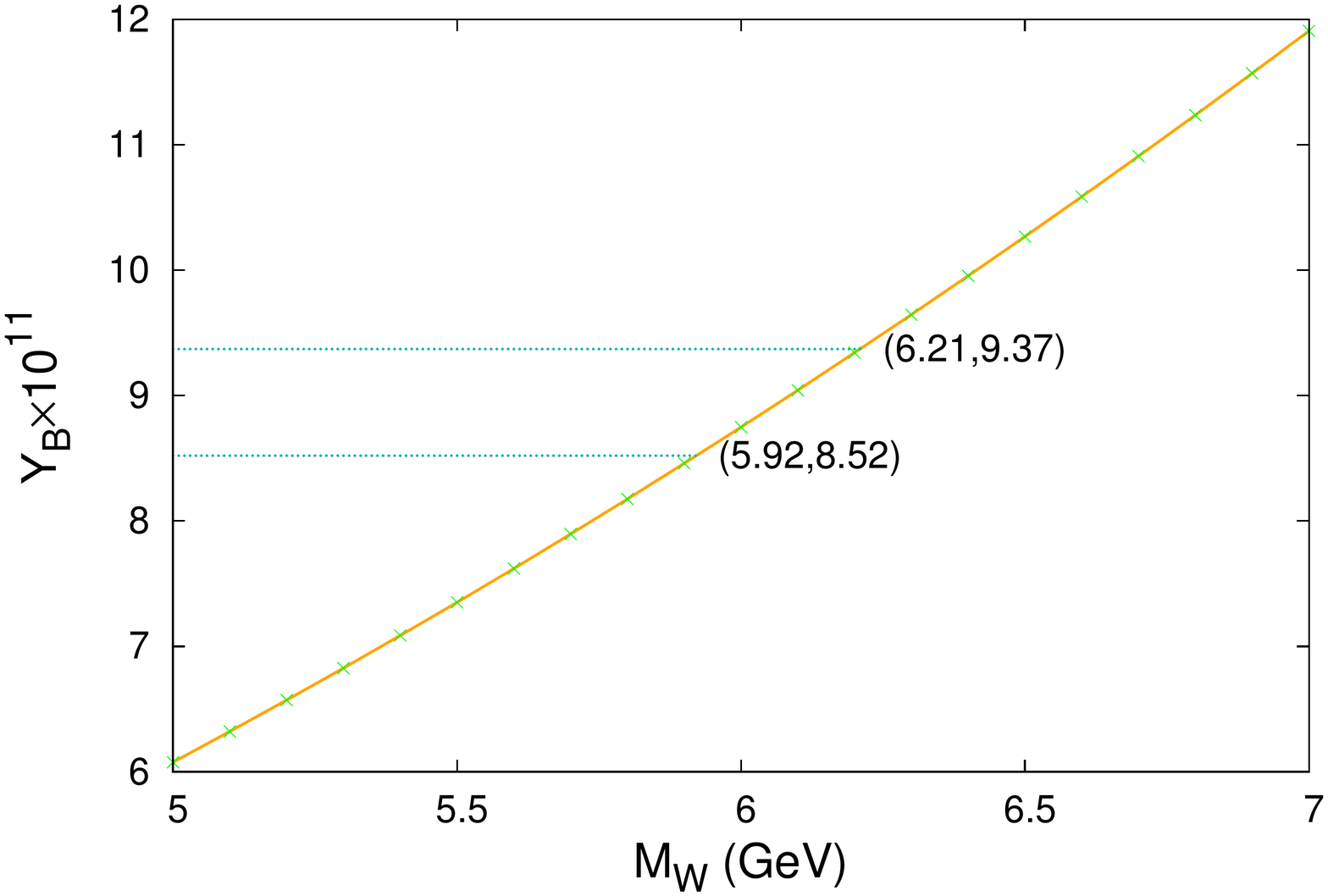}
\includegraphics[width=7cm,height=7cm,angle=0]{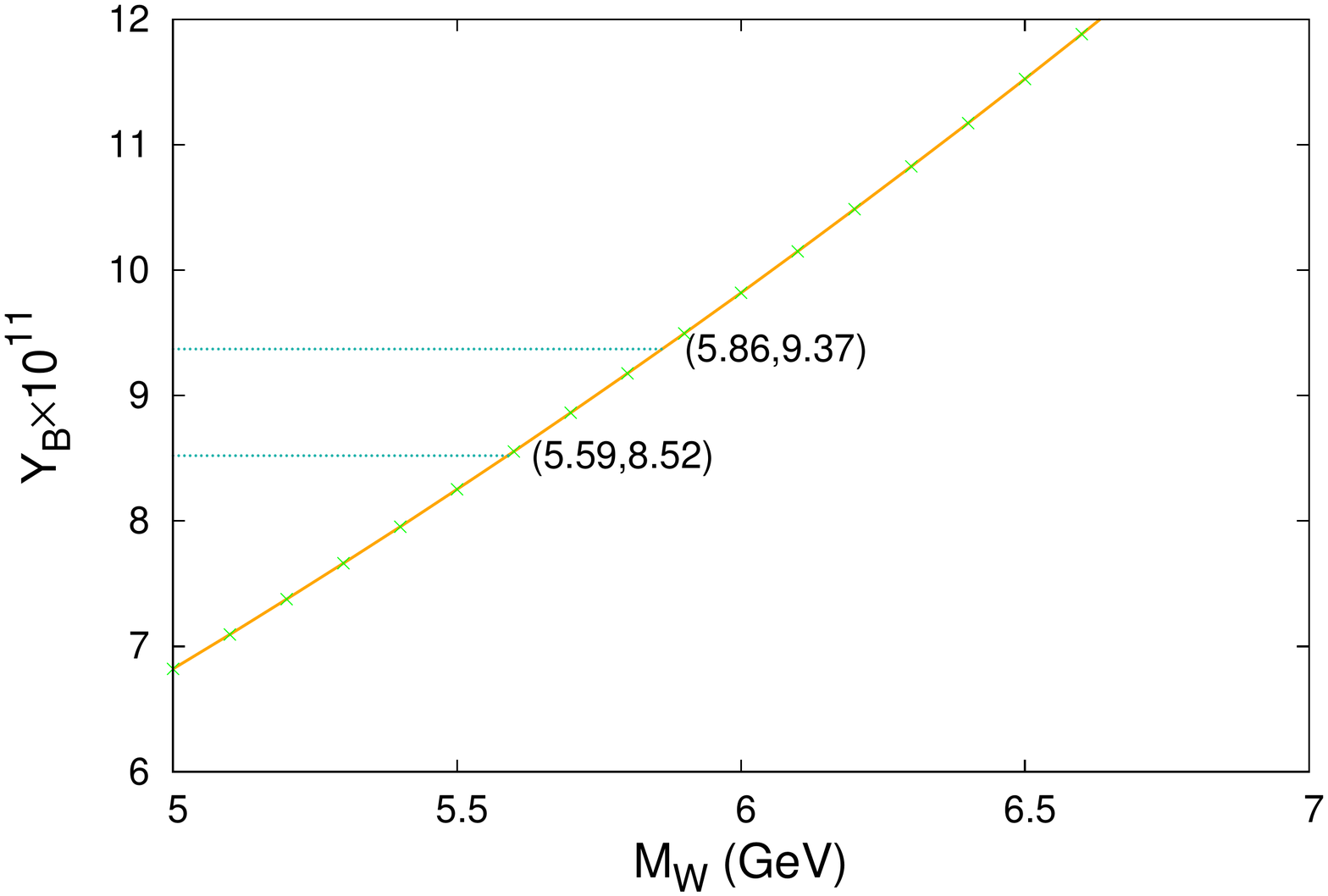}
\caption{The variation of final $Y_B$ with $\mw$ (other Lagrangian parameters fixed at best fit) and determination of bound on 
$\mw$ using the baryon asymmetry bound. Left panel: $Y_B$ is evaluated by direct solution of Boltzmann equations, 
Right panel: $Y_B$ is calculated using $\kappa^f_i$.}
\label{YB_MW}
\end{figure}

Using the set of values of $\{k, \text{Re}(z), \text{Im}(z),\mw^2/\mg\}$ corresponding to least $\chi^2$  (which is sometimes referred to as the best fit 
point) as the bench mark point and keeping $\mw$ fixed at a certain value, we solve the set of Boltzmann equations numerically
and show the variation of $Y_B$ with $z$ in Figure~\ref{YBunflav}. 
The value of $Y_B$ at high $z$ (or equivalently at low temperature) where it doesn't change with $z$ anymore is the final 
value of baryon asymmetry. We have repeated the same procedure for different values of $\mw$ keeping the other lagrangian
parameters fixed at the best fit value and shown the variation of final $Y_B$ with $\mw$  graphically in Figure~\ref{YB_MW} where the left panel is generated by direct numerical 
solution of Boltzmann equations whereas the plot of the right panel is generated using analytic approximation $\kappa^f_i$.
In this plot, we draw two lines
parallel to the $\mw$ axis which represents experimental upper and lower bounds on $Y_B$ respectively. The corresponding $\mw$
coordinates where these straight lines intersects the $Y_B-\mw$ curve represent the upper and the lower bounds on $\mw$ 
respectively. 
To get an idea how precisely the analytical formulas can reproduce the results of the actual Boltzmann solution, 
we show in tabular (Table \ref{ybmw1}) form the final values of $Y_B$ for different values of $\mw$ obtained by 
direct Boltzmann solution as well as using $\kappa^f_i$ and $\kappa_{\rm fit}$. Percentage errors in relation to the 
actual solution of Boltzmann equations are tabulated.
\begin{table}[!ht]
\caption{Comparison between the final baryon asymmetry evaluated by direct numerical solution of Boltzmann equation
(denoted by $(Y_B)_N$ in the table) with that calculated using analytic formulas, $\kappa^f_i$ and $\kappa_{\rm fit}$(corresponding baryon  
asymmetry parameters are denoted by $(Y_B)_{\kappa}$
and $(Y_B)_{\kappa_{\rm fit}}$ respectively).
$E_1$ and $E_2$ are percentage errors given by $E_1\%=\Bigg| \frac{ (Y_B)_{\kappa} - (Y_B)_N}{ (Y_B)_N} \Bigg| \times 100 $ and
$E_2\%=\Bigg| \frac{ (Y_B)_{\kappa_{\rm fit}} - (Y_B)_N}{ (Y_B)_N} \Bigg| \times 100 $.}
\label{ybmw1}
\begin{center}
\begin{tabular}{|c|c|c|c|c|c|c|c|}
\hline
$\mw$  & $M_{1}\times10^{-12}$ & $M_{2}/M_{1}$ & $(Y_B)_N\times 10^{11}$ & $(Y_B)_{\kappa}\times 10^{11}$ & $(Y_B)_{\kappa_{\rm fit}}\times 10^{11}$ & $E_1\%$ & $E_2\%$\\
(GeV)    & (GeV) &  &  &  &  & & \\
\hline
$5.2$ & $1.01$ & $1.90$ & $6.07$ & $6.81$ & $6.53$ & $12.17$& $7.45$\\
$5.4$ & $1.09$ & $1.90$ & $7.08$ & $7.95$ & $7.61$ & $12.28$ & $7.48$\\
$5.6$ & $1.17$ & $1.90$ & $7.62$ & $8.55$ & $8.19$ & $12.20$ & $7.48$\\
$5.8$ & $1.26$ & $1.90$ & $8.17$ &  $9.17$ & $8.78$ & $12.23$ & $7.46$\\
$6.0$ & $1.34$ & $1.90$ & $8.74$ & $9.81$ & $9.40$ & $12.24$ & $7.55$\\
$6.2$ & $1.44$ & $1.90$ & $9.34$ & $10.48$ & $10.04$ & $12.20$ & $7.49$\\
$6.4$ & $1.53$ & $1.90$ & $9.95$ & $11.17$& $10.70$ & $12.26$ & $7.53$\\
$6.6$ & $1.63$& $1.90$  & $10.58$ & $11.88$ & $11.37$ & $12.28$ & $7.46$\\
\hline
\end{tabular}
\end{center}
\end{table}
It is clear from Table \ref{ybmw1} that analytical formulas produce results very close to the actual solution of Boltzmann 
equations. The error, if we use $\kappa_{\rm fit}$, is even less than $10\%$. Thus we can readily use the analytical formulas  
to scan the whole parameter space (Figure~\ref{unflav_para}) and estimate the bound on $\mw$ for all the points belonging to the parameter
space constrained by $3\sigma$ oscillation data with positive $Y_B$ bound. 
\begin{figure}[!h]
\centering
\includegraphics[width=5.0cm,height=5.0cm,angle=0]{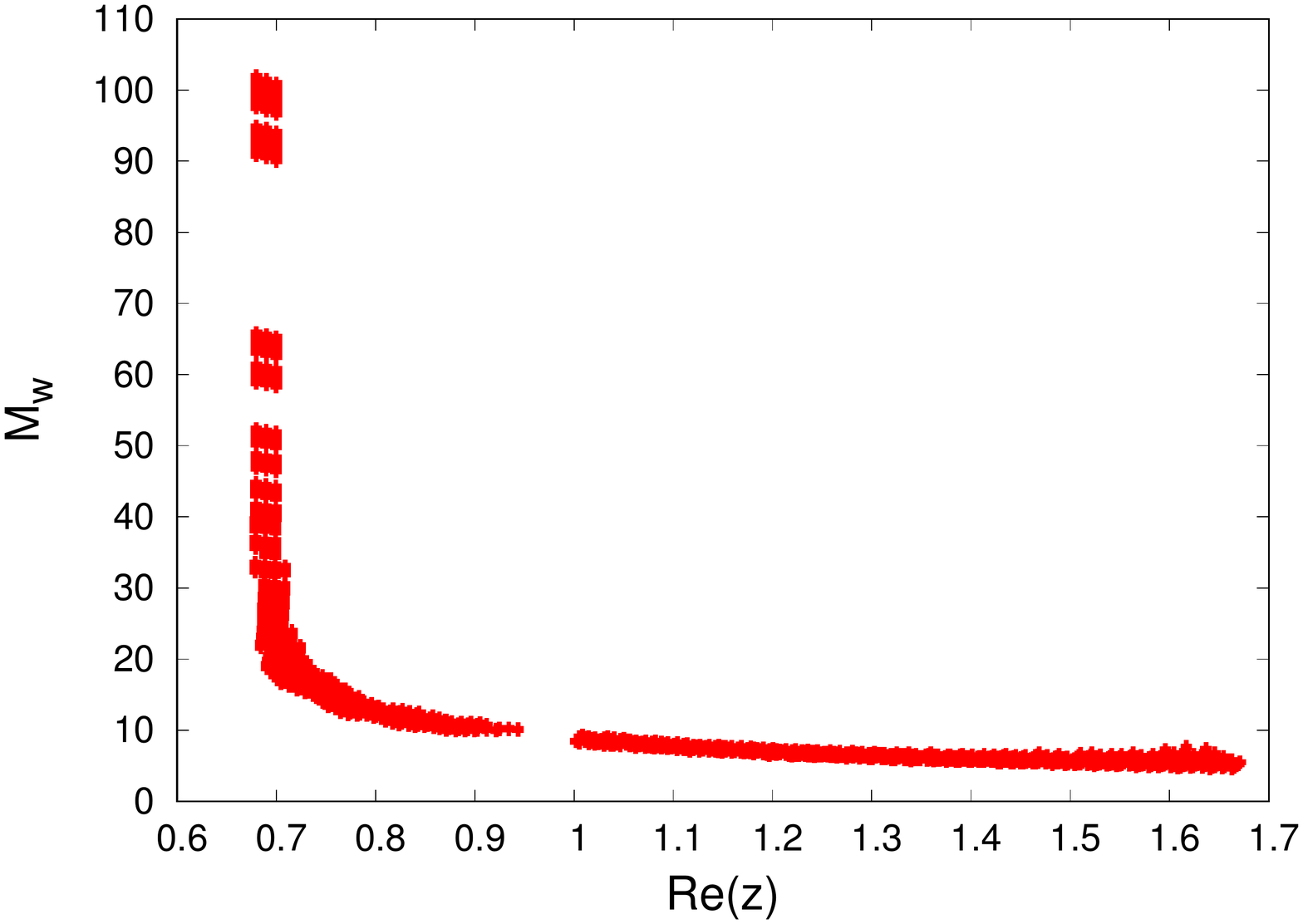}
\includegraphics[width=5.0cm,height=5.0cm,angle=0]{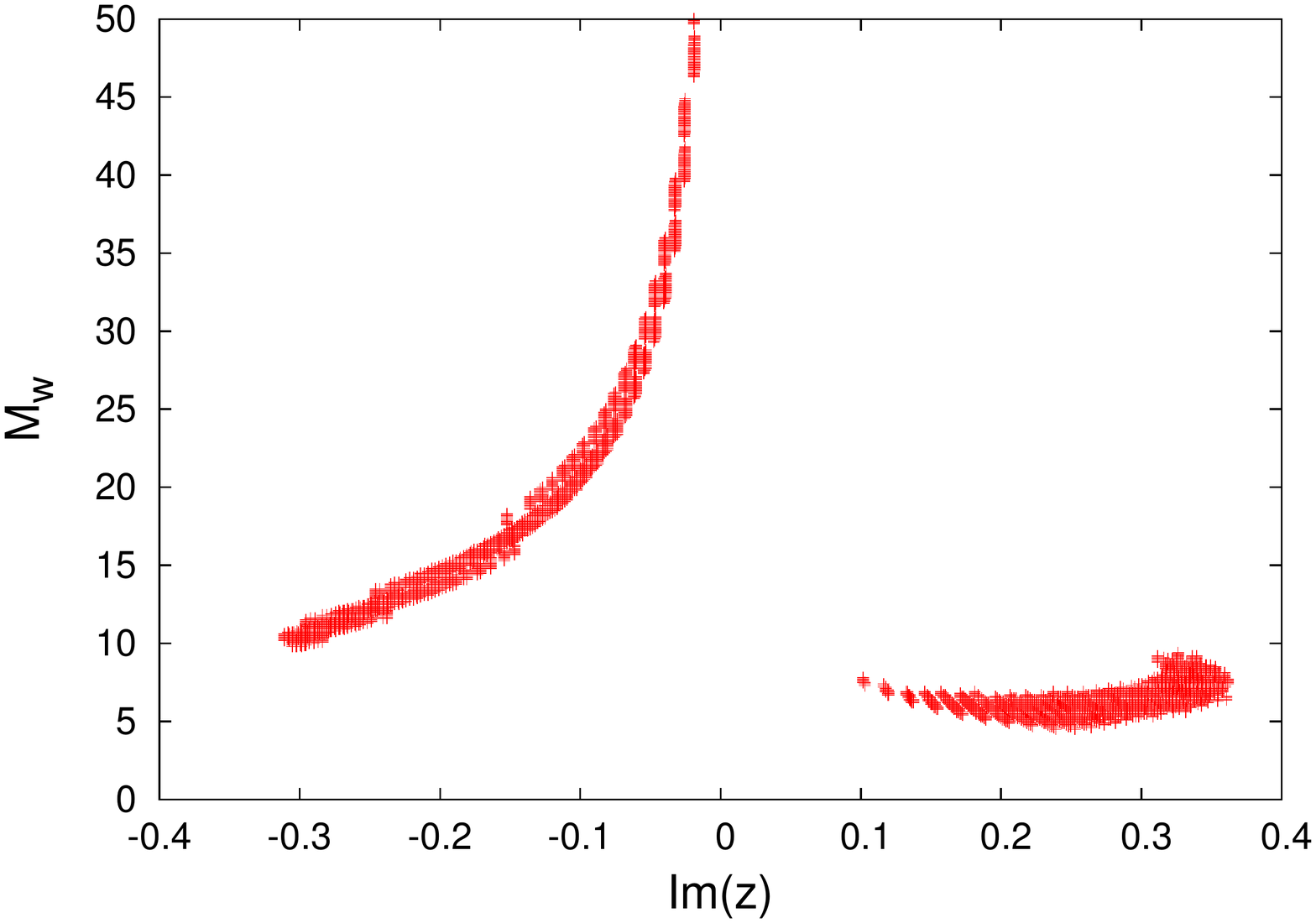}
\includegraphics[width=5.0cm,height=5.0cm,angle=0]{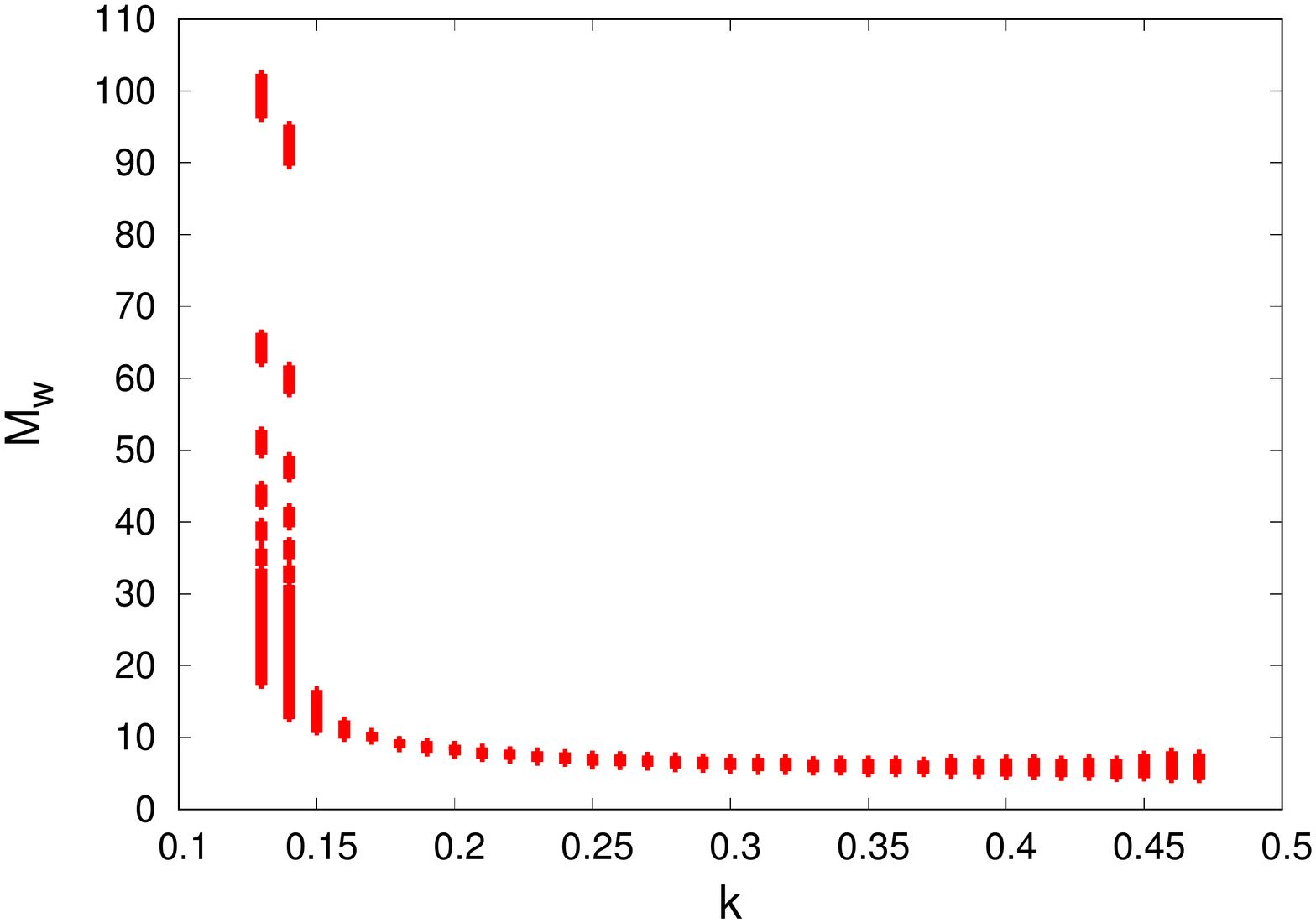}
\caption{Allowed $\mw$ constrained by the baryon asymmetry bound for the whole $3\sigma$ parameter space of oscillation data.
($Y_B$ has been calculated using $\kappa^f_i$. The result will not vary much with that of $\kappa_{\rm fit}$ ). }
\label{MWrange}
\end{figure}
In Figure~\ref{MWrange}, we show graphically the variation of $\mw$ with the allowed range of the Lagrangian parameters. The spread in 
$\mw$ for each value of the variable along the abscissa is the corresponding bound on $\mw$ imposed by the $Y_B$ bound.

\subsubsection{$\tau$-flavoured regime} It has been noticed during the analysis of unflavoured regime that (specially from the Table~\ref{ybmw1}) the experimentally observed 
value of baryon asymmetry is obtained for such value of $M_1$ (lightest right handed neutrino) which lies at the lower edge of the unflavoured regime. 
So, there is a possibility of generating adequate asymmetry in the $\tau$-flavoured regime (since washout in flavoured leptogenesis is less than that of unflavoured case). 
In this context we want to remind the reader about the pure $N_2$ contribution (asymmetry generated by $N_2$ decay along $\tau^\perp_\perp$). Since this asymmetry does not suffer
washout of lightest RH neutrino inverse decay it has a significant contribution to the final asymmetry (its typical value is comparable to that of lightest RH neutrino). 
It has been examined that the CP asymmetry $(\varepsilon^\perp_{2\tau^\perp})$ associated to this contribution is always negative 
throughout the whole parameter space which ascertain that this asymmetry will always have a positive contribution 
towards final baryon asymmetry. 

\begin{figure}[!h]
\centering
\includegraphics[width=7.5cm,height=7.5cm,angle=0]{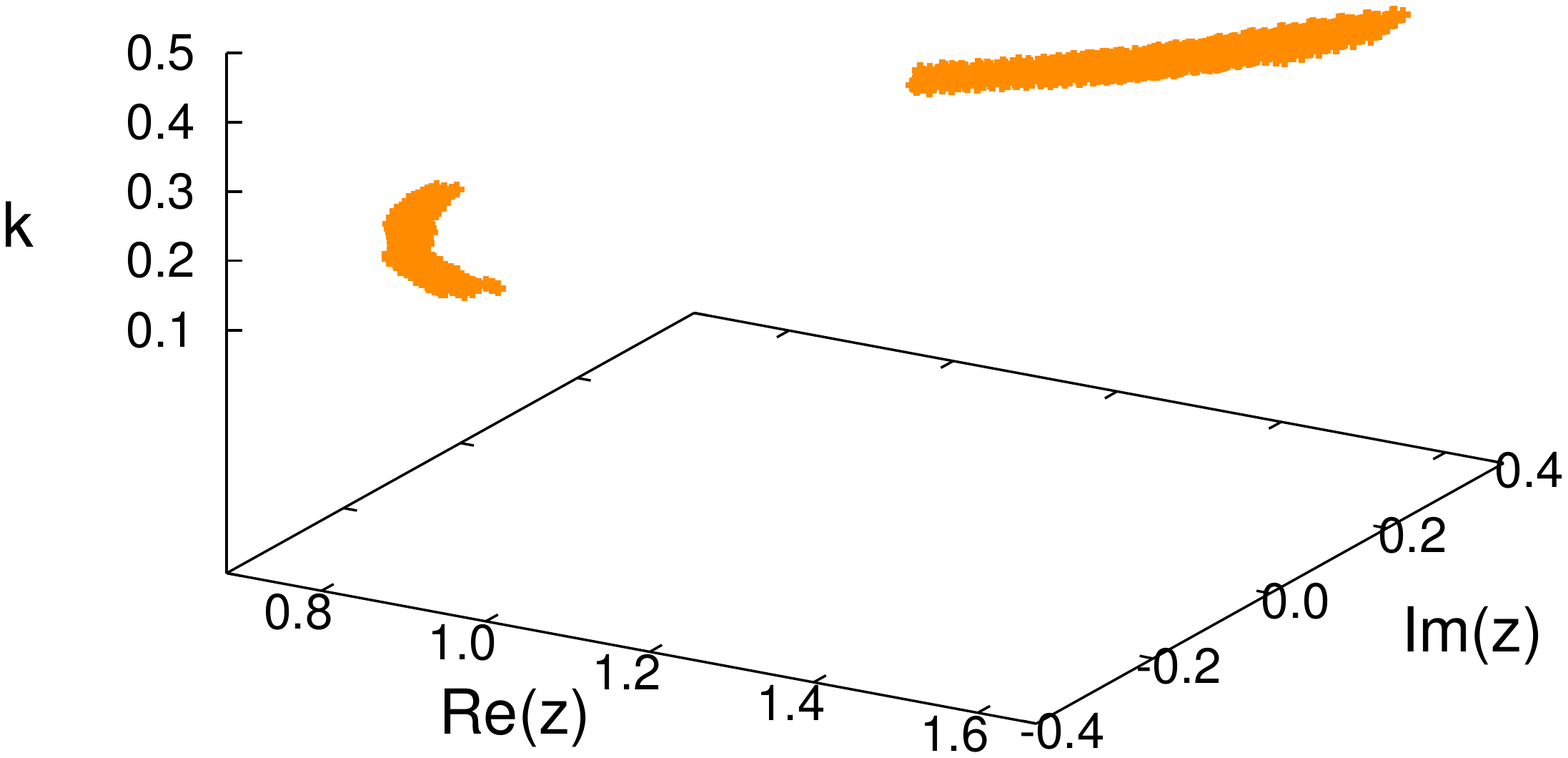}
\includegraphics[width=7.5cm,height=7.5cm,angle=0]{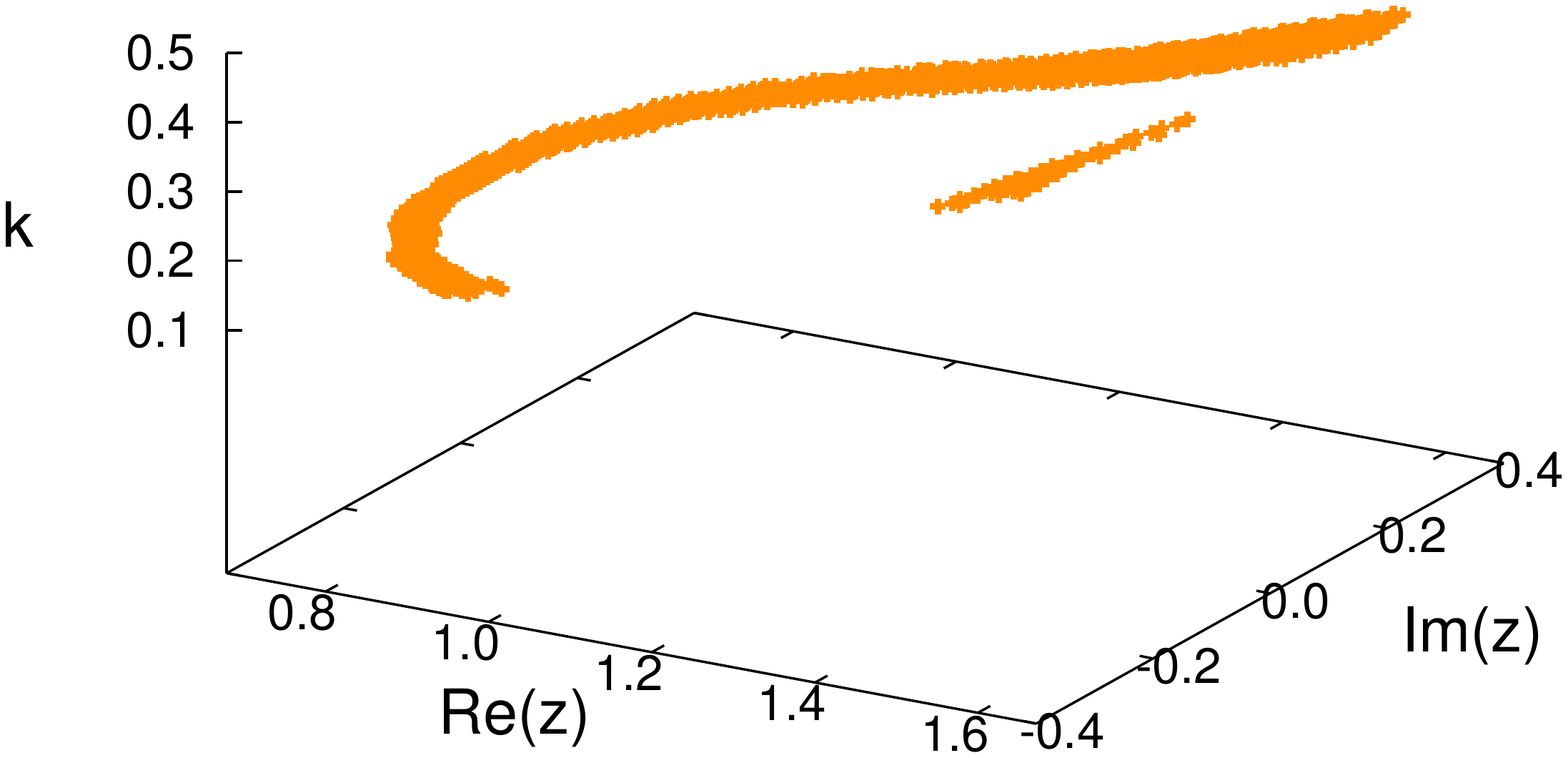}
\caption{The parameter space constrained by $3\sigma$ oscillation data as well as the requirement of positive baryon asymmetry. Left panel: $Y_B$ estimated without pure $N_2$ contribution. Right panel: $Y_B$ is calculated taking into account this pure $N_2$ contribution.}
\label{para_2flav}
\end{figure}

\begin{figure}[!h]
\centering
\includegraphics[width=9.5cm,height=8.5cm,angle=0]{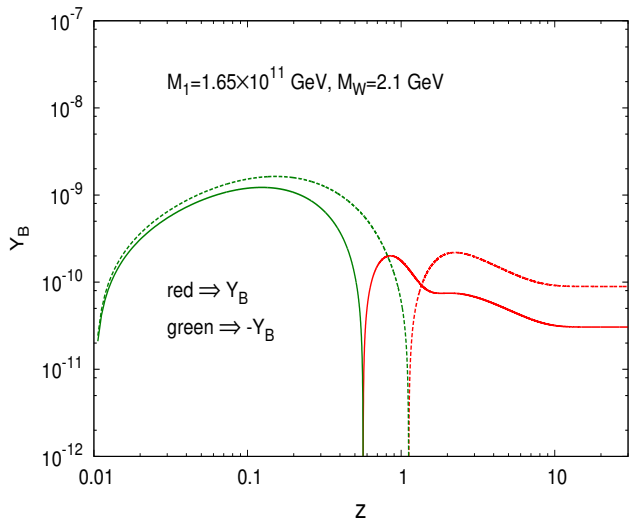}
\caption{Variation of $Y_B$ with $z$, where the dashed line represents baryon asymmetry with pure $N_2$ contribution and 
the solid line represents $Y_B$ excluding it. Lagrangian parameters are fixed at best fit and $\mw$ is chosen to 
be $2.1$ GeV. At this value of $\mw$, we can get experimentally allowed asymmetry only if the pure $N_2$ contribution is 
taken into account.}
\label{YB2flav}
\end{figure}

We follow exactly the same line of analysis as we have done in the case of unflavoured leptogenesis. We constrain the parameter space (already restricted by $3\sigma$ oscillation data) by the requirement of positive baryon asymmetry. This exercise is repeated twice first without taking into account the pure $N_2$ contribution and then using it in the formula for total $Y_B$. The corresponding restricted parameter spaces are shown in the left and right panels of Figure~\ref{para_2flav}.
The parameter space without pure $N_2$ contribution is very much similar to its unflavoured counterpart, whereas the inclusion of pure $N_2$ contribution results in a larger parameter space. This happens because of the remarkable positive pure $N_2$ contribution which drives the $Y_B$ value to the positive side and thus a few more points appear in the parameter space of the right panel (which was absent in the left panel). The effect of pure $N_2$ contribution is shown vividly in Figure~\ref{YB2flav} where we show the variation of $Y_B$ with $z$ (while the set $\{k, \text{Re}(z),\text{Im}(z),\mw^2/\mg\}$ is kept fixed at its best fit value and a definite value of $\mw$ is used such that the final $Y_B$ freezes to the experimentally admissible range).
$Y_B$ with or without pure $N_2$ contribution is denoted by dashed or solid lines respectively. Addition of the new contribution results in a 
three fold gain in the final asymmetry. It is clear from the plot that, for $\mw=2.1$ GeV the final baryon asymmetry generated by the usual components of asymmetry falls short of the experimental lower limit, whereas the inclusion of the new contribution easily drives this value within the admissible range. 
\begin{table}[!ht]
\caption{Comparison between the final baryon asymmetry evaluated by direct numerical solution of Boltzmann equations
with those calculated using the analytical formulas (i.e. using $\kappa^f_{i\alpha}$). $(Y_B)_{1}$ and $(Y_B)_{2}$ denote the results produced by the solution of Boltzmann equations with and without pure $N_2$ contribution respectively. Similarly $(Y_B)_{\kappa_1}$ and $(Y_B)_{\kappa_2}$ are the values of final baryon asymmetry evaluated using the analytical formula. 
$E$ is the percentage errors given by $E\%=\Bigg| \frac{ (Y_B)_{\kappa_1} - (Y_B)_{1}}{ (Y_B)_{1}} \Bigg| \times 100 $.}
\label{ybmw2}
\begin{center}
\begin{tabular}{|c|c|c|c|c|c|c|c|}
\hline
$\mw$  & $M_{1}\times10^{-12}$ & $M_{2}/M_{1}$ & $(Y_B)_{1}\times$ & $(Y_B)_{2}\times$ & $(Y_B)_{\kappa_1}\times$ & $(Y_B)_{\kappa_2}\times$ & $E\%$\\
(GeV)    & (GeV) &  & $10^{11}$ & $10^{11}$ & $10^{11}$ & $10^{11}$ & \\
\hline
$2$    & $1.49$ & $1.90$ & $8.09$ & $2.76$ & $8.87$ & $3.07$ & $9.6$\\
$2.04$ & $1.55$ & $1.90$ & $8.42$ & $2.87$ & $9.23$ & $3.19$ & $9.61$\\
$2.08$ & $1.62$ & $1.90$ & $8.75$ & $2.99$ & $9.6$ & $3.32$ & $9.71$\\
$2.12$ & $1.68$ & $1.90$ & $9.09$ &  $3.10$ & $9.97$ & $3.45$ & $9.68$\\
$2.16$ & $1.74$ & $1.90$ & $9.44$ & $3.22$ & $10.35$ & $3.58$ & $9.63$\\
$2.20$ & $1.81$ & $1.90$ & $9.79$ & $3.34$ & $10.73$ & $3.71$ & $9.60$\\
\hline
\end{tabular}
\end{center}
\end{table}
We show a tabular (Table \ref{ybmw2}) comparison of numerical solution of full Boltzmann equations versus the analytical
approximations (Eq.~(\ref{kff}), Eq.~(\ref{ybf})). The unconstrained $\mw$ has been varied from $2$ to $2.2$~GeV, whereas $\{k, \text{Re}(z), \text{Im}(z), \mw^2/\mg \}$ is fixed at the best fit (i.e. the point corresponding to least $\chi^2$).
The $\%$ errors of the analytical approximations in comparison to the actual solution of Boltzmann equations are also shown in the same table.

\begin{figure}[!h]
\centering
\includegraphics[width=7.5cm,height=7.5cm,angle=0]{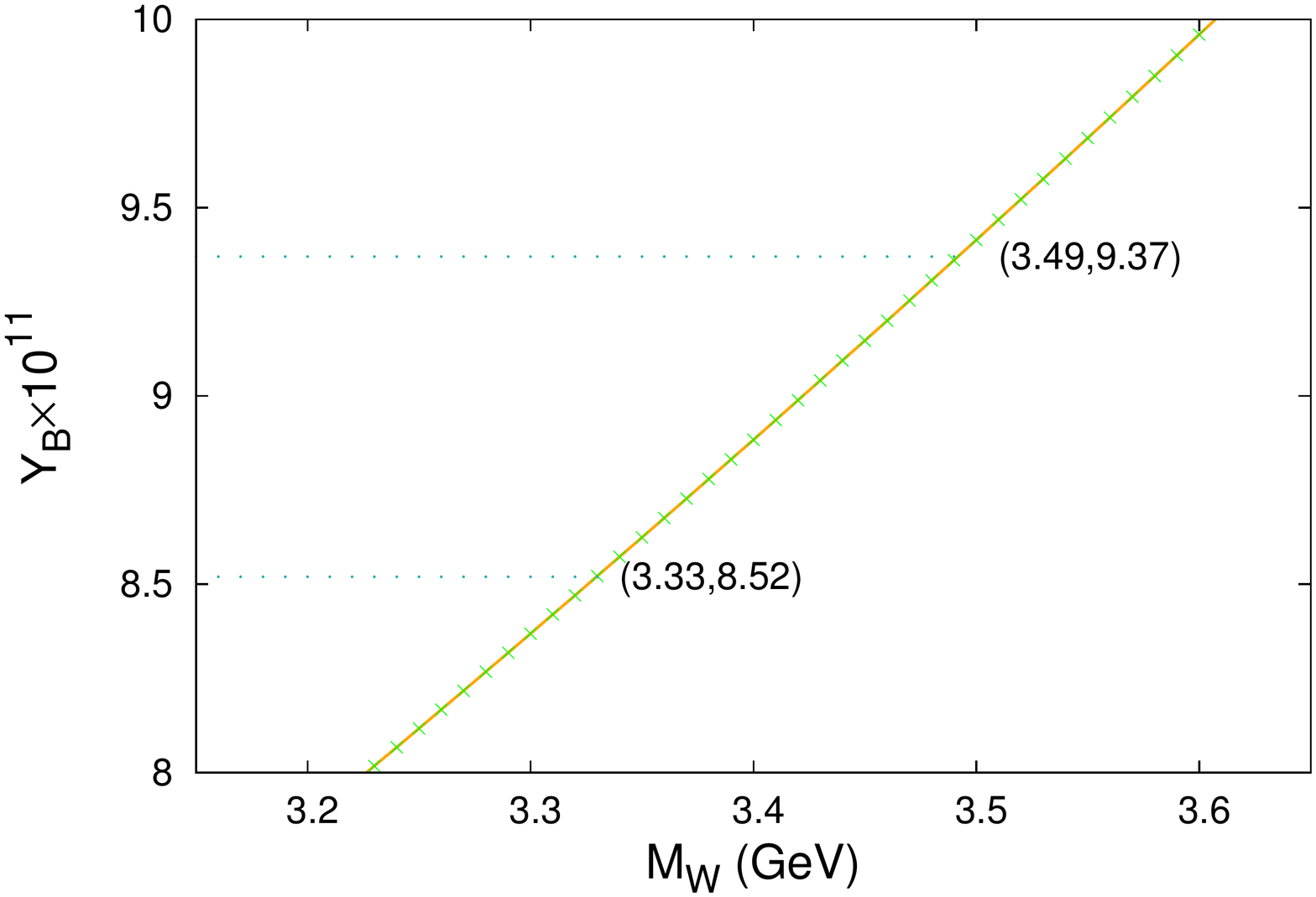}
\includegraphics[width=7.5cm,height=7.5cm,angle=0]{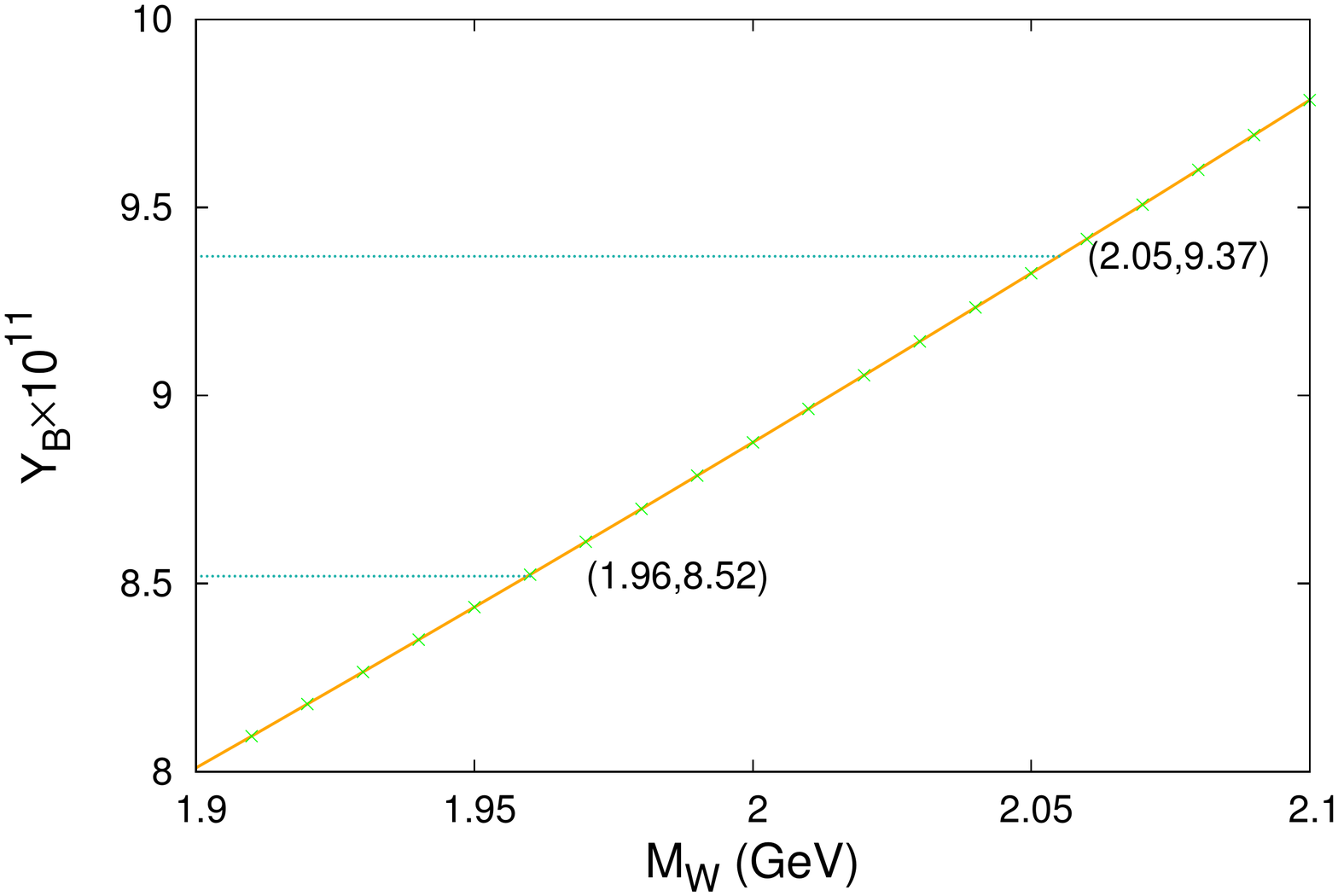}
\caption{Variation of final $Y_B$ with $\mw$ where the set of parameters $\{k, \text{Re}(z), \text{Im}(z) \mw^2/\mg \}$ kept 
fixed at their best fit value. In the left panel, pure $N_2$ contribution has not been considered whereas the plot of 
right panel takes into account this new contribution.}
\label{YB_MW2}
\end{figure}

It is found that the error is always less than $10\%$ which allows us to 
scan the whole $3\sigma$ parameter space using analytical 
formulas instead of solving the chain of Boltzmann equations. The allowed range of $\mw$ obtained
by imposing the experimental bound on $Y_B$ is depicted in Figure~\ref{YB_MW2}. 
It is clearly visible from the plots
that use of the new asymmetry component shifts this required $\mw$ (to generate $Y_B$ within the experimental
range) to a lower value. This analysis is carried out for a fixed set (corresponding to least $\chi^2$) of values 
of the Lagrangian parameters $\{ k, \text{Re}(z), \text{Im}(z), \mw^2/\mg \}$ picked from $3\sigma$ parameter space. For a fixed
value of the ratio $\mw^2/\mg$, $\mg$ increases linearly with square of $\mw$. Thus the bound on $\mw$ dictates a restriction on 
$M_1$ (the lightest RH neutrino) too. So, we can say that the inclusion of the new asymmetry component helps us 
to get the required baryon asymmetry at a lower RH neutrino mass. In the present work, although we are successful in getting the flavoured leptogenesis even excluding the pure $N_2$ contribution, situations may arise where adequate asymmetry in the 
flavoured regime can be obtained only if we consider this new contribution. 

\begin{figure}[!h]
\centering
\includegraphics[width=5.0cm,height=5.0cm,angle=0]{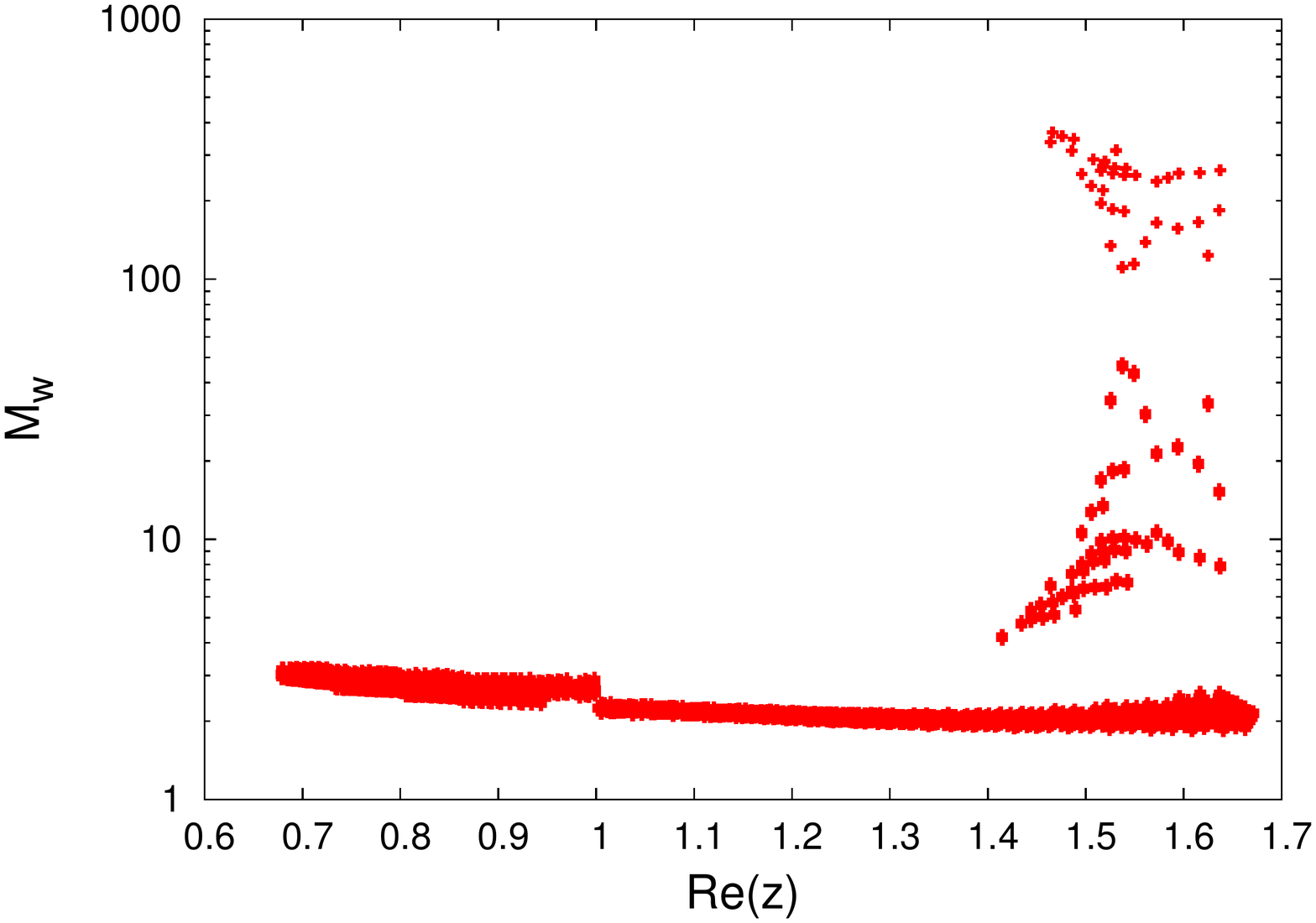}
\includegraphics[width=5.0cm,height=5.0cm,angle=0]{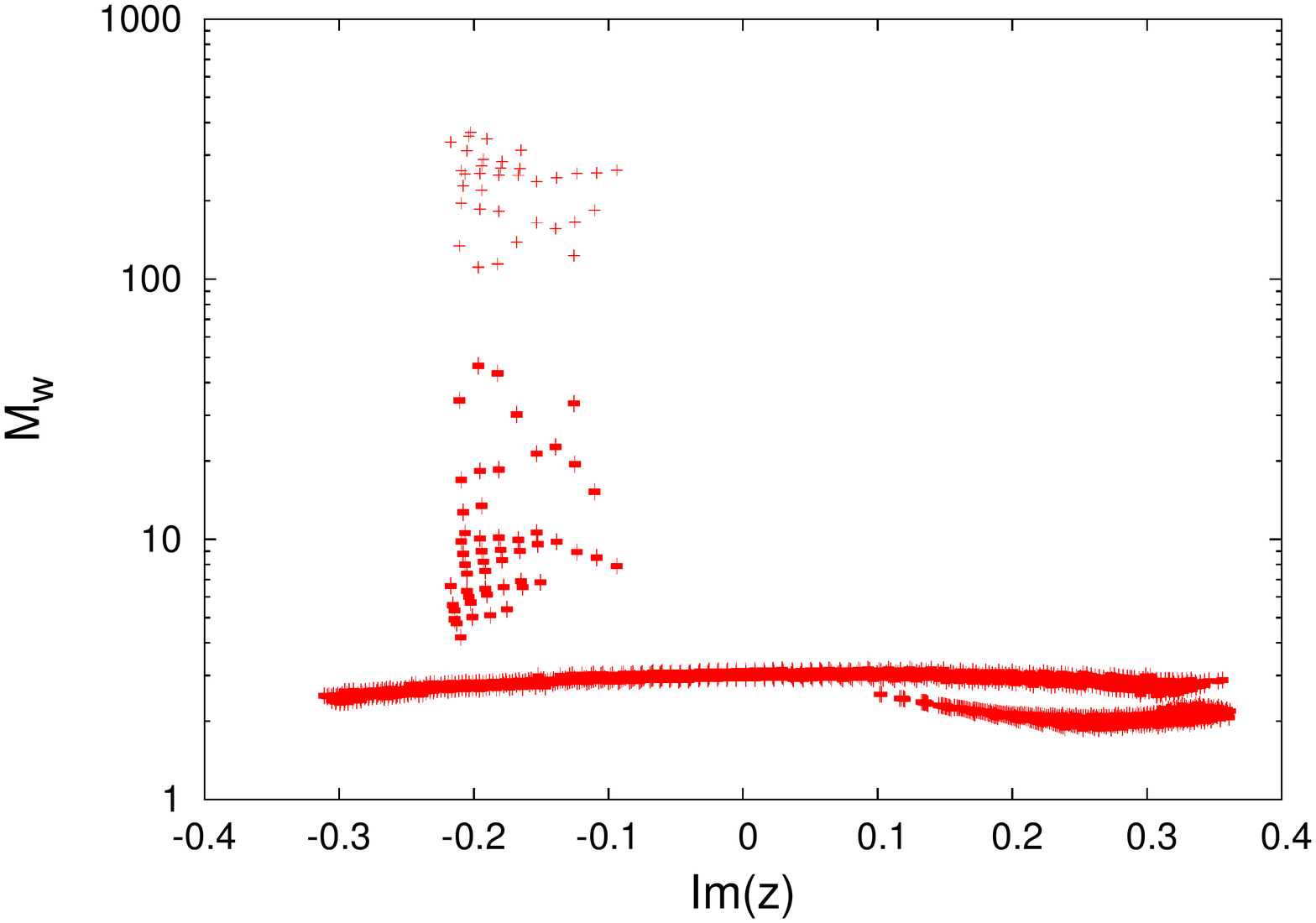}
\includegraphics[width=5.0cm,height=5.0cm,angle=0]{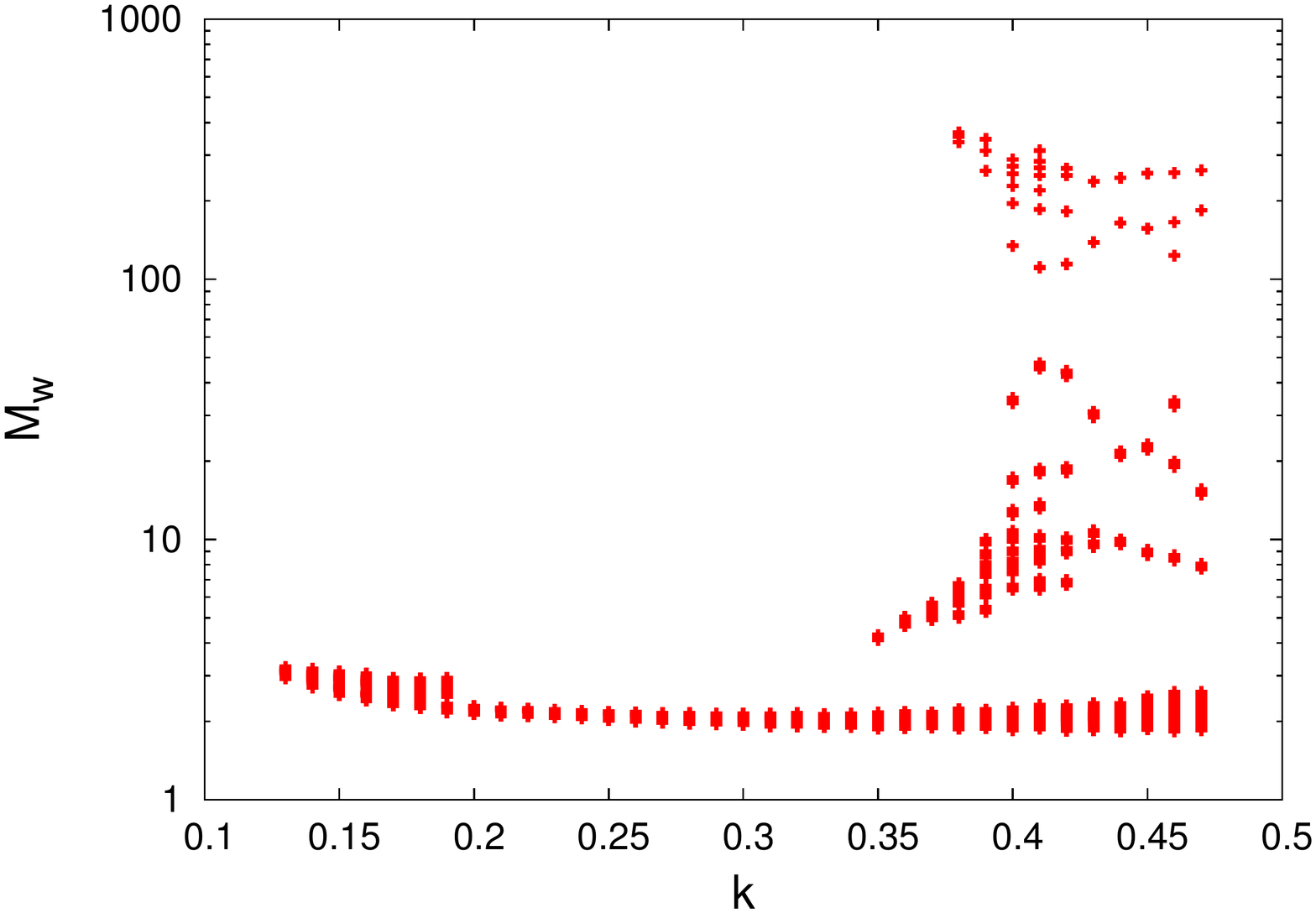}
\caption{The allowed range of $\mw$ constrained by the baryon asymmetry bound for the whole $3\sigma$ parameter space of 
oscillation data. $Y_B$ is calculated using $\kappa^f_{i\alpha}$ and pure $N_2$ contribution is also taken into account. }
\label{MWrange1}
\end{figure}

The final $Y_B$ for each and every point belonging to the whole $3\sigma$ parameter space has been calculated using the 
$\kappa^f_{i\alpha}$ formula. In Figure~\ref{MWrange1}, we show\footnote{It is to be noted that $Y_B$ 
is calculated considering the pure $N_2$ contribution.} the graphical representation of the variation
of $\mw$ with $k$, $\text{Re}(z)$, $\text{Im}(z)$. The spread in $\mw$ for a definite value of the abscissa signifies the bound on $\mw$
obtained due to the imposition of baryon asymmetry bound.

\section{Concluding summary}\label{summ}
We have investigated a flavon model based on Standard Model with $S_4$ discrete symmetry group adhering to Type-I seesaw 
mechanism. Our model leads to $\text{TM}_1$ mixing through the incorporation of appropriate flavon fields.
In our model neutrino oscillation phenomenology is described using four parameters. We carry out a chi-squared analysis
fitting these parameters with the three mixing angles, the CP phase and the two mass-squared differences. The fact that the 
$TM_1$ mixing has two inbuilt constraints which are consistent with the data enables us to successfully carry out this fit.
One of these constraints leads to near-maximal breaking of the CP symmetry, $-1< \sin \delta <-0.9$. The specific texture
of the seesaw mass matrix in the model results in the prediction of the light neutrino masses. We show that their
values are consistent with the $\sum m_i$ and $m_{\beta\beta}$ bound. Furthermore, we have studied baryogenesis
via leptogenesis in two different ways and have also shown their equivalence. Lagrangian parameters already constrained by 
the $3\sigma$ limit of oscillation data are used in the calculations of leptogenesis. 
We have successfully generated baryon asymmetry within the experimentally observed range through flavoured and unflavoured leptogenesis.
Only high energy parameters contribute to the unflavoured leptogenesis whereas the asymmetry in the 
flavoured case gets non trivial contribution from both high energy as well as low energy (Dirac and Majorana type)
CP Phases.
The estimation of final 
baryon asymmetry has been carried out by solving the network of coupled Boltzmann equations as well as using appropriate 
analytical fits. Equivalence between these two methods has been shown clearly with corresponding numerical results. 
Moreover we have also shown the substantial contribution from $N_2$ leptogenesis in the context of flavoured leptogenesis.


\appendix

\section{Construction of the flavon potentials} \label{pots}

\subsection{Charged-lepton sector}

The flavon $\fc=(\fcx,\fcy,\fcz)^T$ couples in the charged-lepton sector. Besides being a triplet ($\rt$) under $S_4$, $\fc$ also transforms as $\om$ under $C_3$, Table~\ref{tab:flavourcontent}. Therefore, its components have complex degrees of freedom. Using Eq.~(\ref{eq:tp1}-\ref{eq:tp3}), we construct the following multiplets that are quadratic in $\fc$:
\begin{align}
\left(\fc^*\fc\right)_{\rs}& = \fcx^* \fcx + \fcy^* \fcy + \fcz^* \fcz\,,\label{eq:fd1}\\
\left(\fc\fc\right)_{\rs}& = \fcx^2 + \fcy^2 + \fcz^2\,,\\
\left(\fc\fc\right)_{\rd}& = \left(2 \fcx^2 - \fcy^2 - \fcz^2, \sqrt{3} (\fcy^2 - \fcz^2)\right)^T\,,\\
\left(\fc\fc\right)_{\rtp}& = 2 \left(\fcy \fcz, \fcz \fcx, \fcx \fcy\right)^T\,.\label{eq:fd3}
\end{align}

Note that $\left(\fc\fc\right)_{\rt}$, which is antisymmetric under the exchange of the two constituent triplets, vanishes. The only invariant term at the quadratic order is
\begin{equation}\label{eq:fcquad}
{\mathcal{T}(\fc^2)} = \left(\fc^*\fc\right)_{\rs}.
\end{equation}
We do not have any invariant at the cubic order. We construct the following invariants at the quartic order:
\begin{align}
{\mathcal{T}_1(\fc^4)} &= \left(\fc\fc\right)_{\rs}^* \left(\fc\fc\right)_{\rs}, \\
{\mathcal{T}_2(\fc^4)} &= \left(\fc\fc\right)_{\rd}^\dagger\left(\fc\fc\right)_{\rd},\\
{\mathcal{T}_3(\fc^4)} &= \left(\fc\fc\right)_{\rtp}^\dagger\left(\fc\fc\right)_{\rtp}.\label{eq:fcquart}
\end{align}
 $\left({\mathcal{T}(\fc^2)}\right)^2$ does not form an independent invariant since it is related to the above ones,
\begin{equation}
6({\mathcal{T}(\fc^2)})^2 = 2{\mathcal{T}_1(\fc^4)}+{\mathcal{T}_2(\fc^4)}+3{\mathcal{T}_3(\fc^4)}.
\end{equation}
Using Eqs.~(\ref{eq:fcquad}-\ref{eq:fcquart}), we construct the flavon potential,
\begin{align}
\mathcal{V}_C = k_1 {\mathcal{T}_1(\fc^4)}+ k_2 {\mathcal{T}_2(\fc^4)}+ k_3 {\mathcal{T}_3(\fc^4)}-4(3k_2+2k_3) \vc^2 \, {\mathcal{T}(\fc^2)},
\end{align}
where $k_1$, $k_2$, $k_3$ and $\vc$ are real constants; $\vc$ has mass dimension one while the others are dimensionless. $\mathcal{V}_C$, consisting of four independent invariants and four arbitrary constants, is the most general potential that can be constructed using renormalisable terms. The extrema of this potential can be found by calculating its first derivatives with respect to the components of $\fc$. It is straightforward to verify that the set of extrema are given by,
\begin{equation}\label{eq:arbphase}
\fc=e^{i \theta} g_i  \, \vc (1,\om,\ob)^T,
\end{equation}
where $e^{i \theta}$ is an arbitrary phase\footnote{This phase corresponds to the accidental $U(1)$ symmetry, 
$\fc\rightarrow e^{i \theta} \fc$, of our potential which will be broken by the higher-order non-renomalisable terms. 
Since the multiplication of the VEV $\langle\fc\rangle$ with a constant phase has no observable consequence in our model, we ignore it.} 
and $g_i$ represent the elements of the group under which $\fc$ transforms i.e.~$S_4\times C_3$. Through SSB one of these 
extrema\footnote{Whether the extrema are minima, maxima or saddle points is determined by the values of 
the parameters $k_1$, $k_2$ and $k_3$. We have verified that the extrema corresponds to minima in a large region of
this parameter space.} becomes the VEV, Eq.~(\ref{eq:vc}). 

\subsection{Dirac neutrino sector}
 
The singlet $\nd$ and the triplet ($\rtp$) $\fd=(\fdx,\fdy,\fdz)^T$ couple in the neutrino Dirac mass terms. They transform as $-1$ under
the $C_2$ subgroup of the $C_6$ group, Table~\ref{tab:flavourcontent}, and they are real fields. Using $\nd$, we obtain the following 
quadratic and quartic invariants:
\begin{equation}\label{eq:ndinvs}
{\mathcal{T}(\nd^2)} = \nd^2, \quad {\mathcal{T}(\nd^4)} = \nd^4.
\end{equation}
Using two $\fd$s, we construct a singlet, a doublet and a triplet:
\begin{align}
\left(\fd\fd\right)_{\rs} &= \fdx \fdx + \fdy \fdy + \fdz \fdz\,, \\
\left(\fd\fd\right)_{\rd} &= \left(2 \fdx^2 - \fdy^2 - \fdz^2, \sqrt{3} (\fdy^2 - \fdz^2)\right)^T\,,\\
\left(\fd\fd\right)_{\rtp} &= 2 \left(\fdy \fdz, \fdz \fdx, \fdx \fdy\right)^T\,.
\end{align}
The singlet forms the quadratic invariant,
\begin{equation}\label{eq:fdquadinv}
{\mathcal{T}(\fd^2)} = \left(\fd\fd\right)_{\rs}~.
\end{equation}
At the quartic order, we obtain the invariants,
\begin{align}
{\mathcal{T}_1(\fd^4)} &= \left(\left(\fd\fd\right)_{\rs}\right)^2,\label{eq:fdquart1} \\
{\mathcal{T}_2(\fd^4)} &= \left(\fd\fd\right)_{\rd}^T\left(\fd\fd\right)_{\rd},\label{eq:fdquart2}\\
{\mathcal{T}_3(\fd^4)} &= \left(\fd\fd\right)_{\rtp}^T\left(\fd\fd\right)_{\rtp}.
\end{align}
They satisfy the relation,
\begin{equation}
4{\mathcal{T}_1(\fd^4)}-{\mathcal{T}_2(\fd^4)}-3{\mathcal{T}_3(\fd^4)}=0.
\end{equation}
Hence we have only two independent invariants at the quartic order, ${\mathcal{T}_1(\fd^4)}$ and ${\mathcal{T}_2(\fd^4)}$. 
At the cubic order, we construct the $S_4$ invariant,
\begin{equation}
\fd^T \left(\fd\fd\right)_{\rtp} = 6 \fdx \fdy \fdz,
\end{equation}
which transforms as $-1$ under $C_2$. This term is coupled with $\nd$ to form the invariant,
\begin{equation}\label{eq:ndfd}
{\mathcal{T}(\nd \fd^3)} = 6 \nd \fdx \fdy \fdz ~.
\end{equation}

We use Eqs.~(\ref{eq:ndinvs}, \ref{eq:fdquadinv}, \ref{eq:fdquart1}, \ref{eq:fdquart2}, \ref{eq:ndfd}) to construct the most general
renormalisable potential involving $\nd$ and $\fd$,
\begin{align}
\begin{split}
\mathcal{V}_D = & k_\eta {\mathcal{T}(\nd^4)}+k_1 {\mathcal{T}_1(\fd^4)}+ k_2 {\mathcal{T}_2(\fd^4)}+ k_c\frac{\vds}{\vdt}{\mathcal{T}(\nd \fd^3)}\\
&-(2 k_\eta \vds^2 +3 k_c \vdt^2){\mathcal{T}(\nd^2)}-(3k_c \vds^2+6k_1 \vdt^2){\mathcal{T}(\fd^2)},
\end{split}
\end{align}
where $k_\eta$, $k_1$, $k_2$, $k_c$ are dimensionless constants whereas $\vds$, $\vdt$ are constants with mass dimension one. 
Note that the total number of arbitrary constants (6) matches with the total number of invariants in the potential. 
By calculating the first order derivatives of $\mathcal{V}_D$, we can show that its extrema correspond to
\begin{equation}
\nd = \pm \vds, \quad \fd = \pm g_i \vdt (1,1,1),
\end{equation}
where $g_i$ are the elements of $\rtp$ of $S_4$ and $\pm$ corresponds to the $C_2$ group acting on $\nd$ and $\fd$. 
One set of alignments among these extrema is chosen as the VEVs, Eq.~(\ref{eq:vd1}, \ref{eq:vd3}).

\subsection{Majorana neutrino sector}

In the Majorana sector, we have the flavons $\nm$ and $\fm$ which form a singlet and a triplet ($\rtp$) respectively under $S_4$. They also transform as $\om$ under the $C_3$ subgroup of $C_6$, Table~\ref{tab:flavourcontent}. Using $\nm$, we construct the quadratic and the quartic invariants,
\begin{equation}\label{eq:nminvs}
{\mathcal{T}(\nm^2)} = \nm^*\nm, \quad {\mathcal{T}(\nd^4)} = (\nm^*\nm)^2.
\end{equation}
Similar to Eqs.~(\ref{eq:fd1}-\ref{eq:fd3}), we construct quadratic multiplets in terms of $\fm$,
\begin{align}
\left(\fm^*\fm\right)_{\rs}& = \fmx^* \fmx + \fmy^* \fmy + \fmz^* \fmz\,,\\
\left(\fm\fm\right)_{\rs}& = \fmx^2 + \fmy^2 + \fmz^2\,,\\
\left(\fm\fm\right)_{\rd}& = \left(2 \fmx^2 - \fmy^2 - \fmz^2, \sqrt{3} (\fmy^2 - \fmz^2)\right)^T\,,\\
\left(\fm\fm\right)_{\rtp}& = 2 \left(\fmy \fmz, \fmz \fmx, \fmx \fmy\right)^T\,,
\end{align}
We obtain the quadratic and the quartic invariants,
\begin{align}
{\mathcal{T}(\fm^2)} &= \left(\fm^*\fm\right)_{\rs},\label{eq:fmquad}\\
{\mathcal{T}_1(\fm^4)} &= \left(\fm\fm\right)_{\rs}^* \left(\fm\fm\right)_{\rs}, \\
{\mathcal{T}_2(\fm^4)} &= \left(\fm\fm\right)_{\rd}^\dagger\left(\fm\fm\right)_{\rd},\\
{\mathcal{T}_3(\fm^4)} &= \left(\fm\fm\right)_{\rtp}^\dagger\left(\fm\fm\right)_{\rtp}.\label{eq:fmquart}
\end{align}
We also have two quartic invariants involving both $\nm$ and $\fm$,
\begin{align}
{\mathcal{T}(\nm^2\fm^2)} &=\nm^{*2}\left(\fm\fm\right)_{\rs},\\
{\mathcal{T}(\nm\fm^3)} &=\nm^* \fm^\dagger \left(\fm\fm\right)_{\rtp}\label{eq:fmmix}.
\end{align}
Note that these two invariants are complex. Using Eqs.~(\ref{eq:nminvs}, \ref{eq:fmquad}-\ref{eq:fmmix}), we construct the potential,
\begin{align}
\begin{split}
\mathcal{V}_M = & k_\eta {\mathcal{T}(\nm^4)}+k_1 {\mathcal{T}_1(\fm^4)}+ k_2 {\mathcal{T}_2(\fm^4)}+ k_3 {\mathcal{T}_3(\fm^4)}\\
&+ \text{Re[}k_c e^{i(\xi_1-\xi_3)} {\mathcal{T}(\nm^2 \fm^2)}\text{]}+ \text{Re[}k_z {\mathcal{T}(\nm \fm^3)}\text{]}\\
&-(2 k_\eta \vms^2 + k_c \vmt^2 ){\mathcal{T}(\nm^2)}-(k_c \vms^2+(2k_1 +8 k_2)\vmt^2){\mathcal{T}(\fm^2)},
\end{split}
\end{align}
where $\vms$ and $\vmt$ have mass dimension one and $k_\eta$, $k_1$, $k_2$, $k_3$, $k_c e^{i(\xi_1-\xi_3)}$ and $k_z$ are dimensionless. 
Two constants, i.e.~$k_c e^{i(\xi_1-\xi_3)}$ and $k_z$, are complex while the rest, i.e.~$\vms$, $\vmt$, $k_\eta$, $k_1$, $k_2$ 
and $k_3$ are real. $\mathcal{V}_M$ is constructed with two complex and six real invariants which matches with the number of 
arbitrary constants. Therefore, $\mathcal{V}_M$ is the most general potential that can be constructed using the given 
renormalisable terms\footnote{In this potential, we have avoided the terms of the cubic order such 
as $\fm^T\left(\fm\fm\right)_{\rtp}$, $\nm^3$ and $\nm \left(\fm\fm\right)_{\rs}$ to keep the analysis simpler. 
Cubic terms can be forbidden by imposing a $C_2$ symmetry, $(\nm,\fm)\rightarrow -(\nm,\fm)$. This $C_2$ should be 
a subgroup of a larger group, say $(N,L,e_R,\mu_R,\tau_R)\rightarrow i (N,L,e_R,\mu_R,\tau_R)$ (which should be imposed 
in addition to the groups given in Table~\ref{tab:flavourcontent}), so that the construction of the 
Lagrangian, Eq.~(\ref{eq:lagr}), remains unaffected.}.
It can be shown that the extrema of $\mathcal{V}_M$ correspond to
\begin{equation}
\nm = e^{i \theta} \vms, \quad \fm = e^{i \theta} g_i \vmt e^{i (\xi_3-\xi_1)}(1,0,0). 
\end{equation}
where $e^{i \theta}$ is an arbitrary phase\footnote{ This phase is similar to that obtained in Eq.~(\ref{eq:arbphase}) and it 
corresponds to the accidental $U(1)$ symmetry, $(\nm, \fm)\rightarrow e^{i \theta}(\nm,\fm)$, of $\mathcal{V}_M$. 
Higher order terms will break this $U(1)$. However, we do not study the $U(1)$-breaking terms since this phase is not
phenomenologically relevant to us. What is relevant is the relative phase between the VEVs of $\nm$  and $\fm$, i.e.~$\xi_3-\xi_1$, as 
can be inferred from Eq.~(\ref{eq:kdkm}).} and $g_i$ are the elements of $\rtp$ of $S_4$. Through SSB, one set among these extrema 
will become the VEVs, Eqs.~(\ref{eq:vm1}, \ref{eq:vm3}). 
The orientations of the triplet VEVs, i.e.~$\fc\propto(1,\om,\ob)$, $\fd\propto(-1,-1,1)$ and $\fm\propto(0,1,0)$, can be defined 
based on symmetry arguments alone, Section~\ref{sec:massmatrices}. It is not a coincidence that when we constructed the potentials,
we arrived at these VEVs naturally. A framework in which residual symmetries are utilised to fully define the VEVs of the irreducible
representations of flavons was recently proposed~\cite{Krishnan:2019ftw, Krishnan:2019xmk}. An interested reader may go through these 
references.
\section{Dirac neutrino mass matrix in the standard basis} \label{md}
Elements of the neutrino Dirac mass matrix in the standard basis are given by
\begin{eqnarray}
&&({\md^\prime})_{11}=\mw \bigg[ \sqrt{2} \cos \left(\frac{\theta_1}{2}\right)+i \left(\sqrt{2} k \sin \left(\frac{\theta_1}{2}\right)-\sqrt{2} \sin \left(\frac{\theta_1}{2}\right)\right)-\sqrt{2} k \cos
   \left(\frac{\theta_1}{2}\right) \bigg],  \\
&&({\md^\prime})_{12}=\mw ,\\
&&({\md^\prime})_{13}=\mw \bigg[-\sqrt{2} k \cos \left(\frac{\theta_2}{2}\right)+i \sqrt{2} k \sin \left(\frac{\theta_2}{2}\right) \bigg], \\
&&({\md^\prime})_{21}= \mw \bigg[-\frac{1}{2} \sqrt{\frac{3}{2}} \sin \left(\frac{\theta_1}{2}\right)+\frac{\cos \left(\frac{\theta_1}{2}\right)}{2 \sqrt{2}}+\frac{1}{2} \sqrt{\frac{3}{2}} k \sin \left(\frac{\theta_1}{2}\right)-\frac{k
   \cos \left(\frac{\theta_1}{2}\right)}{2 \sqrt{2}}+ \nonumber\\
&&~~~~~~~~~~~~~~i \left(-\frac{\sin \left(\frac{\theta_1}{2}\right)}{2 \sqrt{2}}-\frac{1}{2} \sqrt{\frac{3}{2}} \cos \left(\frac{\theta_1}{2}\right)+\frac{k \sin
   \left(\frac{\theta_1}{2}\right)}{2 \sqrt{2}}+\frac{1}{2} \sqrt{\frac{3}{2}} k \cos \left(\frac{\theta_1}{2}\right)\right) \bigg],\\
&&({\md^\prime})_{22}= \mw \bigg[\frac{3 k}{2}-\frac{1}{2}+i \left(\frac{\sqrt{3} k}{2}+\frac{\sqrt{3}}{2}\right) \bigg],  \\
&&({\md^\prime})_{23}= \mw \bigg[-\frac{1}{2} \sqrt{\frac{3}{2}} \sin \left(\frac{\theta_2}{2}\right)-\frac{3 \cos \left(\frac{\theta_2}{2}\right)}{2 \sqrt{2}}-\frac{3}{2} \sqrt{\frac{3}{2}} k \sin \left(\frac{\theta_2}{2}\right)-\frac{k
   \cos \left(\frac{\theta_2}{2}\right)}{2 \sqrt{2}}+ \nonumber \\
&&~~~~~~~~~~~~~~ i \left(\frac{3 \sin \left(\frac{\theta_2}{2}\right)}{2 \sqrt{2}}-\frac{1}{2} \sqrt{\frac{3}{2}} \cos \left(\frac{\theta_2}{2}\right)+\frac{k \sin
   \left(\frac{\theta_2}{2}\right)}{2 \sqrt{2}}-\frac{3}{2} \sqrt{\frac{3}{2}} k \cos \left(\frac{\theta_2}{2}\right)\right) \bigg],  \\
&&({\md^\prime})_{31}=\mw \bigg[\frac{1}{2} \sqrt{\frac{3}{2}} \sin \left(\frac{\theta_1}{2}\right)+\frac{\cos \left(\frac{\theta_1}{2}\right)}{2 \sqrt{2}}-\frac{1}{2} \sqrt{\frac{3}{2}} k \sin \left(\frac{\theta_1}{2}\right)-\frac{k
   \cos \left(\frac{\theta_1}{2}\right)}{2 \sqrt{2}}+ \nonumber\\
&&~~~~~~~~~~~~~~  i \left(-\frac{\sin \left(\frac{\theta_1}{2}\right)}{2 \sqrt{2}}+\frac{1}{2} \sqrt{\frac{3}{2}} \cos \left(\frac{\theta_1}{2}\right)+\frac{k \sin
   \left(\frac{\theta_1}{2}\right)}{2 \sqrt{2}}-\frac{1}{2} \sqrt{\frac{3}{2}} k \cos \left(\frac{\theta_1}{2}\right)\right) \bigg],   \\
&&({\md^\prime})_{32}= \mw \bigg[\frac{3 k}{2}-\frac{1}{2}+i \left(-\frac{\sqrt{3} k}{2}-\frac{\sqrt{3}}{2}\right)  \bigg] ,\\
&&({\md^\prime})_{33}= \mw \bigg[\frac{1}{2} \sqrt{\frac{3}{2}} \sin \left(\frac{\theta_2}{2}\right)-\frac{3 \cos \left(\frac{\theta_2}{2}\right)}{2 \sqrt{2}}+\frac{3}{2} \sqrt{\frac{3}{2}} k \sin \left(\frac{\theta_2}{2}\right)-\frac{k
   \cos \left(\frac{\theta_2}{2}\right)}{2 \sqrt{2}}+ \nonumber\\
&&~~~~~~~~~~~~~~ i \left(\frac{3 \sin \left(\frac{\theta_2}{2}\right)}{2 \sqrt{2}}+\frac{1}{2} \sqrt{\frac{3}{2}} \cos \left(\frac{\theta_2}{2}\right)+\frac{k \sin
   \left(\frac{\theta_2}{2}\right)}{2 \sqrt{2}}+\frac{3}{2} \sqrt{\frac{3}{2}} k \cos \left(\frac{\theta_2}{2}\right)\right) \bigg]~.
\end{eqnarray}

\section{CP asymmetry parameters in terms of Casas Ibarra parametrization} \label{CSI}
In an alternative approach the well known Type-I seesaw formula can be used to express the Dirac neutrino mass matrix 
in terms of the neutrino (light and heavy) mass eigen values, mixing angles and CP violations (both high energy and low energy).
The light neutrino mass matrix in the diagonal basis of RH neutrinos and charged leptons is obtained using Type-I seesaw mechanism   
as
\begin{equation}
M_{ss}^\prime= -M_D^\prime D_M^{-1}  {M_D^\prime}^T,
\end{equation}
where $M_D^\prime$  (Eq.~(\ref{stndb})) is the Dirac neutrino mass matrix in the diagonal basis of charged leptons and RH neutrinos and 
$V^T M_M V= D_M=diag(M_3,M_2,M_1)$. Using Eq.~(\ref{stndb}) it can be easily shown that 
\begin{equation}
 M_{ss}^\prime= V M_{ss} V^T ~.
\end{equation}
Again using the diagonalization condition of $M_{ss}$  (Eq.~(\ref{eq:msdiag})) it can be easily understood that the $M_{ss}^\prime$ matrix
is diagonalized by $\mathcal{U}=V U_{\rm BM} U_{23}$ which is nothing but the neutrino mixing matrix (conventionally represented using the 
PMNS parametrization $U_{PMNS}$) and the diagonalization equation is
\begin{equation}
\mathcal{U}^\dagger  M_{ss}^\prime \mathcal{U}^\ast =diag(m_1, m_2,m_3)=-D_m ~~({\rm say}) .
\end{equation}
Using the seesaw formula the above equation can be written as
\begin{equation}
\mathcal{U}^\dagger  M_D^\prime D_M^{-1}  {M_D^\prime}^T \mathcal{U}^\ast = D_m~.
\end{equation}
Now multiplying both sides (LHS and RHS) of the above equations (successively from left and right) by the inverse matrix of 
square root of $D_m$ we get 
\begin{equation}
 \left( \sqrt{D_m^{-1}} \mathcal{U}^\dagger M_D^\prime \sqrt{D_M^{-1}} \right) \left( \sqrt{D_M^{-1}} {M_D^\prime}^T \mathcal{U}^\ast 
\sqrt{D_m^{-1}} \right) = I ~~
 {\rm or} ~~R^T R =I,
\end{equation}
where $R$ is the orthogonal matrix given by 
\begin{equation}
R= \sqrt{D_M^{-1}} {M_D^\prime}^T \mathcal{U}^\ast \sqrt{D_m^{-1}} ~. \label{R}
\end{equation}
Therefore the Dirac neutrino mass matrix $(M_D^\prime)$ can be represented in terms of the experimentally measurable low energy
neutrino observables (three mass eigenvalues contained in $D_m$, three mixing angles, one Dirac type CP phase, two Majorana type CP phase 
contained in the $\mathcal{U}$ matrix), three heavy RH neutrino mass eigenvalues (contained in $D_M$) and three 
complex mixing angles\cite{Ibarra:2003xp} (which constitutes the orthogonal $R$ matrix) as 
\begin{equation}
M_D^\prime =  \mathcal{U} \sqrt{D_m} R^T \sqrt{D_M} ~.
\end{equation}
This type of representation of the Dirac neutrino mass matrix is known as Casas-Ibarra\cite{Casas:2001sr} parametrization and now we try to express the CP 
asymmetry parameters (both flavoured and unflavoured) by this modified parametrization in order to understand their dependence on low 
energy or high energy CP phases. The most general form of the flavoured CP asymmetry (Eq.~(\ref{epsi_intro})) parameter in terms of this new parametrization is 
obtained as
\begin{eqnarray}
\varepsilon^\alpha_i &=&\frac{1}{8 \pi v^2 \sum\limits_{n^\prime} m_{n^\prime}| R_{i n^\prime} |^2} 
\sum_{j \neq i}  M_j  \sum_{n,l,k}m_n \sqrt{m_k m_l} Im \left \{ R_{jn} R_{jl} R_{in}^\ast R_{ik}^\ast \mathcal{U}_{ k \alpha}^\dagger 
\mathcal{U}_{\alpha l} \right\} g_1(x_{ij})  \nonumber\\
& + & \frac{1}{8 \pi v^2 \sum\limits_{n^\prime} m_{n^\prime}| R_{i n^\prime} |^2} 
\sum_{j \neq i}  M_j  \sum_{n,l,k}m_n \sqrt{m_k m_l} Im \left \{ R_{in} R_{jl} R_{jn}^\ast R_{ik}^\ast \mathcal{ U}_{ k \alpha}^\dagger 
\mathcal{ U}_{\alpha l} \right\} g_2(x_{ij}), \label{flcp}
\end{eqnarray}
where $g_1(x_{ij})=f(x_{ij})+\frac{\sqrt{x_{ij}}(1-x_{ij})}{(1-x_{ij})^2 +{H_{jj}^\prime}^2/16 \pi^2 v^4}$ and 
$g_2(x_{ij})=\frac{(1-x_{ij})}{(1-x_{ij})^2 +{H_{jj}^\prime}^2/16 \pi^2 v^4}$. When we sum over the flavour index $\alpha$
the second term of Eq.~(\ref{flcp}) vanishes and the first term also gets little bit simplified due to the use of the unitary property of 
mixing matrix $\sum\limits_\alpha \mathcal{ U}_{ k \alpha}^\dagger \mathcal{ U}_{\alpha l} = \delta _{kl}$, as a result the unflavoured CP
asymmetry parameter becomes
\begin{equation}
\varepsilon_i = \frac{1}{8 \pi v^2 \sum\limits_{n^\prime} m_{n^\prime}| R_{i n^\prime} |^2} 
\sum_{j \neq i}  M_j  \sum_{n,k}m_n m_k  Im \left \{ R_{jn} R_{jk} R_{in}^\ast R_{ik}^\ast \right\} g_1(x_{ij})~. \label{uflcp}
\end{equation}
It is evident from Eq.~(\ref{uflcp}) that explicit dependence on the low energy CP phases (both Dirac and Majorana type) is absent 
in the case of the unflavoured leptogenesis. It signifies that CP asymmetry can be generated only by the nonvanishing high 
energy CP phases contained in $R$ even if the low energy CP phases (phases of the $\mathcal{U}$ matrix) turns out to be zero.
Above expression (\ref{uflcp}) dictates that the unflavoured asymmetry parameter will be nonzero
only if the elements of the orthogonal $R$ matrix are general complex numbers. If they are purely real or purely imaginary, the argument
of '$Im$' will be a real quantity which results in a vanishing CP asymmetry parameter. We have already shown that the model under consideration 
is capable of generating nonzero unflavoured CP asymmetry parameter which ensures that the elements of $R$ matrix in our case are
general complex numbers indeed. The argument can be presented in another way. The elements of $R$ will be purely real or purely imaginary 
when there is some specific kind of residual symmetry in $R$ matrix imposed due to invariance of Dirac and Majorana matrices
under some CP transformation (as shown in \cite{Chen:2016ptr}). However in our case although there are some residual symmetries (not due to 
CP kind of transformation ) in Dirac and Majorana type matrices, the orthogonal $R$ matrix does not enjoy such residual symmetry as a whole.
Therefore $R$ can be regarded as made up of three general complex angles. Thus $R$ matrix contributes non trivially to both unflavoured
and flavoured leptogenesis. As it can be understood from Eq.~(\ref{flcp}), in case of flavoured leptogenesis the CP asymmetry gets non zero contribution
from both high energy and low energy CP phases and it is difficult to say conclusively which contribution is more responsible for 
asymmetry generation.
\acknowledgments
Authors would like to thank Rome Samanta for useful discussions regarding Leptogenesis. M.C would like to acknowledge the 
financial support provided by SERB-DST, Govt. of India through the project {\bf EMR/2017/001434}.

\bibliographystyle{JHEP} 
\bibliography{tm1lepto1.bib}
\end{document}